\documentclass{aa}  
\usepackage{graphicx}
\usepackage{gensymb}
\usepackage{verbatim}
\usepackage{siunitx}
\usepackage{amssymb}
\usepackage{amsmath}
\usepackage{setspace}
\usepackage[colorlinks,allcolors=blue]{hyperref}
\usepackage[table,xcdraw]{xcolor}
\usepackage{tabularx}
\usepackage[normalem]{ulem} % for strikethrough using \sout
\usepackage{textcomp} % for \perthousand symbol
\usepackage{txfonts}

\newcommand{\LEt}[1]{\textcolor{red}{{{}}}}

\newcommand{\starforge}{{\small STARFORGE} }

\newcommand{\sedigism}{{\small SEDIGISM} }

\def \HIIdef {H{\,\textsc{ii}}}

\begin{document} 

   % \title{Synthetic molecular clouds in STARFORGE} 
   % \subtitle{I. Cloud Evolution and Properties}

   \title{A multiscale evolutionary study of molecular gas in STARFORGE} 
   \subtitle{I. Synthetic observations of SEDIGISM-like molecular clouds}
   
   \author{K. R. Neralwar\inst{1}\thanks{Member of the International Max Planck Research School (IMPRS) for Astronomy and Astrophysics at the Universities of Bonn and Cologne.},
          D. Colombo\inst{2},
          S. Offner\inst{3},
          A. Karska\inst{4,1},
          M. Figueira\inst{5,1},
          F. Wyrowski\inst{1},
          S. Neupane\inst{1},
          J. S. Urquhart\inst{6},
          A. Duarte-Cabral\inst{7}
          }      

   \institute{Max-Planck-Institut f\"ur Radioastronomie, Auf dem H\"ugel 69, 53121 Bonn, Germany \\ email: \texttt{kneralwar@mpifr-bonn.mpg.de}
    \and
    Argelander-Institut f\"ur Astronomie, Auf dem H\"ugel 71, 53121 Bonn
    \and
    Department of Astronomy, The University of Texas at Austin, Austin, TX 78712, USA
    \and
    Centre for Modern Interdisciplinary Technologies, Nicolaus Copernicus University in Torun, Wileńska 4, 87-100 Toruń, Poland 
    \and
    National Centre for Nuclear Research, Pasteura 7, 02-093, Warszawa, Poland
    \and
    Centre for Astrophysics and Planetary Science, University of Kent, Canterbury, CT2\,7NH, UK
    \and
   School of Physics \& Astronomy, Cardiff University, Queen's building, The parade, Cardiff CF24 3AA, UK
   }

  \date{Received XXX; accepted XXX}

  \abstract 
  {Molecular clouds (MCs) are active sites of star formation in galaxies, and their formation and evolution are largely affected by stellar feedback. This includes outflows and winds from newly formed stars, radiation from young clusters, and supernova explosions.
  High-resolution molecular line observations allow for the identification of individual star-forming regions and the study of their integrated properties. Moreover, state-of-the-art simulations are now capable of accurately replicating the evolution of MCs including all key stellar feedback processes. 
  We present $^{13}$CO(2-1) synthetic observations of the \starforge simulations produced using the radiative transfer code RADMC-3D, matching the observational setup of the \sedigism survey. 
  From these, we identified the population of MCs using hierarchical clustering and analysed them to provide insights into the interpretation of observed MCs as they evolve. The flux distributions of the post-processed synthetic observations and the properties of the MCs, namely radius, mass, velocity dispersion, virial parameter and surface density, are consistent with those of \sedigism. Both samples of MCs occupy the same regions in the scaling relation plots; however, the average distributions of MCs at different evolutionary stages do not overlap on the plots. This highlights the reliability of our approach in modelling \sedigism and suggests that MCs at different evolutionary stages contribute to the scatter in observed scaling relations. 
  We study the trends in MC properties, morphologies, and fragmentation over time to analyse their physical structure as they form, evolve, and are destroyed. MCs appear as small, diffuse cloudlets in early stages, followed by their evolution to filamentary structures, before being shaped by stellar feedback into 3D bubbles and getting dispersed. These trends in the observable properties of MCs are consistent with other realisations of simulations and provide strong evidence that clouds exhibit distinct morphologies over the course of their evolution.}
  
   \titlerunning{STARFORGE synthetic MCs}
   \authorrunning{K. R. Neralwar et al.}
   \maketitle
%%--------------------------------------------------------------------

%-------------------------------------------------------------------

\section{Introduction}

The molecular gas in the interstellar medium (ISM) is hierarchically clustered in the form of molecular clouds \citep{dobbs2014prpl.conf....3D, chevance2023}. 
Molecular clouds are turbulent, magnetically supercritical \citep{crutcher_2012} structures with sizes $\sim 1$--$200$ pc \citep{ballesteros-paredes2020, duarte_cabral2021}, mass M $\sim 10^2$--$10^7 ~ \mathrm{M_\odot}$ \citep{rebolledo_2012}, surface densities $\Sigma \sim 1$--$1000 ~ \mathrm{M_\odot \; pc^{-2}}$ \citep{barnes_2018} and virial parameter ($\alpha_{vir}$) around unity \citep{fukui_2008, roman_duval_2010}. 
They show various morphologies \citep{neralwar2022A&A...663A..56N}, with filaments being the most ubiquitous \citep{andre_2010, ATLASGAL_filaments, ana_2017, arzoumanian_2019, colombo2021A&A...655L...2C, priestley2021}.

The evolution of molecular clouds is often presented in the form of two scenarios, i.e. regulated by either SN-driven turbulence \citep{padoan2016ApJ...822...11P, lu2020ApJ...904...58L} or hierarchical gravitational collapse \citep{ballesteros-paredes2011MNRAS.411...65B, vazquez-semadeni_2019MNRAS.490.3061V}. These mechanisms lead to the fragmentation and formation of dense cores, which further lead to the formation of protostars. Protostars accrete gas from the surrounding ISM, leading to feedback in the form of bipolar outflows \citep{bally2016}. These outflows (jets) are relatively collimated structures that heat and compress the gas as they interact with the surrounding ISM up to pc-scales \citep[e.g., ][]{duarte-cabral2012,skretas2023,karska2025}. Outflows can vary significantly in energetics, with momentum rates between $10^{-5}$ and $10^{-2} ~\mathrm{M}_\odot ~ \mathrm{km} \mathrm{s}^{-1} ~ \mathrm{yr}^{-1}$, for low-mass and O-type stars, respectively \citep[e.g.][]{duarte-cabral2013,maud2015MNRAS.453..645M}.

As the forming protostars move onto the main sequence stage, they begin to drive isotropic stellar winds \citep{vink2024arXiv240616517V}, which help clear out the remaining gas in their envelope. Stellar winds form bubble-shaped cavities around stars \citep{weaver1977ApJ...218..377W, fier16, ali2022MNRAS.510.5592A} by releasing a significant amount of kinetic energy into the ISM ($\sim 10^{48}$~erg over the lifetime of the bubble, \citealt{lui21}). In addition to stellar winds, high-energy radiation from massive stars ($M\ge 8$~$M_{\odot}$) ionises the surrounding gas, releasing thermal energy $\sim 10^{46}$~erg, and forming ionised (\HIIdef) regions \citep{simpson2012MNRAS.424.2442S, fig17, santoro2022A&A...658A.188S}. 
Stellar winds and photoionising radiations are active during the main-sequence phase of stars (few Myrs) and
% The combined effect of stellar winds and photoionisation 
play a major role in dispersing molecular clouds \citep{rosen2020SSRv..216...62R, chevance2020MNRAS.493.2872C}.
The last form of feedback, and perhaps one of the most important in setting the global conditions of the ISM in galaxies, are supernova explosions from massive stars \citep{geen2015MNRAS.448.3248G, lucas2020MNRAS.493.4700L}. Supernovae inject a large amount of energy ($\sim 10^{51}$ ergs) and drive the supersonic turbulence in the ISM on tens to hundreds of parsec scales. \citep{lu2020ApJ...904...58L, dubner2015A&ARv..23....3D}. 

Although various stellar feedback mechanisms have a significant impact on cloud formation, evolution, and dissolution \citep{chevance2023}, observational studies of molecular clouds rarely attempt to distinguish between clouds affected by different forms of feedback. Clouds are often generalised into a single population of quasi-static entities in near equilibrium \citep{colombo2019, duarte_cabral2021}, and are collectively analysed using scaling relation, such as those of \cite{larson1981} and \cite{heyer2009}. However, they are not necessarily hydrostatic structures; as both turbulence support (static clouds) and global hierarchical collapse (GHC; dynamical clouds), cloud formation models produce the same observational signatures \citep{vazquez-semadeni2024arXiv240810406V}. Studying the effects of local feedback events on cloud properties might provide some evidence in support or opposition to these models. 

The last decade has led to several high-resolution surveys of the Milky Way, revealing the intricate structure of molecular clouds and allowing their properties to be resolved in detail (COHRS \citealt{dempsey2013ApJS..209....8D}, CHIMPS \citealt{rigby2016MNRAS.456.2885R}, SEDIGISM \citealt{schuller2021}, OGHReS \citealt{urquhart2024MNRAS.528.4746U, urquhart2025MNRAS.539.3105U}). Complementing such observations, state-of-the-art Giant Molecular Cloud (GMC) simulations are now able to model the ISM at high resolution (e.g. SICLL \citealt{walch2015MNRAS.454..238W, seifried2017MNRAS.472.4797S}, STARFORGE \citealt{grudic2021}, FIRE \citealt{hopkins2023MNRAS.519.3154H}), while including most of the physical phenomena related to star formation. 
Comparison of observations and simulations requires mapping the simulations into observational space using radiative transfer \citep[e.g. RADMC-3D ][]{dullemond2012ascl.soft02015D} to model the emission that the simulated gas and stars would produce.

One goal of the paper is to use state-of-the-art simulations that incorporate all the relevant physics of star formation to produce synthetic observations that closely resemble observational data.
We generate synthetic observations from the \starforge (Star Formation in Gaseous Environments) simulations following the observational setup of the \sedigism survey \citep{duarte_cabral2021}. Using RADMC-3D, we produce the $^{13}$CO(2-1) line-emission cubes from the simulations and use a dendrogram-based cloud identification technique to extract molecular clouds (MCs).  
We thus evaluate the extent to which current simulations can reproduce the observed MCs. 
This work also aims to analyse the observable properties of MCs as they evolve and determine whether their distributions vary across different evolutionary stages, in order to provide some insight into the interpretation of observational molecular cloud surveys. 
As the cloud properties are comparable to {\small SEDIGISM}, we are able to provide a direct comparison of the cloud properties, as they evolve and are affected by stellar feedback. 

The structure of the paper is as follows. Section \ref{sec: data} describes the \starforge simulations used for our analysis and the \sedigism survey, which we use as a observational comparative benchmark. In Section \ref{sec: methods}, we describe the methodology to create the synthetic observations from \starforge (Sect. \ref{sec: methods radiative transfer}), along with the post-processing of the data (Sect. \ref{sec: methods postprocessing}). We then describe the dendrogram algorithm used to extract stuctures from these synthetic observation cubes in Sect. \ref{subsec: Dendrograms}, followed by the definition of MC properties. 
We present our results on the comparison between clouds from the simulations and those from \sedigism in Sect. \ref{sec: SEDIGISM comparison}. We investigate the evolution of the properties and morphologies of clouds over time and connect them to formation of stars in Sect. \ref{sec: time evolution of clouds}.
We analyse the scaling relations to compare the correlation between \starforge and \sedigism cloud properties and understand the time evolution of clouds on these plots in Sect. \ref{sec: scaling relations}.
In Sect. \ref{sec: discussions}, we explain the trends in the properties, morphology, and substructures of MCs over time and connect them to their observed counterparts.
We conclude by summarising our work in Sect. \ref{sec: summary}.

\section{Data} \label{sec: data}

\subsection{STARFORGE}\label{subsec: starforge}

The \starforge\footnote{\url{https://www.starforge.space}}
simulations are three-dimensional radiation magnetohydrodynamic (MHD) simulations that follow the evolution of GMCs, achieving spatial resolutions down to a few tens of AU. These simulations model the formation, evolution, and dynamics of individual stars within a GMC, incorporating all forms of stellar feedback: jets, radiation, stellar winds, and supernovae. 
The simulation framework is built on the {\small GIZMO} code \citep{gizmo}, which uses a Lagrangian meshless finite mass (MFM) method to solve MHD equations \citep{hopkins2016MNRAS.455...51H}. A comprehensive explanation of the numerical methods and validation tests is provided in \cite{grudic2021}.

We use the M2e4a2 suite of \starforge simulations (Table 1 of \citealt{guszejnov2022}). It follows the evolution of a $20~000$ M$_\odot$ GMC\footnote{The GMC hereafter refers to this simulated GMC, as defined by the \starforge collaboration.} to $\sim$ 11 Myr while saving all the properties every 24.7 kyr, thus producing a total of 445 snapshots. Of these, we analyse the 410 snapshots that have significant $^{13}$CO(2-1) emission to identify dendrogram structures. The GMC is initiated as a uniform surface density sphere with $R=10$~pc, $\alpha_{vir}=2$, and T = 10 K, 
surrounded by diffuse medium (density contrast of 1000) in a (100 pc)$^3$ box\footnote{The equations to calculate these properties are described in \cite{guszejnov2021MNRAS.502.3646G}}. 
The gas is initiated as fully atomic, but the ionisation state rapidly converges to local equilibrium, such that the interior of the cloud rapidly becomes fully molecular.
The calculation we analyse includes the heating and cooling contributions from all key molecular, atomic, nebular, and continuum processes \citep[see][for more detail]{Hopkins2018}.  Although the simulation does not explicitly follow the formation and destruction of H$_2$, it computes the molecular gas fraction in each cell using a fitting function that depends on the gas metalicity and surface density \citep{KrumholzGnedin2011}. We then derive the number density of molecular hydrogen by assuming that the molecular gas mass in each cell is composed of molecular hydrogen.

We restricted our analysis to the gas within the inner 60 pc of the simulation box, as that is sufficient to capture the $^{13}$CO(2-1) emission througout the time evolution studied here.
As the simulation progresses, the GMC collapses under self-gravity to form protostars (at 0.8 Myr) with outflows. It then forms stars with photoionising radiation (at 2.7 Myr) and stellar winds (at 3.6 Myr), which disperse most of the GMC before the first supernova (at 9.8 Myr) occurs. This follows a global hierarchichal collapse (GHC) scenario with various stellar feedback mechanisms that can contribute to the injection of local turbulence and provide support against local collapse \citep{grudic2022, guszejnov2022}.

\subsection{SEDIGISM} \label{sec: data sedigism}

The Structure, Excitation, and Dynamics of the Inner Galactic InterStellar Medium (\sedigism; see \citealt{schuller2017, schuller2021} for an overview) survey spans an area of 84 deg$^2$ within the Galactic longitude range of $ -60^\circ \leq l \leq +18^\circ $ and latitude $ |b| \leq 0.5^\circ $ (with some regional variations). It includes multiple molecular tracers, specifically targeting the $ J = 2 $-$ 1 $ transitions of $ ^{13} $CO and C$ ^{18} $O. Observations were conducted from 2013 to 2017 using the 12-meter Atacama Pathfinder Experiment (APEX) telescope \citep{gusten2006}.
The survey provides a contiguous dataset divided into 77 datacubes, each covering approximately $ 2^\circ \times 1^\circ $, with a velocity coverage of $-200$ to 200 $ \mathrm{km \; s}^{-1} $ and a pixel size of $\ang{;;9.5}$. 
The first data release (DR1) features $^{13}$CO observations with a full width at half maximum (FWHM) beam size of $\ang{;;28}$ and a typical 1-$\sigma$ sensitivity of 0.8–1.0 K per 0.25 $ \mathrm{km \; s}^{-1} $.  

Using this dataset, \citet{duarte_cabral2021} constructed a catalogue of 10,663 MCs with their physical properties, further updated by \cite{urquhart2021MNRAS.500.3050U}, \cite{ colombo2022A&A...658A..54C} and \cite{neralwar2022A&A...663A..56N}.
MCs were identified using the Spectral Clustering for Interstellar Molecular Emission Segmentation (\textsc{scimes}) algorithm (v.0.3.2, \citealt{colombo2015, colombo2019}).
\cite{duarte_cabral2021} also defined a science sample that consists of well-resolved clouds with reliable distance estimates. We selected MCs from the science sample that are at distances between 2.5 and 3.5~kpc away for our comparative analysis. 
This distance rages provides a sufficiently large sample size (of 835 clouds), with a consistent physical resolution ($\sim 0.3 - 0.5$ pc) and minimises the effects of different sensitivities at different distances.
In the following, we will refer to these clouds as \sedigism clouds. We use the deconvolved equivalent radius (\text{radius\_dec\_pc}\footnote{Column name in the catalogue in \citet{duarte_cabral2021}}), cloud mass (\text{Mass}), velocity dispersion (\text{sigv\_kms}), virial parameter (\text{alpha\_vir}) and surface density (\text{Surf\_density\_dec\_Mpc2}) to compare \sedigism clouds to synthetic MCs identified in this paper (Sect. \ref{sec: SEDIGISM comparison}).

\section{Methods} \label{sec: methods}

\subsection{Radiative transfer}\label{sec: methods radiative transfer}

We used the RADMC-3D \citep[version 2.0, ][]{dullemond2012ascl.soft02015D} radiative transfer code on \starforge simulation to obtain the position-position-velocity (PPV) $^{13}$CO(2-1) emission cubes that mimic the \sedigism survey. 
RADMC-3D uses the simulation data as input, including the distribution of the density, temperature, composition of the gas, and sources of radiation, and generates a three-dimensional grid of points that are used to sample the environment and calculate the radiative transfer. We briefly outline the procedure in the following.

\subsubsection{Data preprocessing}\label{subsec: methods preprocessing}

We interpolated the \starforge data to a uniform grid with resolution $480 \times 480$ as input for RADMC-3D\footnote{\url{https://github.com/Kartik-Neralwar/gizmo_carver}}, matching the resolution in \sedigism at a distance $\sim$ 3 kpc. The RADMC-3D inputs are gas temperature, velocity, and the number densities of $^{13}$CO and H$_2$, which acts as a collision partner. The collional rate coefficients for $^{13}$CO are provided by the Leiden Atomic and Molecular Database (LAMDA{\footnote{\url{https://home.strw.leidenuniv.nl/~moldata/datafiles/13co.dat}}}). We obtained the $^{13}$CO number density from the H$_2$ number density estimated from the molecular gas fraction using an abundance ratio $\mathrm{H_2/CO} = 10^{4}$ \citep[typical of the inner Milky Way, ][]{Dame2001ApJ...547..792D, bolatto2013ARA&A..51..207B} and a constant isotopic ratio $\mathrm{^{12}CO/^{13}CO} = 42.6$ \citep[at the Galactocentric radius of 5 kpc]{jacob2020A&A...640A.125J}. We also applied a freeze-out criterion by setting the $^{13}$CO abundance ratio to 50\% in regions with gas temperature below 17 K and gas density above $10^5 \mathrm{cm}^{-3}$ \citep{caselli1999ApJ...523L.165C, lippok2013A&A...560A..41L, roueff2021A&A...645A..26R}.
However, various temperature and density thresholds tested for the freeze-out criterion did not strongly affect the distribution of the $^{13}$CO(2-1) emission studied in this work.
RADMC-3D calculates line profiles based on thermal broadening by default, but allows the inclusion of turbulent widths (microturbulence\footnote{{\url{https://www.ita.uni-heidelberg.de/~dullemond/software/radmc-3d/manual_radmc3d/lineradtrans.html\#input-the-local-microturbulent-broadening-optional}}}).
We calculated the microturbulence as the product of the velocity gradient in a cell and the size of the cell, where the velocity gradient for the cells is obtained using the numerical differentiation functionality of the \texttt{MESHOID}\footnote{\url{https://github.com/mikegrudic/meshoid}} script.

\subsubsection{RADMC-3D} \label{subsec: radmc3d}

RADMC-3D calculates line emission by obtaining the level population in each grid cell based on a line transfer model. We used the non-LTE approximation together with the Large Velocity Gradient (LVG) mode and the Escape Probability method\footnote{RADMC-3D lines\_mode = 3}. 
The $^{13}$CO(2-1) transition emits at 220.398\,GHz or 1360.227\,$\mu$m. We used this as the central wavelength to obtain PPV cubes with 65 channels separated by 0.25 kms$^{-1}$, leading to a velocity coverage of $\pm$ 8\,kms$^{-1}$. 
We used the RADMC-3D \texttt{image} option with \texttt{loadlambda} to create PPV cubes of size $65 \times 448 \times 448$ pixels 
at a distance of 3 kpc 
and a pixel size of 9.5\arcsec (corresponding to 60 pc), in order to be directly comparable to the \sedigism clouds (see Sect. \ref{sec: data sedigism})
We also made use of second-order integrations while creating the image to avoid pixelation. For each snapshot, the $^{13}$CO(2-1) cubes were constructed for three different lines of sight, along the $x$, $y$ and $z$ directions (App. \ref{app: projection axes}), using the phi=\{90, 0, 0\} and incl=\{90, 90, 0\} parameters respectively. %This procedure was performed for each of the 410 snapshots resulting in a total of 1230 synthetic datacubes.

\begin{figure}
    \centering
    \includegraphics[width=0.49\linewidth]{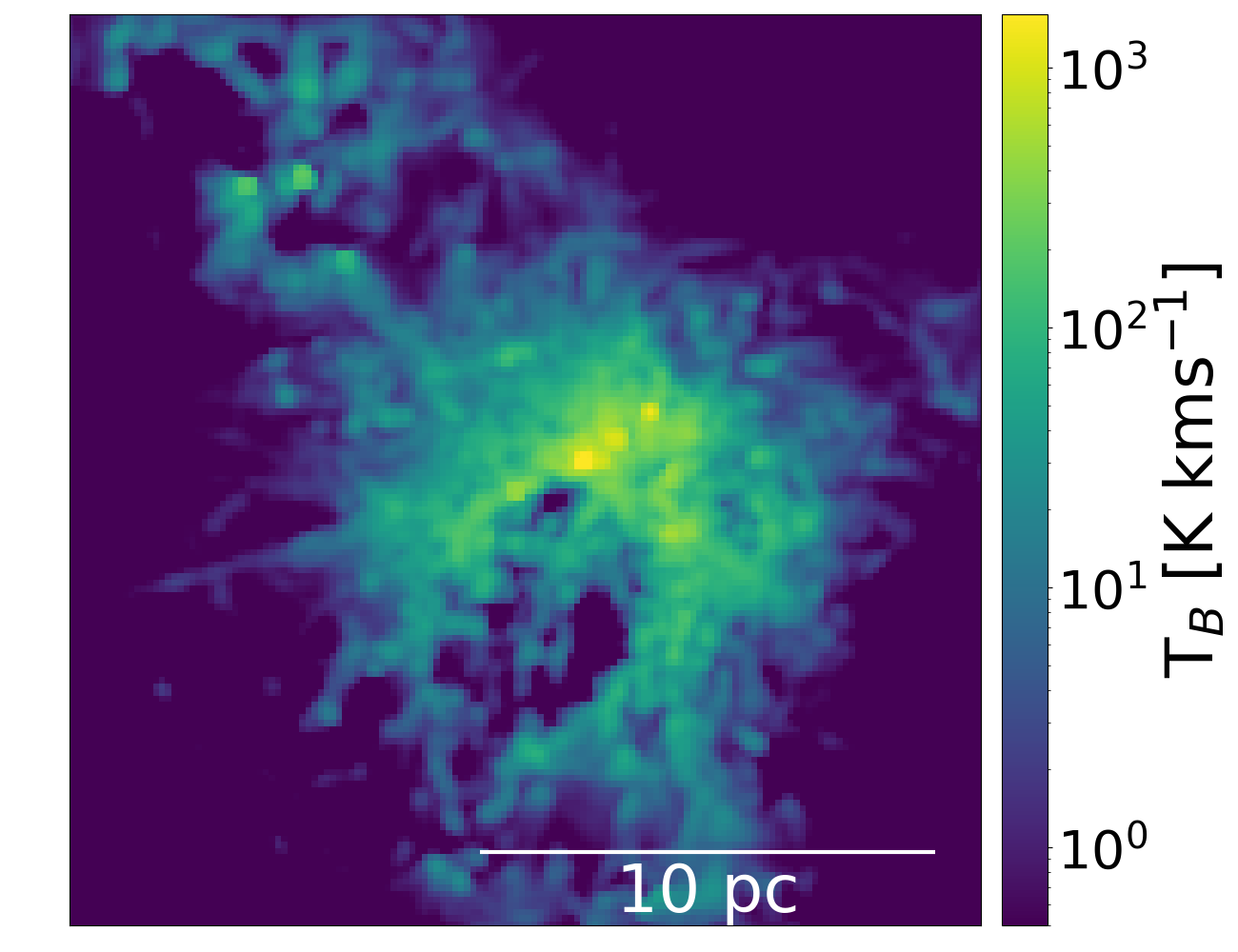}
    \includegraphics[width=0.49\linewidth]{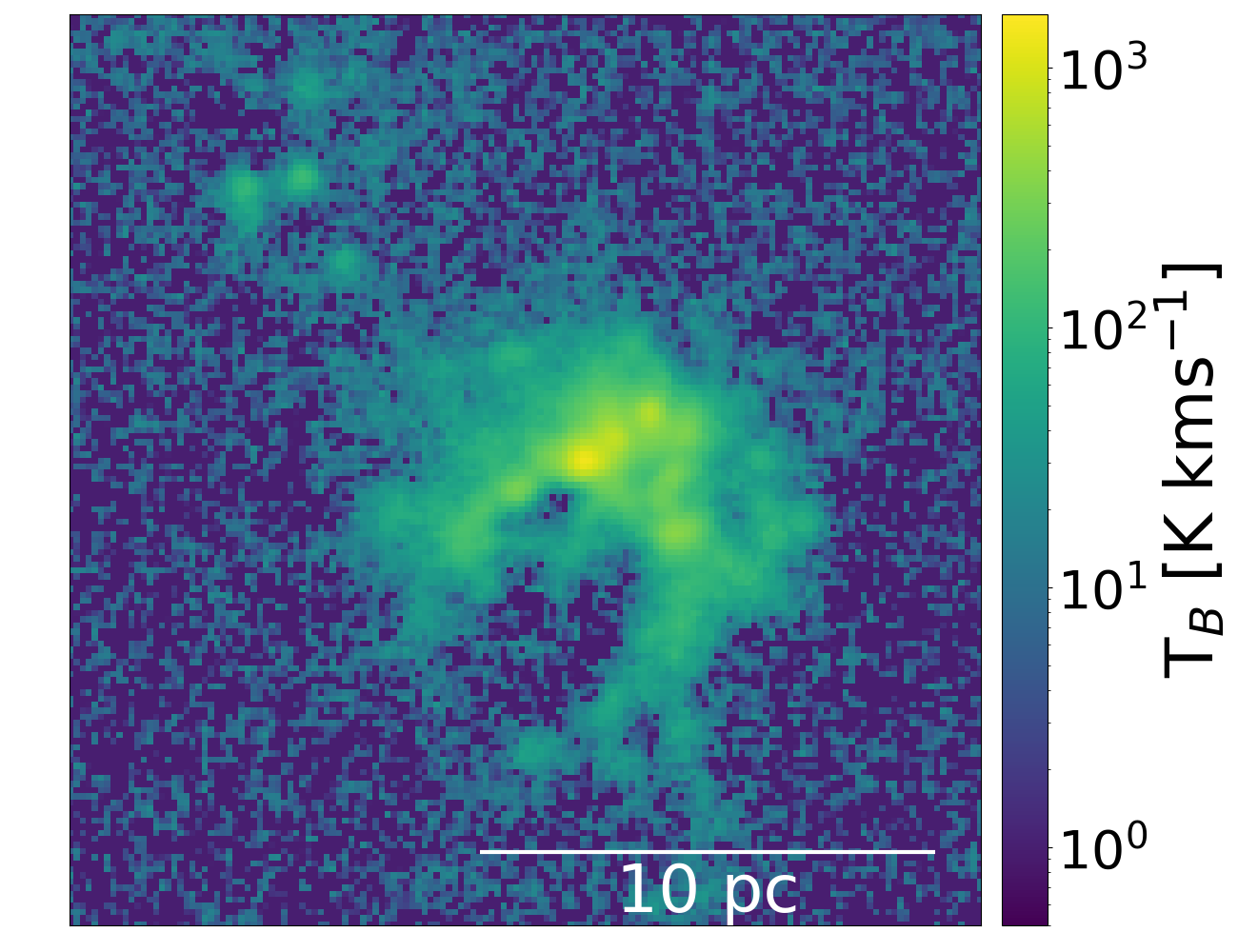}
    \caption{Left: A synthetic integrated $^{13}$CO(2-1) emission map created with RADMC-3D. Right: The same map after convolution and with noise of sigma = $0.2 \pm 0.02$ K.}
    \label{fig: raw cube}
\end{figure}

\subsection{PPV cube post-processing} \label{sec: methods postprocessing}

We post-processed the output PPV cube from RADMC-3D (Fig. \ref{fig: raw cube}, left) to replicate the characteristics of \sedigism observations using the following procedure. First, the cube is convolved with a 2D Gaussian kernel with a full width at half maximum (FWHM) of 28\arcsec, corresponding to the APEX telescope beam at 220.398 GHz. The data from the noisy (first and last 50) channels of the \sedigism G305 cube is to added to the cube as noise. %. 
To improve the signal-to-noise ratio, the data is spectrally smoothed at 0.5 kms$^{-1}$ resolution and resampled to 0.25 kms$^{-1}$ \citep[Sect. 3.1.1 in][]{duarte_cabral2021}. An example of an original and post-processed cube is shown in Fig. \ref{fig: raw cube}. 

To suppress the noise and prepare the data for the analysis of dendrograms, we applied the dilated masking method described in \cite{grishunin2024A&A...682A.137G}, resulting in clean masked cubes (App. \ref{app: GMC and MC imshow images}). 
We perform two masking iterations as \cite{grishunin2024A&A...682A.137G} for robust data cleaning, with s2n\_low = 2, s2n\_high = 4, and s2n\_vel = 3 for the first iteration and s2n\_low = 2, s2n\_high = 5, and s2n\_vel = 3 for the second. The s2n\_low follows the min\_val parameter in \cite{duarte_cabral2021}, setting the value of pixels whose emission is lower than 2$\sigma$ of the local noise to zero.
The higher value of s2n\_high in the second iteration provides a stricter constraint to remove any spurious sources.
These data processing steps allow us to closely replicate the observed data set in \sedigism (Sect. \ref{sec: SEDIGISM comparison}).
Of the 445 snapshots, the first 35 snapshots (up to $\sim$ 1 Myr) have no detectable $^{13}$CO(2-1) above the s2n\_low limit. Our analysis is therefore limited to 410 snapshots, resulting in a total of 1230 synthetic datacubes for the three projections. Excluding the early snapshots minimises potential biases in the reproduction of the synthetic line emission that could arise from the initial settling of the cloud. By $\sim$ 1 Myr, the gas density distribution reaches a quasi-equilibrium state, satisfying a log-normal distribution, which is expected for supersonically turbulent gas \citep{padoan1997ApJ...474..730P, lane2022MNRAS.510.4767L}. %ensures that the gas in our identified MCs is indeed mostly molecular and unaffected by the residual effects of the initial conditions in the simulation.}

\subsection{Dendrograms and cloud properties} \label{subsec: Dendrograms}
% \subsection{Cloud extraction and properties} \label{subsec: Dendrograms}

Dendrograms describe the distribution and nesting of isosurfaces in data cubes and have long been used to discretise molecular gas emission at different scales in observations \citep{colombo2019,duarte_cabral2021} and simulations \citep{offner2022MNRAS.517..885O}.
We used dendrograms \citep{rosolowsky2008ApJ...679.1338R} to segment the molecular gas in each snapshot into leaves, branches, and trunks. 
Leaves are structures formed by single local maxima in the gas distribution and are nested in branches, which in turn are nested in trunks. The trunks can be isolated structures without substructures or hierarchical structures with multiple substructures (App. \ref{app: hier isolated}). 
Dilated masking sets the value of all noisy voxels to zero, setting the lower threshold of the dendrograms. We built the dendrograms using min\_val = 0 and n\_delta = 2 $\sigma_{rms}$, where $\sigma_{rms} = 0.87$ represents the mean rms noise in our data (Sect. \ref{sec: methods postprocessing}). We set the min\_pix using n\_area = 3 and n\_vchan =2, such that all structures are both spatially (i.e. at least 3 beam) and spectrally (span at least 2 velocity channels) resolved.
We find a total of 3710 hierarchichal trunks and refer to them as MCs throughout this work. The entirety of detectable $^{13}$CO(2-1) emission in a snapshot is referred to as the ``molecular gas complex'' or the ``entire $^{13}$CO(2-1) emission''.  
Additionally, the dendrograms store the information about the nested structures (descendants) for every structure. We perform a quantitative analysis of these substructures (descendants) in each MC in Sect. \ref{sec: fragmentation}.

% \begin{table*}[]
%     \centering
%     \caption{Definitions and counts of gas structures in the paper.}
%     \begin{tabular}{|l|l|l|}
%     \hline
%     Giant molecular cloud (GMC) & Entire $^{13}$CO(2-1) emission in a snapshot cube  & 1230       \\ \hline
%     Molecular clouds (MCs)       & Hierarchichal trunks of dendrogram & 3710   \\ \hline
%     Substructures                & Descendants (branches) of MCs & 27928 \\ \hline
%     \end{tabular}
%     \label{tab: gmc mc substructures}
% \end{table*}

% \subsection{Cloud properties}\label{subsec: cloud properties}

The dendrogram analysis returns a catalogue of structures\footnote{\texttt{ppv\_catalog}: \url{https://dendrograms.readthedocs.io/en/latest/api/astrodendro.analysis.ppv_catalog.html}} with their properties \citep{rosolowsky2006PASP..118..590R}, which we use to derive other MCs properties assuming a heliocentric distance of 3 kpc. The most useful directly-measured properties are the projected footprint area ($A$), the velocity dispersion ($\sigma_r$), and the total brightness temperature (T$_B$). 
The effective radius is defined as $R_{eff}$ = $\sqrt{A/\pi}$ and the deconvolved radius as $R$ = $\sqrt{R_{eff}^2 - R_{beam}^2}$, where $R_{beam} = 0.2$ pc is the physical size of the beam\footnote{\sedigism beam (28\arcsec) FWHM at 3 kpc is 0.4 pc}. 
The luminosity mass is estimated from the luminosity ($L$) as $M_{lum} ~[\mathrm{M}_\odot]= \alpha_\mathrm{CO} ~ L ~ [\mathrm{L}_\odot]$, where $\alpha_\mathrm{CO}$ = 22.43 M$_\odot$ (K km s$^{-1}$)$^{-1}$ pc $^{-2}$, estimated using $X_{^{13}\mathrm{CO}(2-1)} = 1^{+1}_{-0.5} \times 10^{21} \mathrm{cm}^{-2}$ (K km s$^{-1})^{-1}$ to be consistent with \sedigism. 
From this we derive the surface mass density $\Sigma = M_{lum}/A$ and the virial parameter $\alpha_{vir} = 5\sigma_v^2R/GM_{lum}$, assuming a spherical and uniform cloud.
% This formulation of $\alpha_{vir}$ only takes into account the balance between kinetic and gravitational energies, assuming a spherical geometry and uniform density.
The analysis presented in this paper uses deconvolved properties, although global trends are virtually unchanged by this choice.

We obtained the true molecular gas mass ($M$) directly from the simulation by projecting the H$_2$ masses of gas cells onto the RADMC-3D and summing over the pixels associated with a given dendrogram structure. 
The true molecular gas mass is not subject to observational biases or limitations (e.g. $\alpha_{\mathrm{CO}}$ factor) and includes the CO-dark and freeze-out regions.
% This represents the true mass and is not subject to observational biases or limitations (e.g. $\alpha_{\mathrm{CO}}$ factor).
% Therefore, it also includes the mass from CO-dark regions and the mass that is omitted because of the CO freeze-out. 
We used this mass to study the time evolution of clouds in Sect. \ref{sec: prop vs snap}, but their derived properties are calculated using $M_{lum}$ to be consistent with the observations.
\starforge also tracks the positions, mass, ages, and evolutionary stages of individual stars and protostars \citep{grudic2021}. We record the number of newborn stars (protostars and stars younger than 250 kyr) and the main-sequence stars (with $M > 2 \, \mathrm{M}_\odot$) in each snapshot and use them to investigate the MC fragmentation in Sect. \ref{sec: fragmentation}.
% In \starforge, main-sequence stars with masses above $2 \, M_\odot$ produce stellar winds; we refer to these as main-sequence stars hereafter. The two populations are used in Sect. \ref{sec: fragmentation} to investigate the fragmentation of MCs.

% \subsection{RJ plots}\label{subsec: rj plots}

We analyse cloud morphologies using the $RJ$ plots \citep{clarke2022MNRAS.516.2782C} algorithm, which is an extension of $J$ plots \citep{jaffa2018MNRAS.477.1940J}. These algorithms classify pixelated structures into different morphologies based on the relationship between their mass distribution and moment of inertia. 
% We use the $RJ$ plots \citep{clarke2022MNRAS.516.2782C} algorithm to study the morphological properties of our sample of clouds across the entire time evolution of the simulation. 
% $RJ$ plots is a method to classify and quantify a pixelated structure into different morphologies based on the degree of elongation and central concentration. It is an improved version of the automated morphological classification technique $J$ plots \citep{jaffa2018MNRAS.477.1940J}, aimed at a more unbiased classification. These algorithms classify structures using the relationship between the mass distribution in a structure and its moment of inertia.
The $J$ moments $J_1$ and $J_2$ are obtained by compairing the principal moments ($I_{1} \, \& \, I_2$) with those of a uniform surface density disk $\left(I_0 = \mathrm{\frac{AM}{4 \pi}} \right)$ of the same area and mass; 
$
    J_i = \frac{I_0 - I_i}{I_0 + I_i}\; %, \quad
    \{i = 1, 2\}.% \; .
$
% The $J_2 = J_1$ and $J_2 = -J_1$ lines represent the degree of elongation and the degree of central concentration of the structures, respectively. 
The $RJ$ moments $R_1$ and $R_2$ are obtained by rotating the $J$ plot 45 degrees in the anticlockwise direction. $R_2$ is further normalised to remove the parameter space constraints given by $|J_i| \leq 1$, resulting in
$
    R_1 = \frac{J_1 - J_2}{2}
    % \label{eqn: rj1}
$
 and 
$
    R_2 = \frac{J_1 + J_2}{\sqrt2 (\sqrt2 - R_1)} .
    % \label{eqn: rj2}
$
% $R_1$ and $R_2$ thus represent the degree of elongation and central concentration respectively.
$R_1 = 0$ corresponds to a perfectly circular structure, with increasing values indicating progressively higher degrees of elongation.
$R_2$ measures the weight distribution of a structure compared to its centre of weight. The positive and negative values of $R_2$ represent centrally overdense and underdense structures, respectively. 

\section{Results}\label{sec: results and disussions}

%In this section, we analyse the properties of the molecular clouds.

\subsection{Validation of the synthetic dataset: comparison with SEDIGISM} \label{sec: SEDIGISM comparison}

\begin{figure}
    \centering
    \includegraphics[width = 0.9\linewidth]{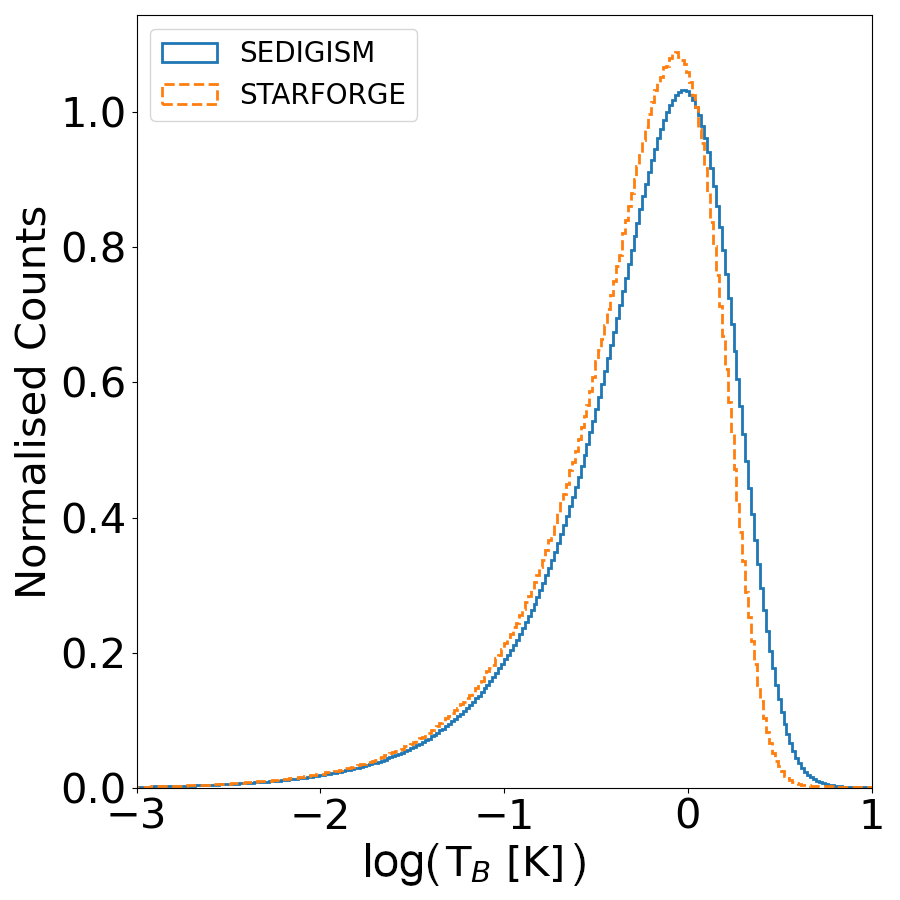}
    \caption{Comparison between the flux values for the noisy pixels from all the \starforge snapshots along all three projections and the complete \sedigism survey ($^{13}$CO(2-1) emission).}
    \label{fig: hist flux compare}
\end{figure}

Before we can study the evolution of MCs using our sample of synthetic clouds, we first need to ensure that our simulated sample is representative of the observed clouds, in terms of properties and parameter space probed. For that purpose, here we compare our synthetic 
%In this section, we compare our synthetic 
MCs with those observed as part of the \sedigism survey \citep{duarte_cabral2021}. Figure \ref{fig: hist flux compare} shows the flux distribution in each pixel for the \sedigism $^{13}$CO cubes %\footnote{This refers to the $2^\circ \times 1^\circ$ PPV cubes centered at $2^\circ$ longitude \url{https://sedigism.mpifr-bonn.mpg.de/cgi-bin-seg/SEDIGISM_DATABASE.cgi}} 
and the emission of all \starforge snapshots along the three projections. The strong agreement between the two flux distributions serves as a validation that our synthetic emission maps replicate the \sedigism data at first order.

\begin{figure}
    \centering
    \includegraphics[width = \linewidth]{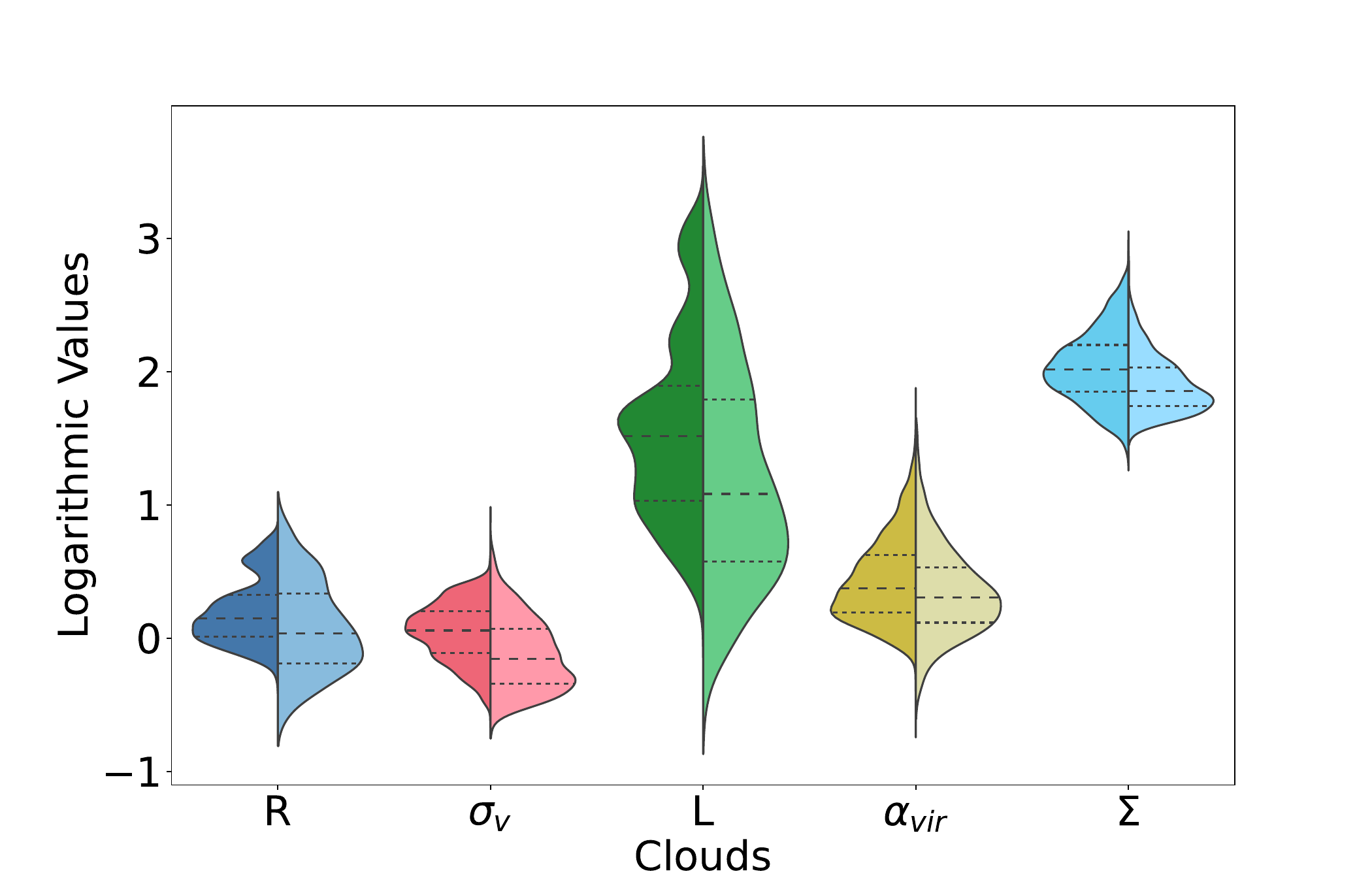}
    \caption{Comparison of MC properties from \starforge (left side violins) and \sedigism (right side violins). The properties are radius (R), velocity dispersion ($\sigma_v$), luminosity (L),  virial parameter ($\alpha_{vir}$) and surface density ($\Sigma$). The horizontal dashed lines represent the quartiles of the distributions.}
    \label{fig: violins prop}
\end{figure}

Figure \ref{fig: violins prop} shows the distribution of the integrated properties for our synthetic MCs and the \sedigism clouds. The good agreement between the two datasets for all properties indicates similar structural characteristics of the clouds in both samples. This further highlights the relability of our approach in not only creating emission cubes similar to \sedigism but also identifying {\small SEDIGISM}-like clouds. However, this should not be confused as a single \starforge simulation replicating the entire diverse sample of clouds in \sedigism. Rather, due to robust data processing, our synthetic MCs occupy the same parameter space as \sedigism clouds.

The slight shift in the \starforge distributions towards higher values compared to \sedigism can be explained as follows. \starforge simulates an isolated GMC and therefore the lowest level dendrogram structures, i.e., hierarchichal trunks, are called MCs. 
\sedigism, on the other hand, traces the larger gas structures in the Galaxy, and every dendrogram structure (e.g. nonoverlapping trunks, branches, or leaves) is considered a cloud if it complies with the clustering criteria set by the segmentation algorithm. Using the same cloud identification criteria as \sedigism, i.e., the {\sc scimes} algorithm \citep{colombo2019} is not suitable for this study. This is mainly due to the relatively small spatial coverage of each snapshot, which causes SCIMES to identify branches within the trunks as MCs, making the extracted clouds significantly smaller than those of SEDIGISM.% This is because {\sc scimes} detects clusters, and it would filter out some of the large trunks that we consider MCs throughout this work.}
% and the branches and leaves of the dendrogram are considered as \sedigism clouds
The exclusion of leaves constraints the lower limit on the properties of our synthetic MCs; however, despite these differences, our MCs are consistent with those in \sedigism (Fig. \ref{fig:sedigism larson} \& \ref{fig:sedigism heyer}).
%a selection effect: our MCs contain only the hierarchichal trunks, as opposed to \sedigism, which limits the lower values of these properties. 

\subsection{Time evolution of MCs}  \label{sec: time evolution of clouds}

\begin{figure*}[htbp]
    \centering    
    \includegraphics[width=0.33\linewidth]{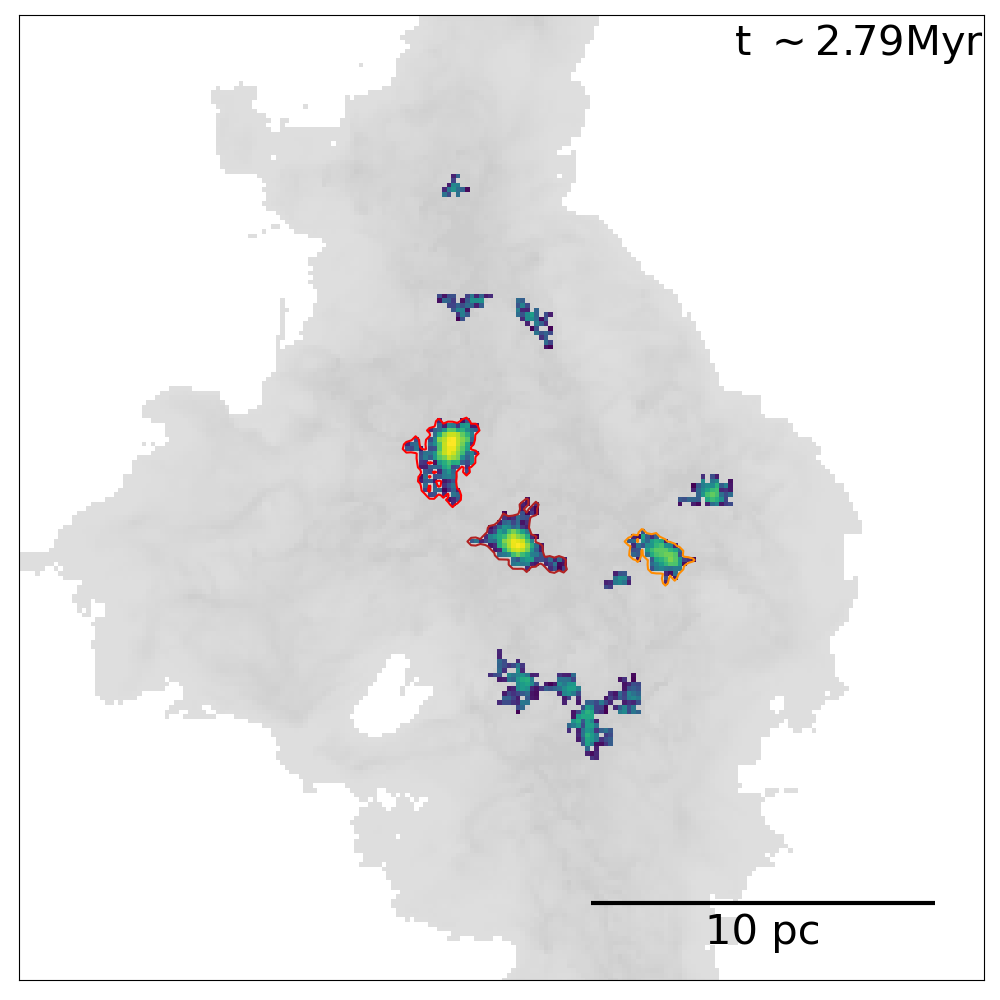}
    \includegraphics[width=0.33\linewidth]{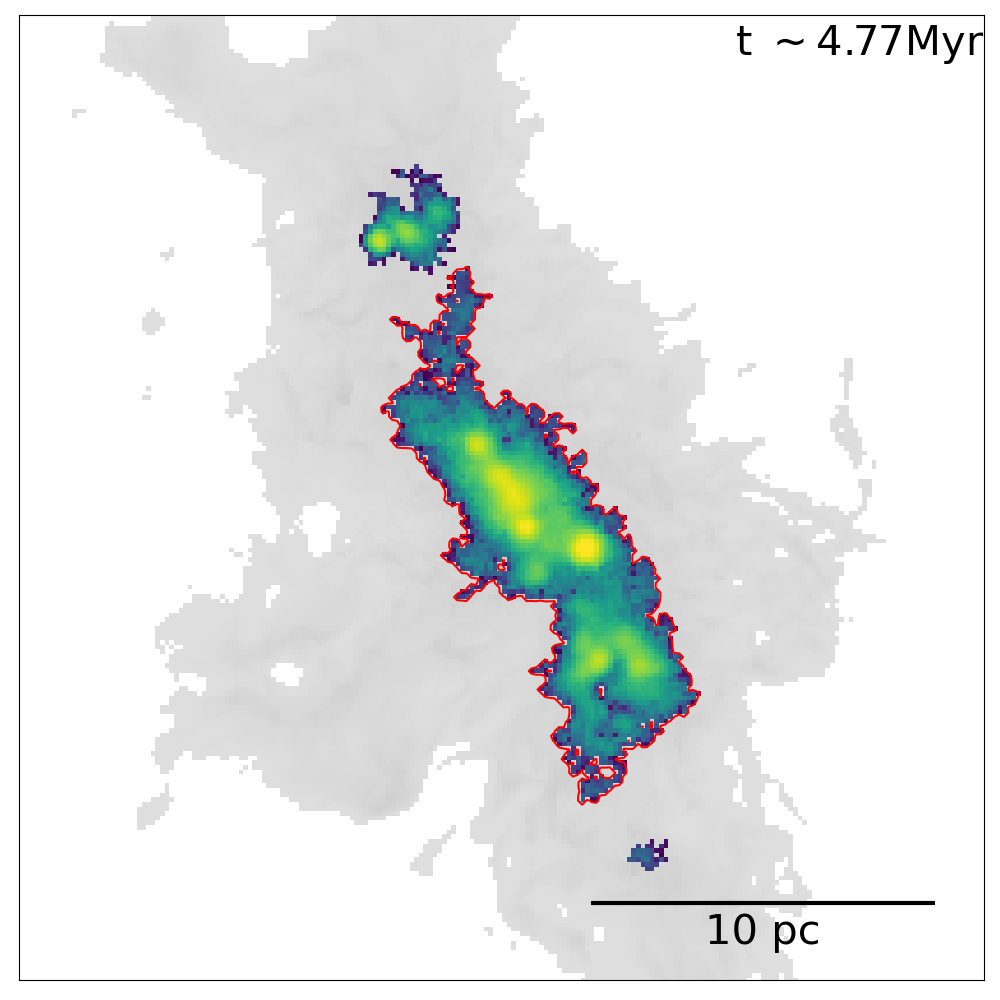}
    \includegraphics[width=0.33\linewidth]{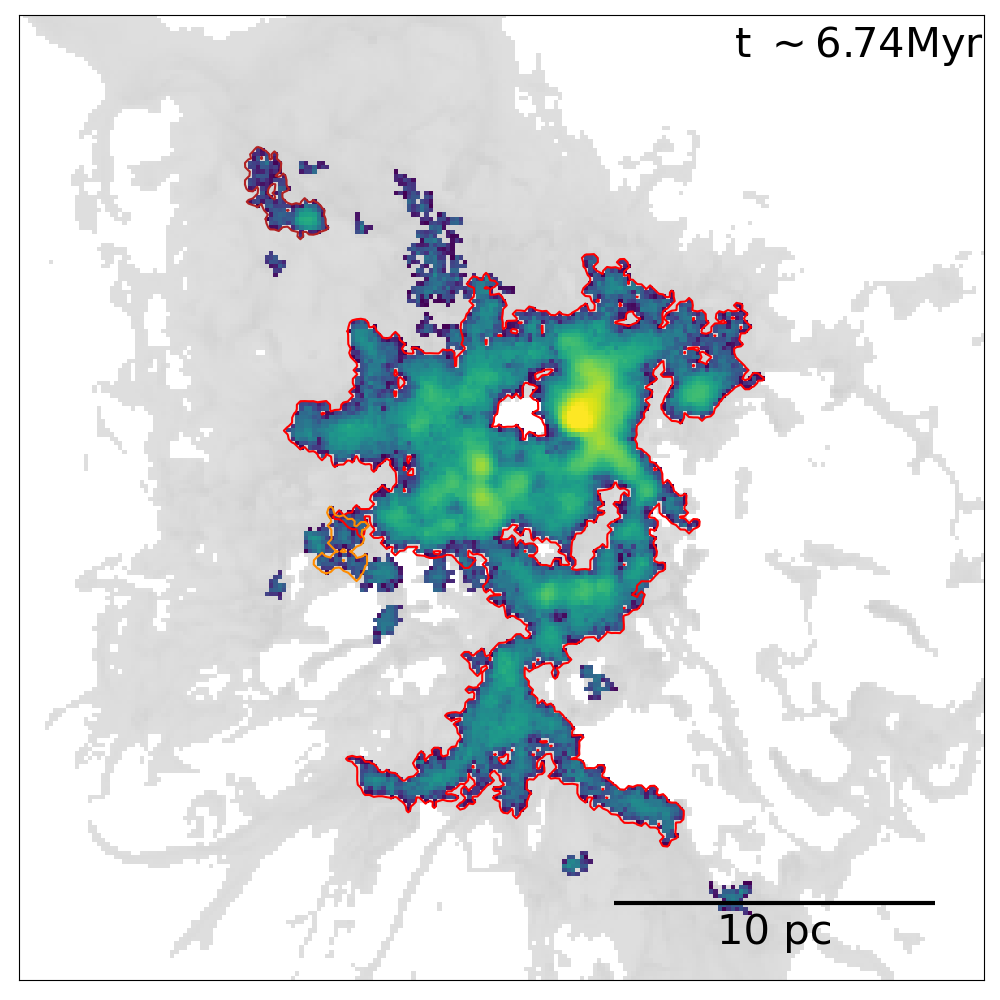}
    \caption{$^{13}$CO(2-1) moment 0 maps at different evolutionary time. The background greyscale represents H$_2$ gas density with $^{13}$CO(2-1) emission overlaid as viridis maps (molecular gas complex), and coloured contours represent different MCs (dendrogram trunks), with red contours representing the largest MCs ($R$) in the cube. The $^{13}$CO(2-1) maps for multiple snapshots along different projections are presented in App. \ref{app: GMC and MC imshow images}.}
    \label{fig: gmc mc proj main}
\end{figure*}

The aim of this section is to understand the changes in the structure and properties of MCs as they evolve and are affected by different stellar feedback mechanisms (Fig. \ref{fig: gmc mc proj main}).
To highlight the broader evolutionary trend, the MCs are binned based on evolutionary time. Each bin corresponds to $\sim 250$ kyr and contains a total of 30 data cubes (three projections per snapshot). The evolution of properties of individual MCs is illustrated in the App. \ref{app: projection axes}. 
The following subsections focus on the general trends in integrated properties of the clouds (Sect. \ref{sec: prop vs snap}), their morphologies (Sect. \ref{sec: cloud morphology}), and substructures (Sect. \ref{sec: fragmentation}). The corresponding figures (Fig. \ref{fig: prop vs snap}, \ref{fig: rj vs snap} \& \ref{fig: frag vs snap}) also show the onset of different stellar feedback mechanisms indicated on the time axis.

\subsubsection{Integrated properties}\label{sec: prop vs snap}

\begin{figure*}
    \centering
    \includegraphics[width=\linewidth]{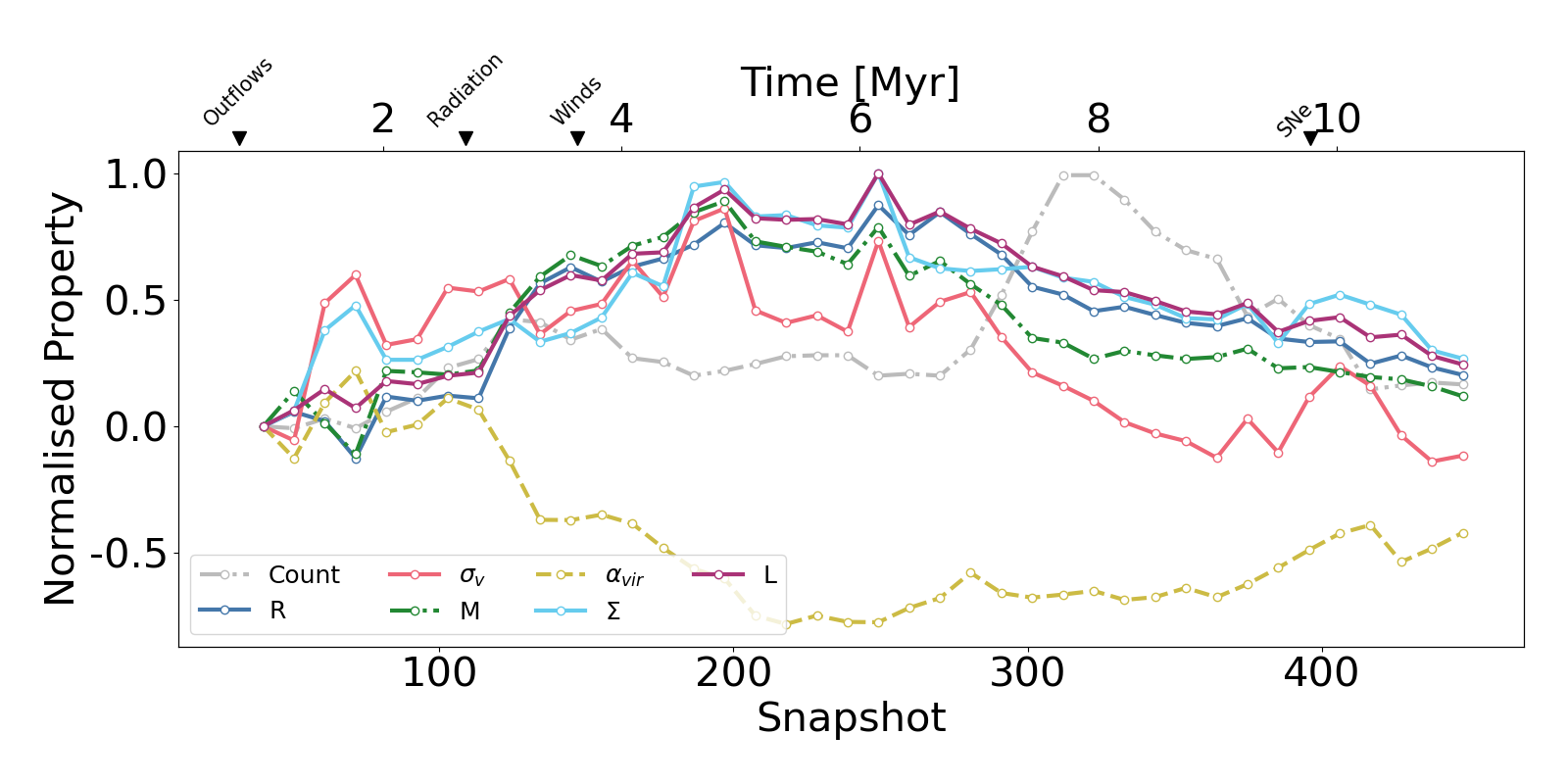}
    \caption{Normalised medians of MC properties (log scales) as a function of time: radius (R), velocity dispersion ($\sigma_v$), molecular gas mass (M), virial parameter ($\alpha_{vir}$), surface mass density ($\Sigma$), and luminosity (L). Count refer to the total number of clouds in a time bin.
    The normalisation includes subtracting the initial value (first bin) of the property from it, followed by a min-max standardisation.
    The symbols on the top represent the times at which outflows, photoionisation radiation, stellar winds and supernovae begin in the simulation.}
    \label{fig: prop vs snap}
\end{figure*}

In this section, we analyse the integrated properties of MCs as a function of time, over the $\sim 11$ Myrs covered by the simulations. 
Figure \ref{fig: prop vs snap} shows that radius, mass, luminosity, and surface density follow each other closely, which is expected because larger MCs are more massive on average \citep{larson1981, kauffmann2010ApJ...712.1137K}. 
The similar trends of molecular gas mass and $^{13}$CO luminosity distributions over time confirm that our results are not significantly affected by the choice of canonical $^{13}$CO abundances (see Sect. \ref{subsec: methods preprocessing}).
The four properties show a steady increase until $\sim 5$, as progressively more $^{13}$CO(2-1) is above the detection limit. The increase represents the active growth of the MCs as they transition from newly formed small diffuse cloudlets to large, massive, and dense MCs (illustrated in Fig. \ref{fig: gmc mc proj main}). 
The increase in mass is also influenced by the transition of the gas from the atomic to the molecular phase and is a key characteristic of the GHC model \citep{shimajiri2019A&A...623A..16S, vazquez-semadeni_2019MNRAS.490.3061V}.
The 6 Myr transition marks the beginning of gas expulsion and cloud dispersal by stellar feedback, as noted by the decrease in the average properties. This is supported by the significant increase in the number of clouds $\sim 7$ Myrs, which suggests that the gas is being removed and eroded from within and around clouds, resulting in a higher number of small cloud fragments (Sect. \ref{fig: frag vs snap}). 
These trends in MC properties are also seen in other simulation sets; however, their onset and duration vary depending on initial conditions (App. \ref{sec: other simulations}).

The velocity dispersion distribution remains relatively flat until $\sim 6$ Myr and then decreases. The initial high values for some MCs are largely a result of the initial supersonic turbulence, with a possible contribution from the momentum injected by the protostellar outflows. The average virial parameter decreases from $\sim 10$ to $\sim 2$ during the first 5 Myr and remains relatively constant throughout most of the evolution (Fig. \ref{fig: prop vs snap}). These distributions are not significantly affected by the growth and dispersal of MCs.
However, they show peaks at $\sim 2$ Myr and $\sim 10$ Myr, which are also seen in the surface density distribution.
The peaks correspond to the formation of turbulent gas structures and the onset of the first supernova, respectively. Supernovae inject turbulence into the ISM and produce relatively dense MCs in an environment that has been mostly dispersed by stellar winds and radiation. However, they do not significantly alter the general trends in the MC properties \citep{grudic2022}.

\subsubsection{Morphology}\label{sec: cloud morphology}

\begin{figure*}
    \centering
    \includegraphics[width=\linewidth]{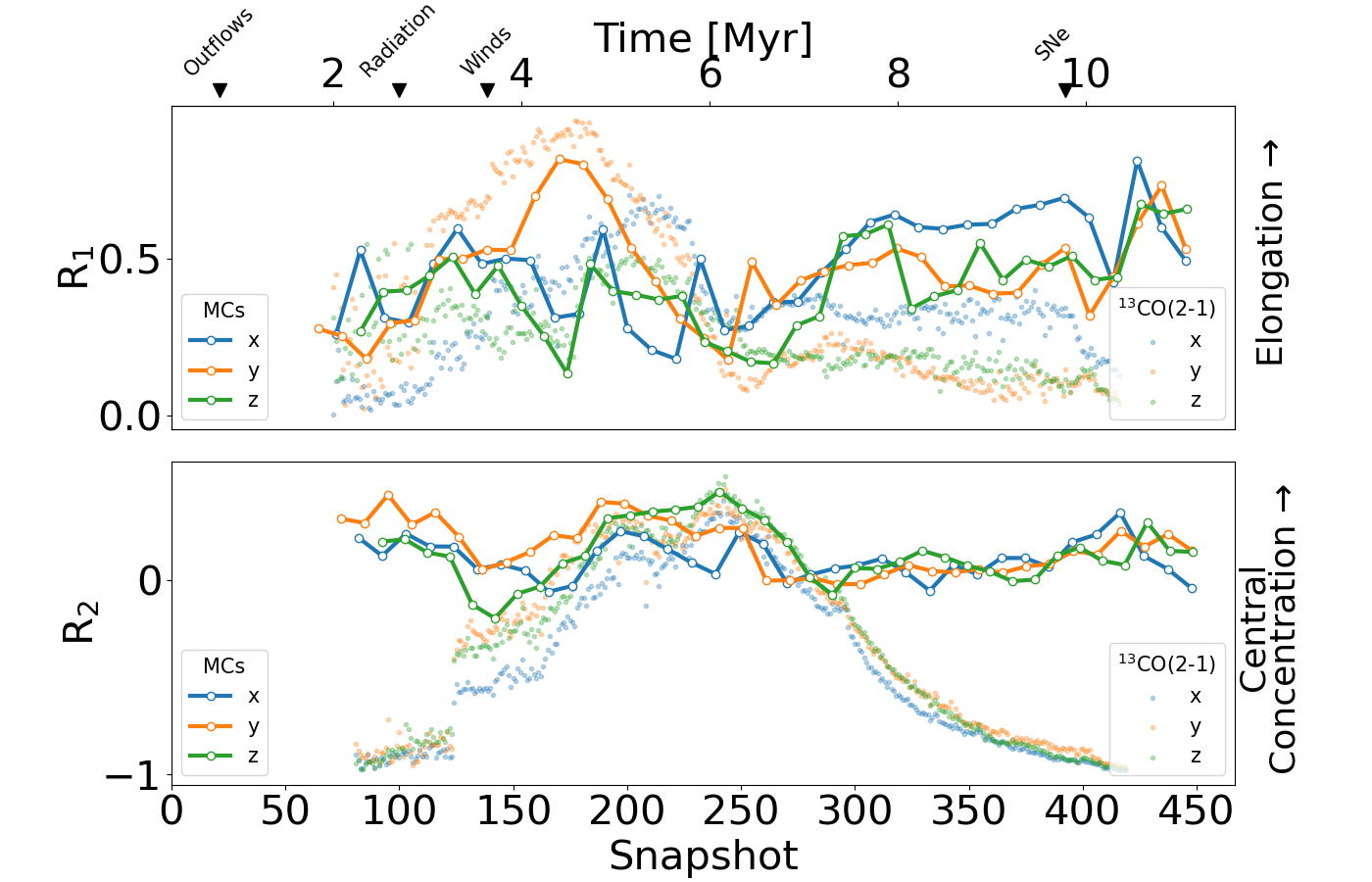}
    \caption{Morphological analysis of the MCs (solid line) and molecular gas complexes (points) in different snapshots. $R_1$ represents the degree of elongation and $R_2$ represents the degree of central concentration. The lines represent the $R_1$ and $R_2$ values for the MCs, and the scatter points represent the same for the molecular gas complexes (entire $^{13}$CO(2-1) emission in snapshot) obtained using the projections along the three axes. The color scheme is the same as in Fig. \ref{fig: violins prop}.}
    \label{fig: rj vs snap}
\end{figure*}

In this section, we study how the morphology of the MCs vary over time under the effect of various stellar feedback mechanisms.
We use the $RJ$ plots algorithm (Sect. \ref{subsec: Dendrograms}) to analyse the morphology of the entire $^{13}$CO(2-1) emission in the simulation box, as well as the individual MCs as they evolve over time (Fig. \ref{fig: rj vs snap}). MCs show a large scatter in $R_1$ and $R_2$ at all times, as cloud structures present a continuous spectrum rather than discrete classes. We therefore bin the MCs following the same criteria as Sect. \ref{sec: prop vs snap} to analyse the overall morphological changes and present the scatter plots in App. \ref{app: projection axes}.
We also analyse the $R_1$ and $R_2$ moments for molecular gas complexes showing the structure of the entire emission in a snapshot.

Figure \ref{fig: rj vs snap} shows that on average the clouds (both the individual MCs and the molecular gas complexes) have different elongations ($R_1$) along the different projections. This is expected as they evolve in a unisotropic environment, which causes them to evolve asymmetrically. However, this difference is not very significant for individual MCs, as seen by the large scatter in Fig. \ref{app: projection axes} (bottom left).
The trends in $R_2$ are similar along the three projections, which highlights that the internal structure of the clouds appears the same regardless of the viewing angle.

The large average values of $R_1$ for the MCs throughout most of the simulation highlight the ubiquitousness of filamentary structures (solid lines in Fig. \ref{fig: rj vs snap}, top). 
A notable trend in the distribution of $R_1$ is a peak during 3-6 Myr, suggesting an elongated shape for CO emission in snapshots (Fig. \ref{fig: gmc mc proj main}, centre). 
The increase in elongation is more dominant in one projection direction, suggesting that the molecular gas complex is collapsing faster along one axis. 
Although the average value of $R_1$ for MCs increases after 6 Myr, this is not the case for molecular gas complexes.
It points towards the emergence of filamentary MCs within feedback-affected spherical bubble-like regions, potentially indicating the formation of intra-cloud filaments. However, this increase should be interpreted with caution due to the large scatter in $R_1$ among individual MCs.
The distribution of $R_2$ for the molecular gas complexes shows a constant increase until $\sim 6$ Myr, followed by a constant decrease. The rise in central concentration ($R_2$) is in agreement with the gravitational collapse of the simulated GMC. The decrease signifies the dispersion of the molecular gas due to stellar feedback, leading to centrally underdense bubble-like structures (Fig. \ref{fig: gmc mc proj main}, right). The increase in $R_2$ along all three projections suggests that these MCs represent 3D bubbles rather than 2D rings (illustrated in App. \ref{app: GMC and MC imshow images}). %The spherical nature of these bubbles is also highlighted by the decrease in $R_1$ at $\sim 6$ Myr. 

The different morphologies of the clouds for different projections reveal the anisotropy of the MCs. The anisotropy is driven by a combination of the initial turbulence, magnetic fields, and various stellar feedback events. A caveat of \starforge is that it assumes an isolated cloud and does not account for galactic-scale processes, such as gas accretion and the galactic potential. However, we expect the galactic potential to have a minor effect on the cloud scale over a timescale of 10 Myr. %, and the stellar associations in \starforge are in agreement with the observed ones \citep[e.g. local gaia associations, ][]{farias2025MNRAS.541..101F}. 
% While the morphology of the GMC at a given time depends on the initial conditions of the simulation \citep[Fig. 4 in ][]{guszejnov2022}, the elongated structure of the actively growing molecular gas complexes and the bubble-like structure of the feedback-affected molecular gas complexes are consistent across simulations with different initial turbulence ($R_1$ and $R_2$ distributions for the molecular gas complexes in Fig. \ref{fig: prop vs snap a1} \& \ref{fig: prop vs snap a4}).} 
Although we see the trends in the morphology for the entire $^{13}$CO emission in snapshots, the individual MCs on average remain elongated centrally concentrated structures throughout the simulation. Visual analysis further confirmed their filamentary and clumpy nature (App. \ref{app: GMC and MC imshow images}) during most of their lifetime ($\sim$ 3-7 ) Myr.

\subsubsection{MC substructures}\label{sec: fragmentation}

\begin{figure*}
    \centering
    \includegraphics[width=\linewidth]{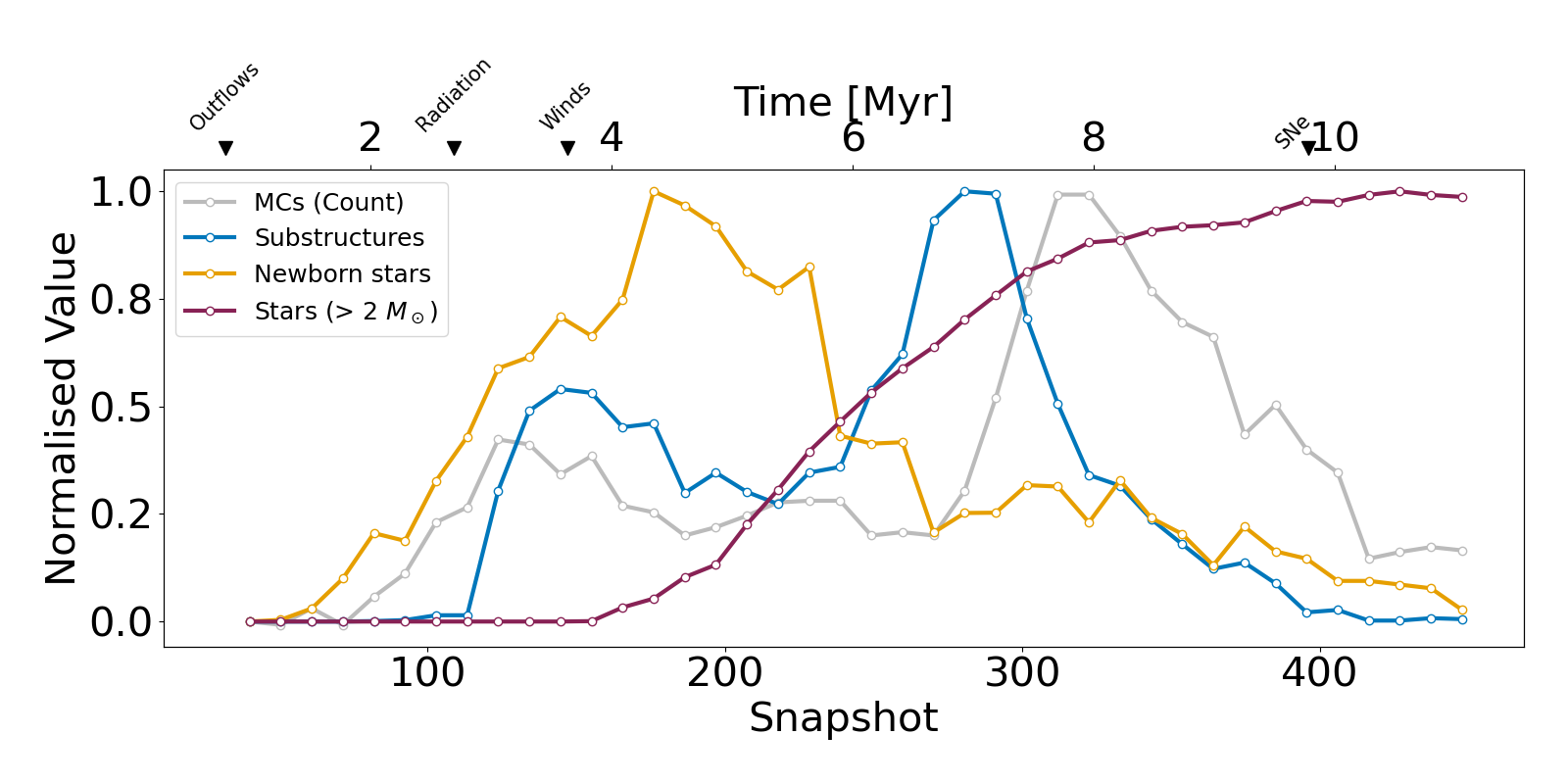}
    \caption{Quantitative representation of MCs (counts), their substructures (dendrogram descendants), the newborn stars (age < 250 kyr), and the main sequence stars more massive than $2 M_\odot$ as a function of time. The values are normalised using a min-max standardisation.}
    \label{fig: frag vs snap}
\end{figure*}

The fragmentation of MCs is often attributed to various stellar feedback mechanisms \citep{mazumdar2021A&A...656A.101M, grishunin2024A&A...682A.137G}, however, fundamental questions about the physics responsible for fragmenting MCs are still open. 
Analysing the number of MCs, their substructures, and stars informs us about how the MCs fragment to form dense substructures that lead to star formation, which in turn produce stellar feedback and disperse these gas structures.
Here, we present a quantitative analysis of the substructures within our MCs. The substructures are stored by dendrogram algorithm as descendants and represent subparsec-scale compact structures, typically referred to as clumps. 
Figure \ref{fig: frag vs snap} shows an increase in the number of MCs and their substructures around ($\sim$ 3 Myr), representing progressively more emission that is above the noise level. 
The second peaks ($\sim 7-8$ Myr) in both distributions are the result of gas dispersion by stellar feedback.

We also analyse the number of stars and protostars with ages $ < 250$ kyr and collectively refer to these as newborn stars. These reflect the instantaneous star formation rate of the clouds, with the 3-7 Myr period representing the peak of star formation activity. The newborn stars evolve to the main sequence, producing stellar winds and photoionising radiation. The stellar evolution is a strong function of their mass \citep{hosokawa2011ApJ...738..140H} and a large number of low-mass stars remain in the main-sequence phase throughout the lifetime of the simulated GMC, seen as a constant increase in the number of stars over time (Fig. \ref{fig: frag vs snap}). 

The significant increase in the number of stars $\sim$ 5 Myr is followed by peaks in the number of substructures (7 Myr) and MCs (8 Myr). 
Stellar winds and radiation from individual stars disperse and expels gas in their neighbourhood, fragmenting dense gas structures (6-7 Myr, Fig. \ref{fig: gmc mc proj main}). 
Over time, the feedback becomes stronger and erodes these fragmented clumps, decreasing in their number (> 7 Myr). This strong gas dispersal affects the entire MC leading to the removal of gas between dense MCs, i.e. the entire ${^13}$CO emission is identified as multiple small MCs instead of a single continuous structure/trunk (7-8 Myr). These smaller MCs are dispersed over time as a result of continuous feedback events (>8 Myr). The lack of a dense molecular gas further decreases the number of embedded stars. 

\section{Scaling relations}\label{sec: scaling relations}

The $-$ Larson's and Heyer's $-$ scaling relations show the correlations between the physical properties of clouds. Larson's first relation, originally derived from the analysis of numerous MCs by \citet{larson1981}, was later refined by \citet{solomon1987}, resulting in the relation $\sigma_v = 0.74 \, L^{0.5}$ \citep[discussed in ][]{colombo2019}. The spatial and velocity structures of MCs following power laws are often considered a proof of universal cloud turbulence \citep{padoan2016ApJ...822...11P}.
It is a simplification of Kolmogorov's law for turbulence, indicating that larger clouds exhibit broader line-widths. Larson's laws individually do not provide information about the virial state of a cloud. To take this into account, \cite{heyer2009} combined Larson's second and third laws, creating Heyer's relation, which compares the surface density of a cloud with its scaling parameter ($\sigma_v^2 / R \propto \Sigma$).

\begin{figure*}
    \centering
    \includegraphics[width=\linewidth]{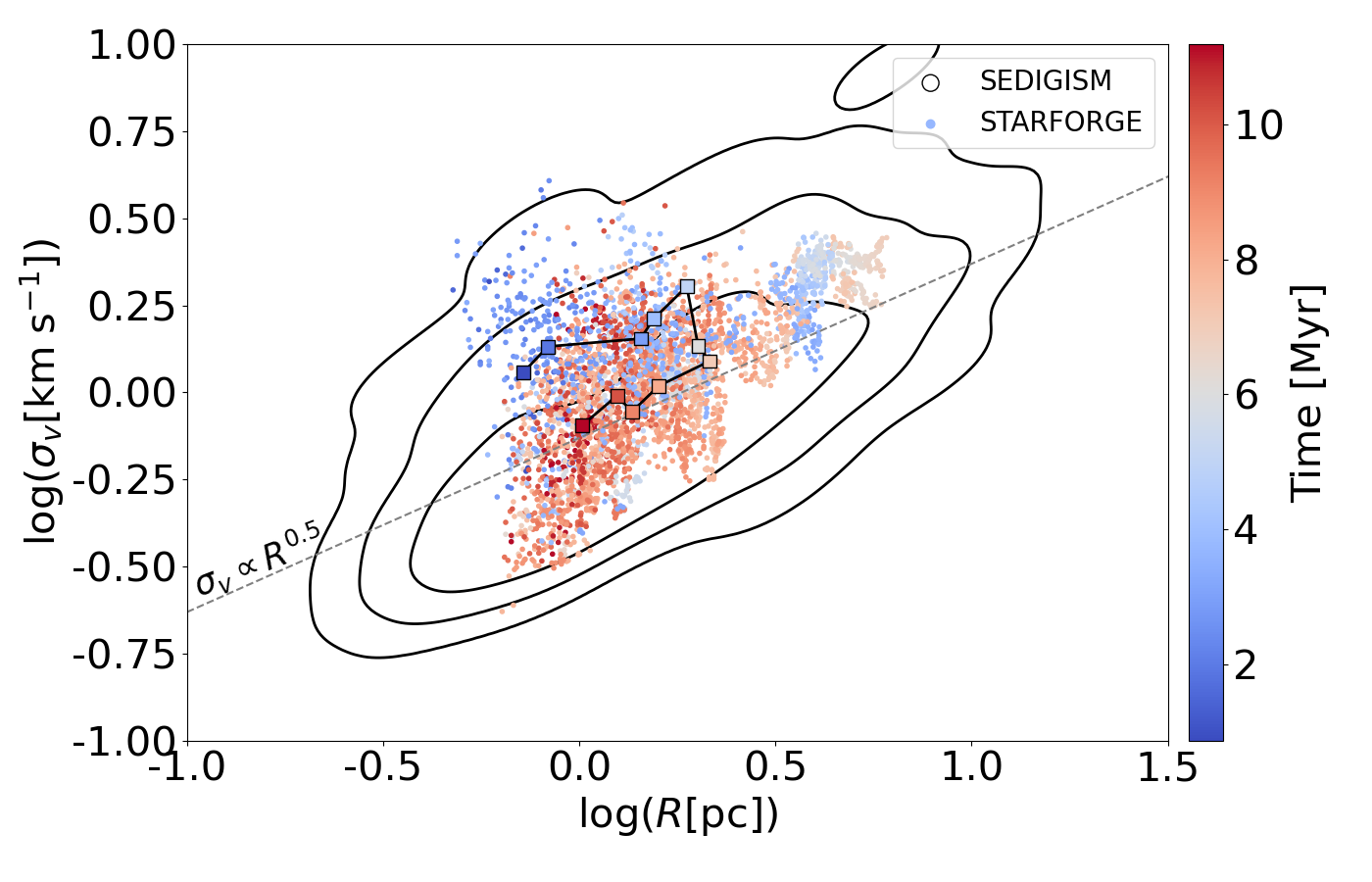}
    \caption{Size-linewidth relation ($\sigma_\varv$ versus $R$) for our MCs (scatter points), color coded with respect to the time elapsed (in Myr) since the start of the simulation. The squares represent medians of distributions in $\sim 1$\,Myr bins.
    The black contours represent the 1$\sigma$, 2$\sigma$, 3$\sigma$ levels for the \sedigism clouds. The dashed line represents Larson's first relation \citep{larson1981, solomon1987}.}
    \label{fig:sedigism larson}
\end{figure*}

\begin{figure*}
    \centering
    \includegraphics[width=\linewidth]{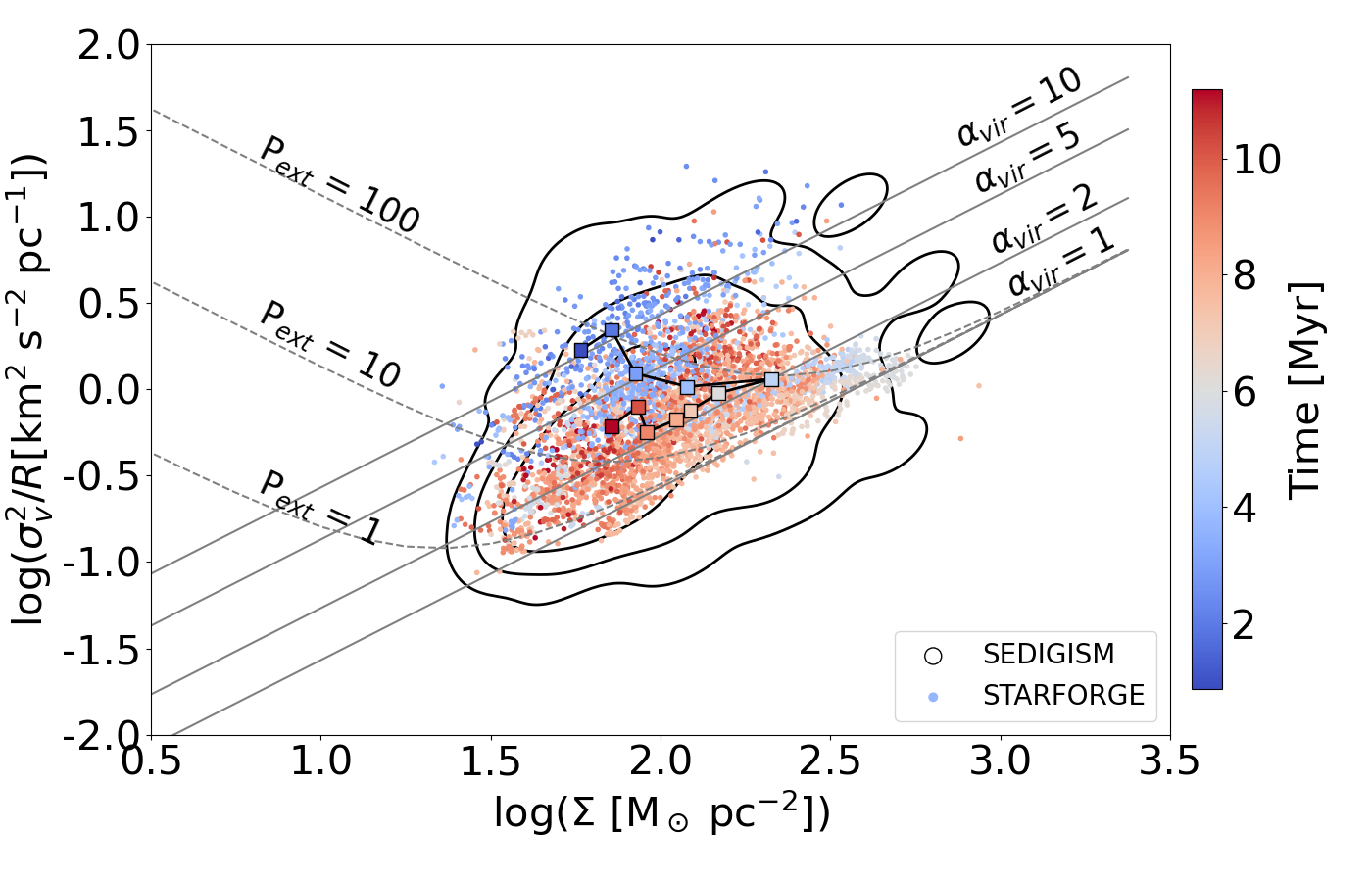}
    \caption{Scaling relation between $\sigma_\varv^2$/$R$ and surface mass density ($\Sigma$). The symbols and notations follow Fig. \ref{fig:sedigism larson}. The solid gray lines represent isocontours of virial parameters. The dashed lines represent $\alpha_\mathrm{vir} = 1$ when including an external pressure P$_\mathrm{ext}$ = 1, 10, 100 $\mathrm{M_\odot \; pc ^{-3} \; km^2 \; s^{-2}}$. }
    \label{fig:sedigism heyer}
\end{figure*}

We present the two scaling relations for our MCs and compare them with the \sedigism clouds in Fig. \ref{fig:sedigism larson} \& \ref{fig:sedigism heyer}. 
The synthetic MCs distribution shows almost a complete overlap with the 3-$\sigma$ KDE (kernel density estimator) for \sedigism clouds. This provides strong evidence that the MCs from our synthetic observations have global properties similar to those of real clouds. Moreover, this shows that the correlation of properties is consistent across both samples, e.g. similar sized MCs have similar velocity dispersions, leading to their overlap on the scaling relation plots.

Figure \ref{fig:sedigism larson} shows an increase in the average size and linewidth of the clouds up to $\sim 6$ Myr. This is largely a result of the formation of dense gas structures that merge and result in progressively more $^{13}$CO(2-1) emission being detectable. 
In addition, stellar feedback mechanisms drive the velocities in the MCs and expand them, resulting in larger sizes and linewidths over time.  
At evolutionary times beyond $\sim 6$ Myr, stellar feedback mechanisms begin to disperse the gas significantly, resulting in the identification of smaller MCs. This appears as a sharp drop in the average velocity dispersion followed by a gradual decrease in their size. The higher values in the average velocity dispersion at early times (< 6 Myr) for similar sized structures are most likely due to the initial supersonic turbulence injected in the simulation.

Figure \ref{fig:sedigism heyer} highlights an initial trend of MCs as they transform from underdense and highly supervirial structures to denser virialised structures. The decrease in the average scaling ratio ($\sigma^2/R$) is due to the significant increase in the MC radius compared to the velocity dispersion. The erosion of MCs due to feedback after 6 Myr causes a horizontal shift in Fig. \ref{fig:sedigism heyer} toward lower surface densities and higher virial parameters. 

Molecular clouds are often analysed collectively in scaling relation plots, which typically show a large scatter \citep{colombo2019, duarte_cabral2021}. 
\cite{neralwar2022b} show that the cloud morphology and internal substructures influence their distribution in these relations and hypothesised that the different morphologies might correspond to different evolutionary stages.
Figures \ref{fig:sedigism larson} \& \ref{fig:sedigism heyer} show that MC populations at different evolutionary times occupy different positions in the scaling relation plots. This suggests that the large scatter in the scaling relations could be due to the emsemble of observed MCs being at different stages of their evolution.
\sedigism clouds could have undergone through a diversity of physical conditions, as they are influenced by the larger Galactic environment and feedback events and follow various evolutionary paths. The gas flows and the effects of external factors are not simulated in \starforge. However, our MCs lie in the same parameter space as \sedigism and the simulation traces all relevant physics of star formation at parsec scales. Therefore, at least some of the \sedigism clouds follow an evolutionary path similar to that of our MCs. 
When combined with the fact that MC morpholgies evolve over time (Sect. \ref{sec: cloud morphology}), our results support the hypothesis proposed by \cite{neralwar2022b}.

\section{Discussion} \label{sec: discussions}

\subsection{The lifecycle of synthetic MCs and their observed counterparts}

The distribution of integrated properties, morphology, and fragmentation show that MCs are evolving from small, diffuse structures to dense filamentary structures before being dispersed by stellar feedback. They appear as filamentary and clumpy structures throughout most of their lifetimes, being consistent with other simulations \citep[e.g., ][]{clarke2017MNRAS.468.2489C}. 
The smaller structures they host collapse and form stars, even though the parent MC appears unbound ($\alpha_{vir}\gg 1$). The stars produce stellar feedback that disperses the molecular gas resulting in smaller, less massive, and diffuse structures. The presence of MCs as small clumps, filamentary and bubble-like structures is also supported by observations \citep{neralwar2022A&A...663A..56N}.

The initial turbulence in the simulations produces overdensities that become denser over time because of gravitational collapse. These MCs (< 3 Myr) appear as small, diffuse, low mass, gravitationally unbound ($\alpha_{vir}\sim 10$) and approximately spherical structures. We refer to them as MCs following their definition as hierarchichal trunks to be consistent throughout the paper; however, they are closer to starless molecular gas clumps in observations (e.g., starless clumps in \citet{traficante2018MNRAS.477.2220T} and quiescent clumps in \citet{urquhart2022MNRAS.510.3389U}).

Molecular clouds in 3-7 Myr appear as large filamentary structures with dense clumps. The entire $^{13}$CO(2-1) emission in the simulation box appears as a single (or a few) large MC(s), since most of the gas in the simulation domain is molecular\footnote{as shown by the molecular gas fraction \url{https://starforge-tools.readthedocs.io/en/latest/data.html\#gas-data-fields} values stored in \starforge for each snapshot.}. These represent a majority of the MCs detected in observational surveys \citep{molinari2010A&A...518L.100M, arzoumanian2011A&A...529L...6A, colombo2021A&A...655L...2C, neralwar2022A&A...663A..56N, ge2023A&A...675A.119G}.
The long lives of filamentary MCs are often attributed to continuous gas flows from the larger environment onto small-scale clumps through the filaments \citep{gomez2014ApJ...791..124G}. 
\cite{peretto2023MNRAS.525.2935P} show that the substructures within the MCs produce a deep gravitational potential and accrete the gas from the filament, thus dynamically decoupling from the MCs to grow faster. 
The formation of these dense clumps\footnote{These central overdensities in MCs are visible in Fig. \ref{fig: gmc mc 2 proj} and is evident from the large values of $R_1$ (Sect. \ref{sec: cloud morphology}).} leads to a central infall of gas along the filament, which feeds the clumps, forms new small clumps, and causes turbulent movements \citep[previously discussed in ][]{gong2018A&A...620A..62G, lu2018ApJ...855....9L, williams2018A&A...613A..11W, krumholz2020MNRAS.494..624K}. 
The higher number of dense clumps results in an accelerated formation of protostars and stars (Fig. \ref{fig: frag vs snap}; 4-6 Myr). 

Stellar winds and radiation become more effective throughout the simulation domain after $\sim$ 6 Myr, resulting in gas expulsion and dispersion. 
These phenomena result in the $^{13}$CO(2-1) emission appearing as centrally underdense structures with a shell-like morphology. These 3D bubble-like clouds are widely studied as wind- and radiation-driven bubbles \citep{churchwell2004ApJS..154..322C,palmeirim2017A&A...605A..35P, tiwari2021ApJ...914..117T}, associated with {H~\sc{ii}} regions \citep{neupane2024A&A...692A.114N} and classified as the last evolutionary stage of clouds \citep{kawamura2009ApJS..184....1K}. 
Feedback disperses most of the $^{13}$CO emission by $\sim$ 8 Myr, causing the broken shells to be identified as individual MC. The formation of massive stars at $\sim$ 3 Myr and most of the $^{13}$CO(2-1) emission being dispersed by $\sim$ 8 Myr agrees with the fast dispersal of molecular clouds by feedback \citep[up to $\sim$5 Myr,][]{kruijssen2019, chevance2020MNRAS.493.2872C,figueira2020A&A...639A..93F, knutas2025arXiv250508874K}.

\subsection{Caveats and Outlooks} \label{subsec: caveats outlooks}

\starforge simulates an isolated GMC within a closed box with a fixed total gas mass $(2 \times 10^4 M_\odot)$, restricting the upper mass limit of the MCs.  
Moreover, the simulation does not track the real-time abundance of CO, so the use of canonical abundance values and ad hoc freeze-out prescriptions is a simplification. This results in under- or overestimation in the real abundances, thus introducing an uncertainty on the measured M$_{lum}$ from the synthetic observations, potentially skewing these distributions.
To minimise this error, we set the CO abundance to zero in regimes where it can freeze out. We also set strict constraints while performing dilated masking (Sect. \ref{sec: methods postprocessing}) and choose only the hierarchichal trunks (Sect. \ref{subsec: Dendrograms}) as MCs to avoid spurious sources. 
Inclusion of a chemical network in the simulations or radiative transfer could improve the accuracy of property estimates, but this goes beyond our current scope and does not significantly affect our overall analysis (App. \ref{app: uclchem}).

Predicting the evolutionary stages of \sedigism MCs based on their properties and morphology might be possible, since our MCs share the same parameter space as \sedigism (Sect. \ref{sec: scaling relations}). However, the degeneracy in these distributions on either side of the 6 Myr peak, visible as the large scatter in the scaling relation plots, makes this task extremely challenging using solely $^{13}$CO(2-1) observations. The early MCs show a H$_2$ envelope (Fig. \ref{fig: gmc mc 2 proj}) which could be traced with diffuse gas tracers such as $^{12}$CO, thus separating them from the feedback-affected MCs (Fig. \ref{fig: gmc mc 3 proj}). Moreover, an analysis of the dense gas structures within MCs using tracers such as N$_2$H$^+$ and NH$_3$ could reveal the fragmentation trends and aid in the evolutionary classification of MC. However, such a multiwavelength study is beyond the scope of this work.

Observational works often perform multiwavelength studies using tracers of dense gas, young stellar objects, and HII regions to classify molecular clouds and clumps into various evolutionary stages \citep{kawamura2009ApJS..184....1K, traficante2018MNRAS.477.2220T, urquhart2022MNRAS.510.3389U, watkins2025MNRAS.536.2805W}. 
In a follow-up paper, we will study how clumps (dendrogram branches) and cores (dendrogram leaves) within our MCs are affected by various stellar feedback mechanisms  \citep[similar to ][]{neralwar2024A&A...690A.345N}. 
This will improve our understanding of the evolution of the molecular gas structures from clouds to core scales.
We will also compare these gas structures with their observed counterparts in observations at various evolutionary stages \citep{urquhart2022MNRAS.510.3389U}, to understand the degree to which such multiwavelenth analysis are able to predict the evolutionary stages of gas structures.

\section{Summary and Conclusions}\label{sec: summary}

In this paper, we have created synthetic observations from a $20~000$ M$_{\odot}$ \starforge simulation modelled after the \sedigism survey.
We used the RADMC-3D radiative transfer code to convert the gas density cubes into $^{13}$CO(2-1) emission maps and performed a dendrogram analysis to identify MCs. We analysed this sample of synthetic MCs and investigated the trends in properties, morphology, and substructures to understand how MCs evolve under the effects of different stellar feedback mechanisms.

The flux distributions of the \sedigism ppv cubes and our synthetic data cubes are in strong agreement, validating the replication of the \sedigism data to first order. The properties of synthetic MCs show good agreement with the \sedigism clouds and the two sample fill the same parameter space in the scaling relation plots, which further confirms the robustness of our approach. Although the two cloud show overall good agreement, the sythetic MCs reproduce only a subset of the diversity seen in the observations.
Moreover, synthetic MCs at different evolutionary stages occupy distinct regions of the scaling relation plots, suggesting that evolutionary time plays a significant role in driving the observed scatter.

We study the formation, evolution, and destruction of MCs through variation in their observable properties. The initial turbulence in the simulations creates gas overdensities that collapse under self-gravity and are detected as $^{13}$CO(2-1) emission. These reflect the early cloudlets in observations that are accreting gas from the larger environment to appear as moderately dense gas structures. Gas flows from large to small scales shape MCs into elongated filamentary structures with multiple substructures. The fractal substructures in MCs form stars, which eject matter and radiation into the surrounding environment, driving the formation of gas bubbles. These 3D bubble-like MCs are often associated with stellar winds, radiation, and {H~\sc{ii}} regions.
Our analysis presents MCs as evolving from small, diffuse structures to dense filamentary MCs followed by 3D gas bubbles, and these evolutionary trends are consistent with simulations initialised differently. This confirms the key hypothesis from our previous observational work that MCs evolve from concentrated to elongated to ring-like structures.

In conclusion, we have produced $^{13}$CO(2-1) synthetic observations modelling the \sedigism survey using the \starforge simulations that include all the relevant physics for star formation. Analysing the properties, morpholgies, and fragementation trends of MCs, we show that they evolve from small, diffuse structures to dense filamentary structures to bubble-like structures. The distributions of MCs occupy different parameter spaces in the scaling relation plots, suggesting that they drive the scatter in the observed scaling relations. In an upcoming paper, we will study the effect of individual feedback mechanisms -- outflows, stellar winds, radiation, supernovae -- on these MCs and their substructures. We will also explore the possibility of compairing the structures at different evolutionary stages in simulations and observations.

\begin{acknowledgement}
The authors thank the anonymous referee for a constructive report, which has significantly improved the quality of the manuscript.
KN thanks Prof. Stefanie Walch-Gassner, Dr. Daniel Seifried, and Dr. Piyush Sharda for helpful discussions.
AK acknowledges support from the Polish National Science Center SONATA BIS grant No. 2024/54/E/ST9/00314. MF acknowledges support from the Polish National Agency for Academic Exchange grant No. BPN/BEK/2023/1/00036/DEC/01 and from the Polish National Science Centre SONATA grant No. 2022/47/D/ST9/00419. S.N. gratefully acknowledges the Collaborative Research Center 1601 (SFB 1601 sub-project B1) funded by the Deutsche Forschungsgemeinschaft (DFG, German Research Foundation) – 500700252.
\end{acknowledgement}

\footnotesize{
\bibliographystyle{aa}
\bibliography{reference}
}

\appendix

\section{Effects of projection angles} \label{app: projection axes}

The ppv cubes for this study were produced using RADMC-3D by projecting along three orthagonal axes. This is achieved using three combinations of incl-phi: 0-0, 90-0 and 90-90 in the RADMC-3D script for spectral line imaging (\texttt{radmc3d image})\footnote{\url{https://www.ita.uni-heidelberg.de/~dullemond/software/radmc-3d/manual_radmc3d/imagesspectra.html}}. The simulation box is thus projected along the $z$, $y$, and $x$ axes, respectively. MCs identified in different projections have similar properties, which is consistent with previous similar works \citep[e.g.][]{priestley2023}. This is largely due to the fact that $^{13}$CO(2-1) emission is optically thin and thus the entire MC is traced along all projections.
We conclude that using a specific projection does not alter the MC properties and provides a sanity check that the simulations and RADMC-3D produce model clouds reasonably well. 

\begin{figure*}
    \centering
    \includegraphics[width=\linewidth]{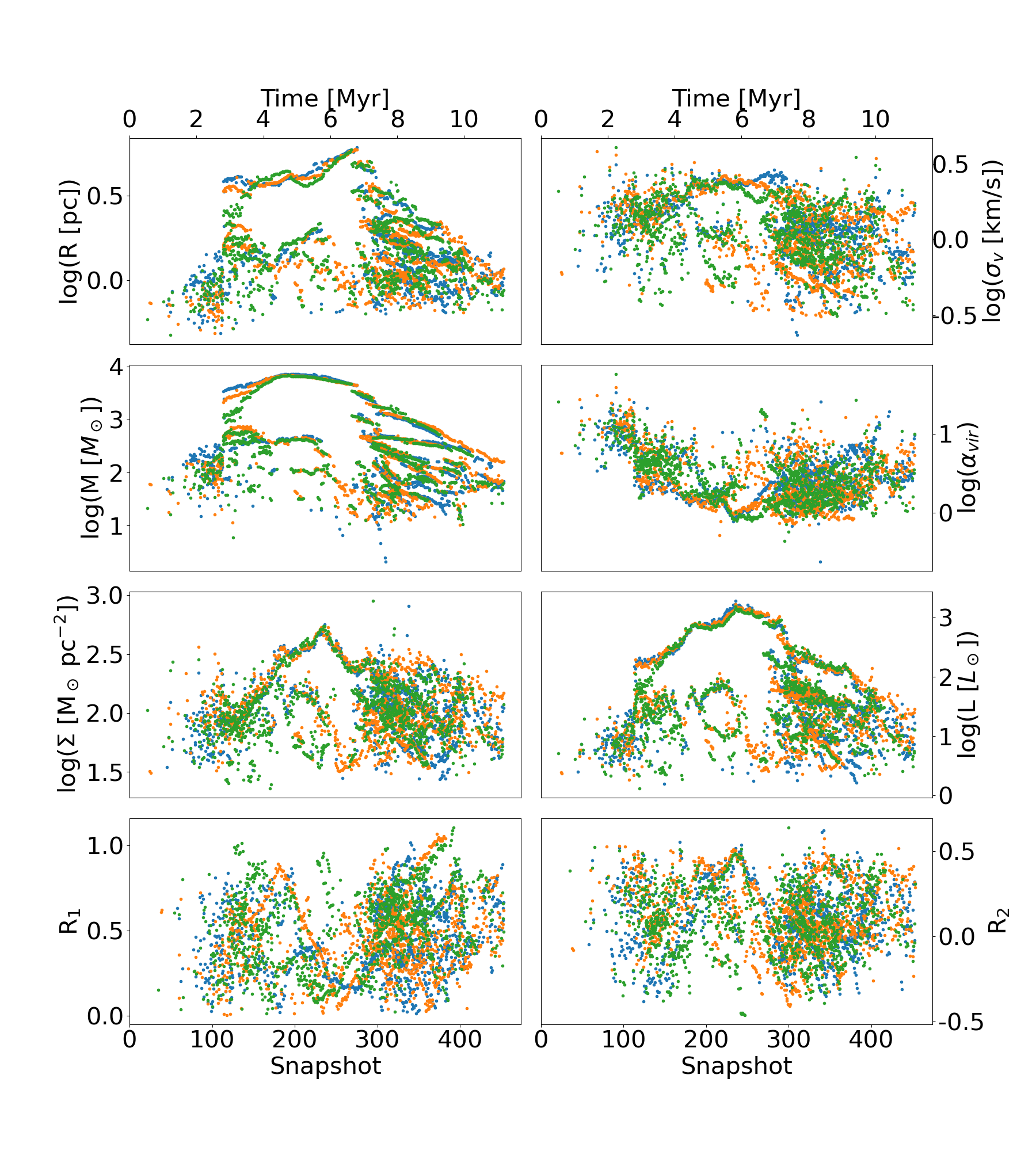}
    \caption{The three colors represent the axes along which the cube is projected. The color scheme follows fig. \ref{fig: rj vs snap}.}
    \label{fig: prop vs snap scatter config}
\end{figure*}

\section{$^{13}$CO(2-1) emission maps} \label{app: GMC and MC imshow images}

\begin{figure*}[htbp]
    \centering    
    \includegraphics[width=0.33\linewidth]{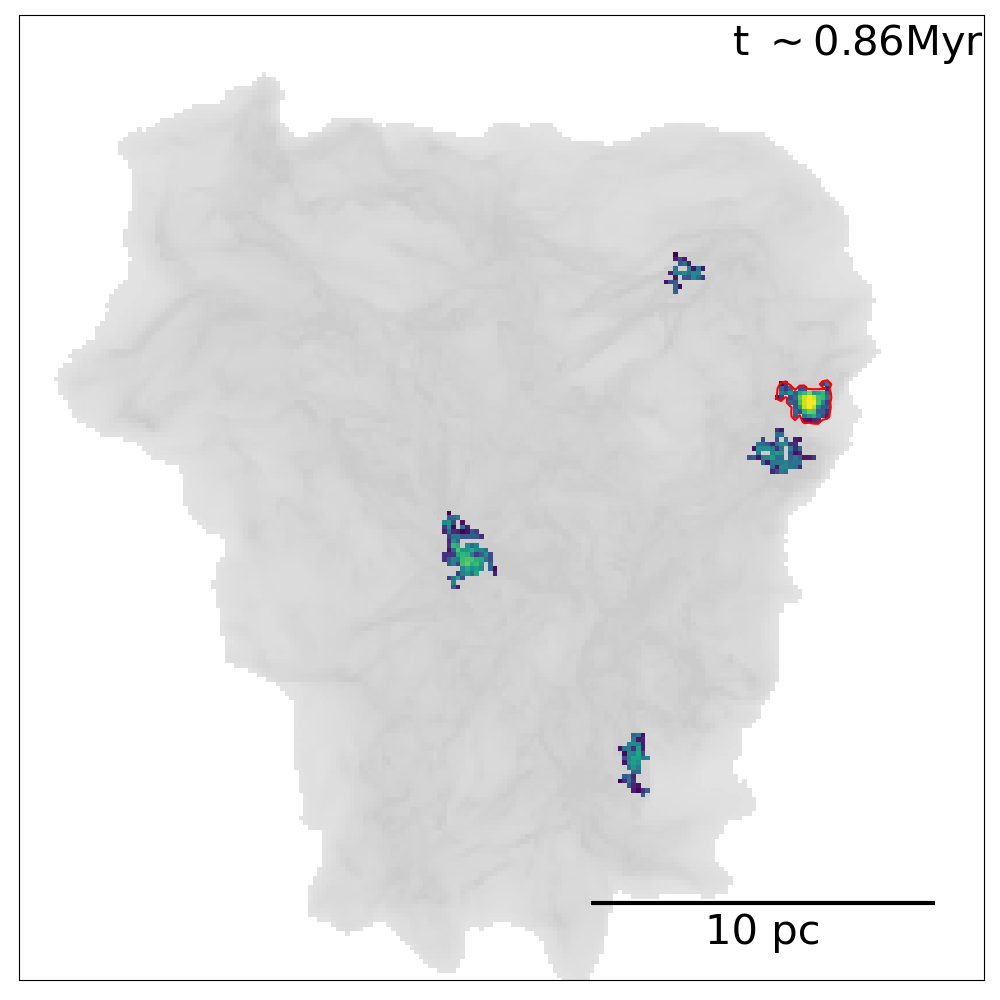}
    \includegraphics[width=0.33\linewidth]{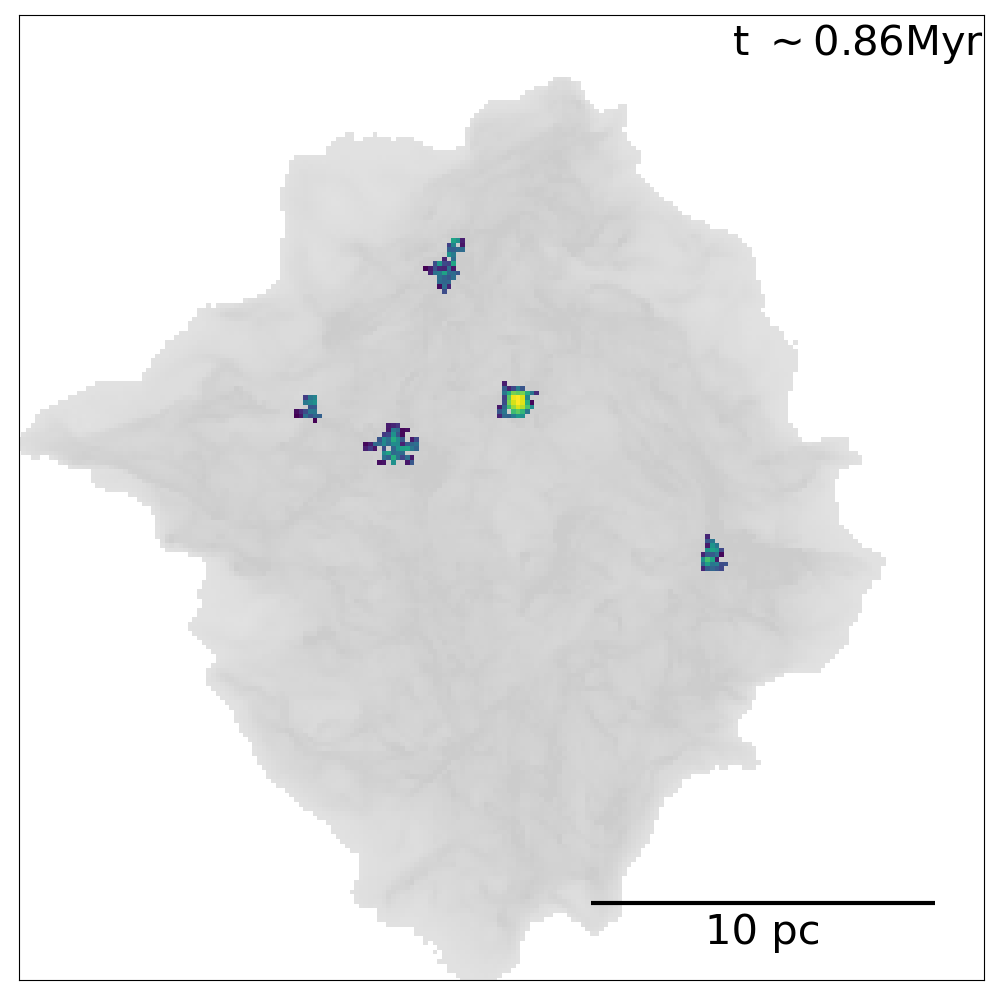}
    \includegraphics[width=0.33\linewidth]{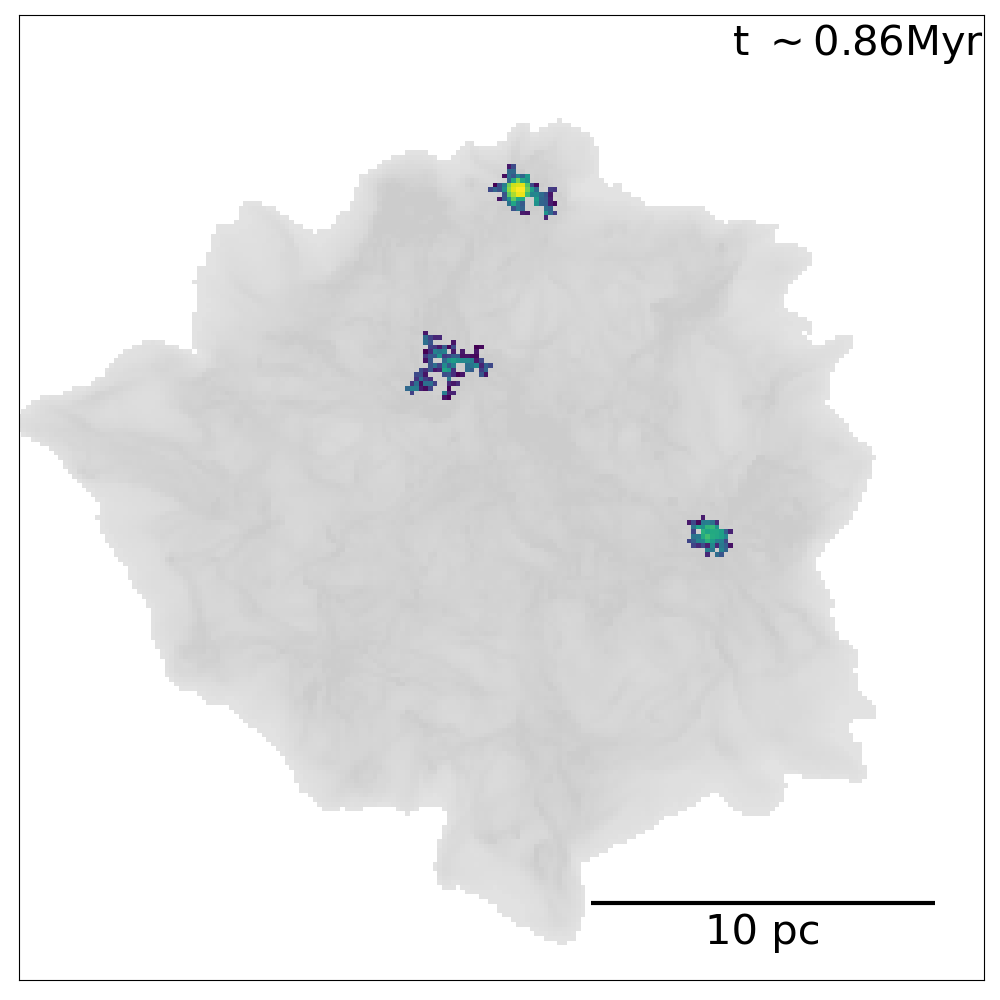}

    \includegraphics[width=0.33\linewidth]{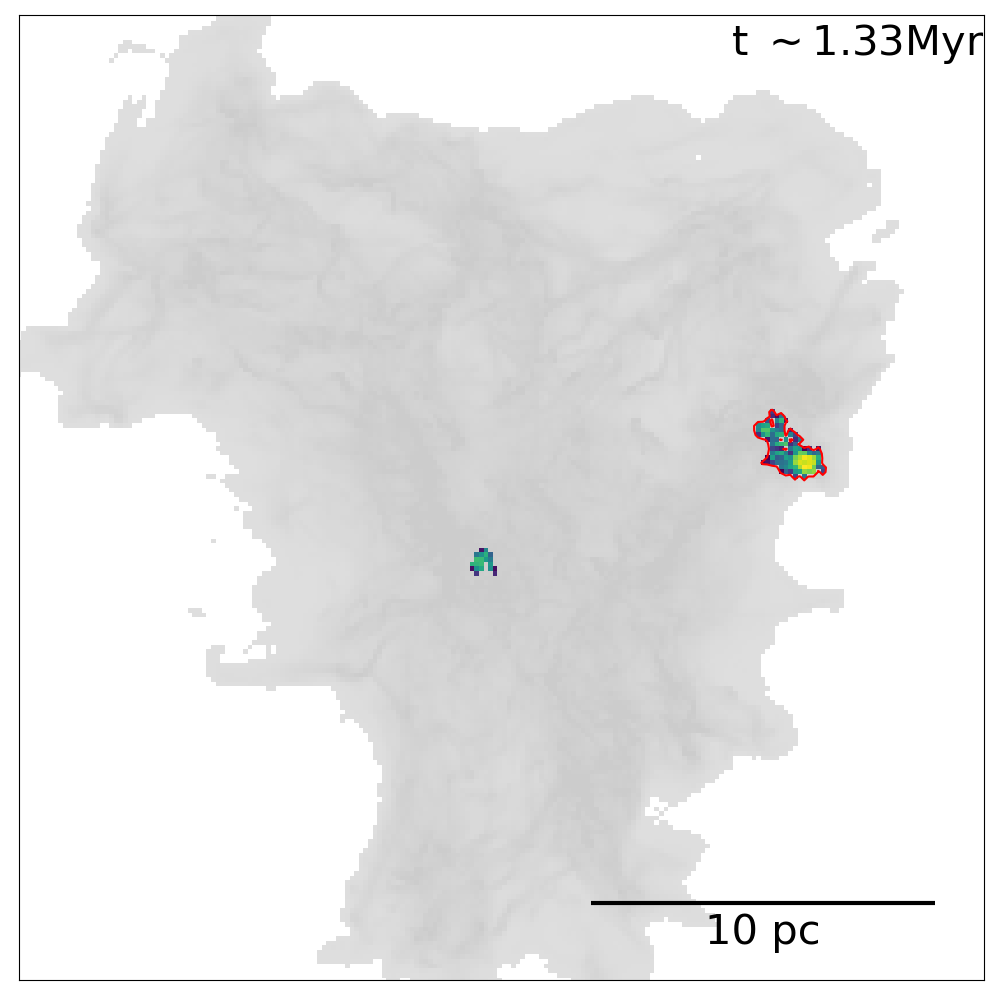}
    \includegraphics[width=0.33\linewidth]{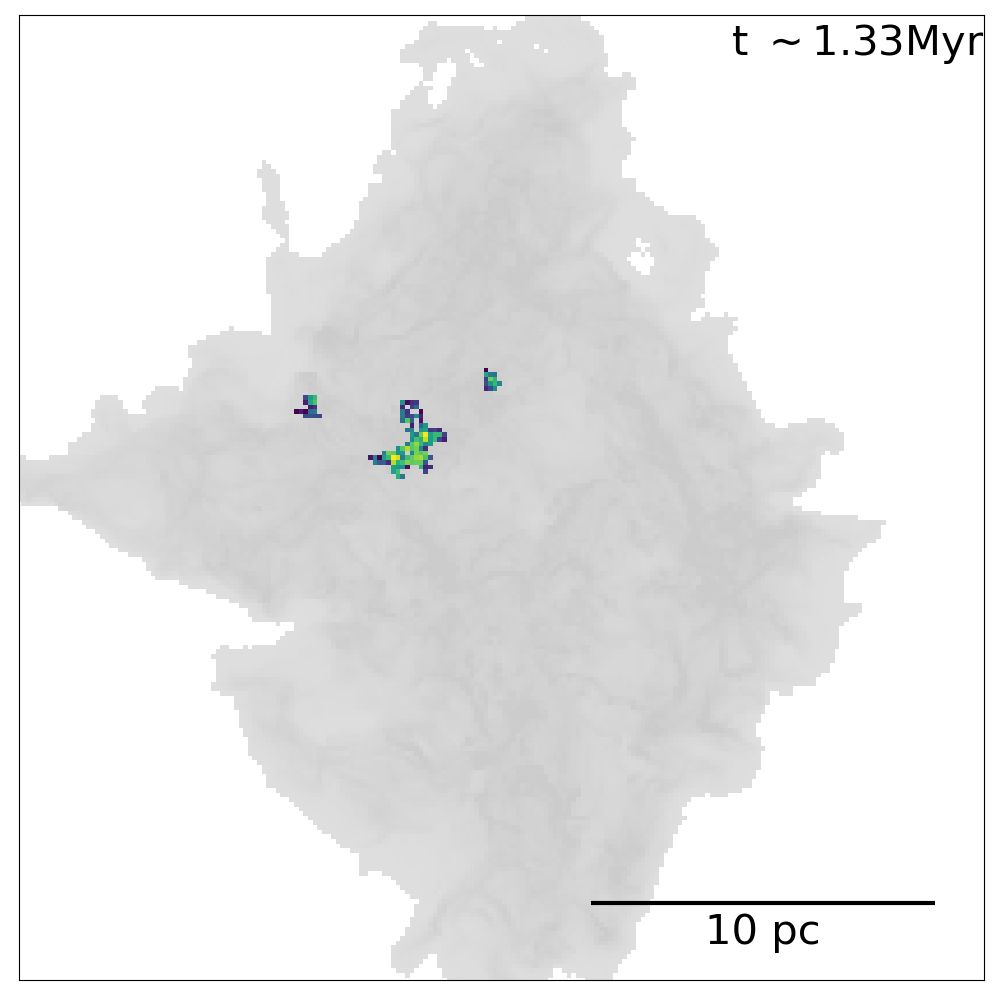}
    \includegraphics[width=0.33\linewidth]{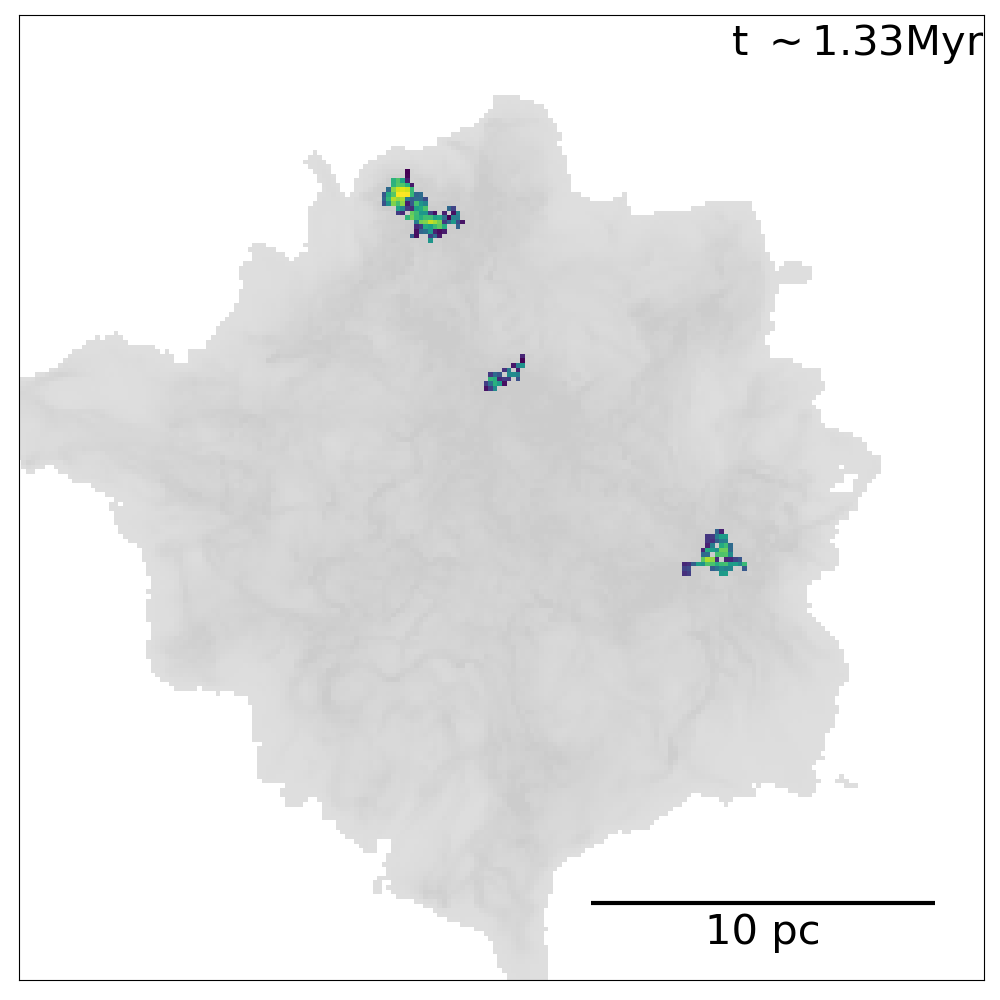}

    \includegraphics[width=0.33\linewidth]{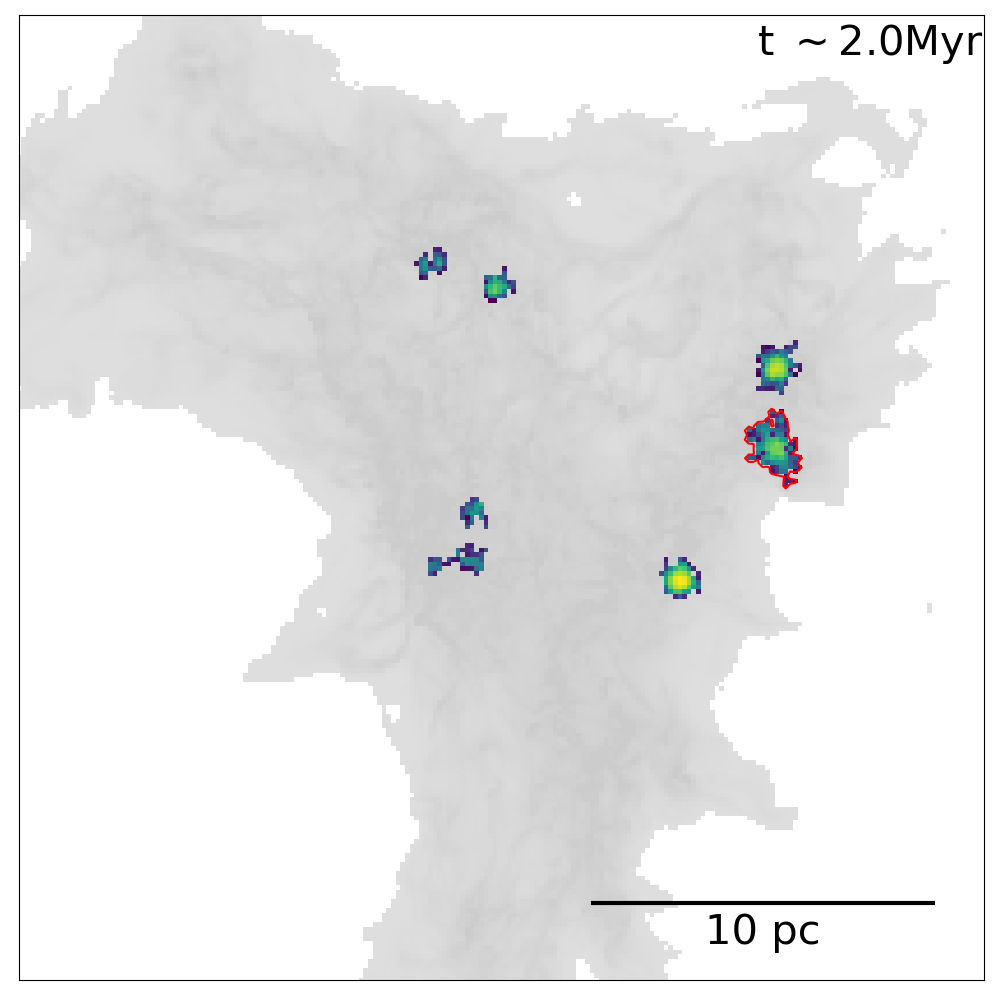}
    \includegraphics[width=0.33\linewidth]{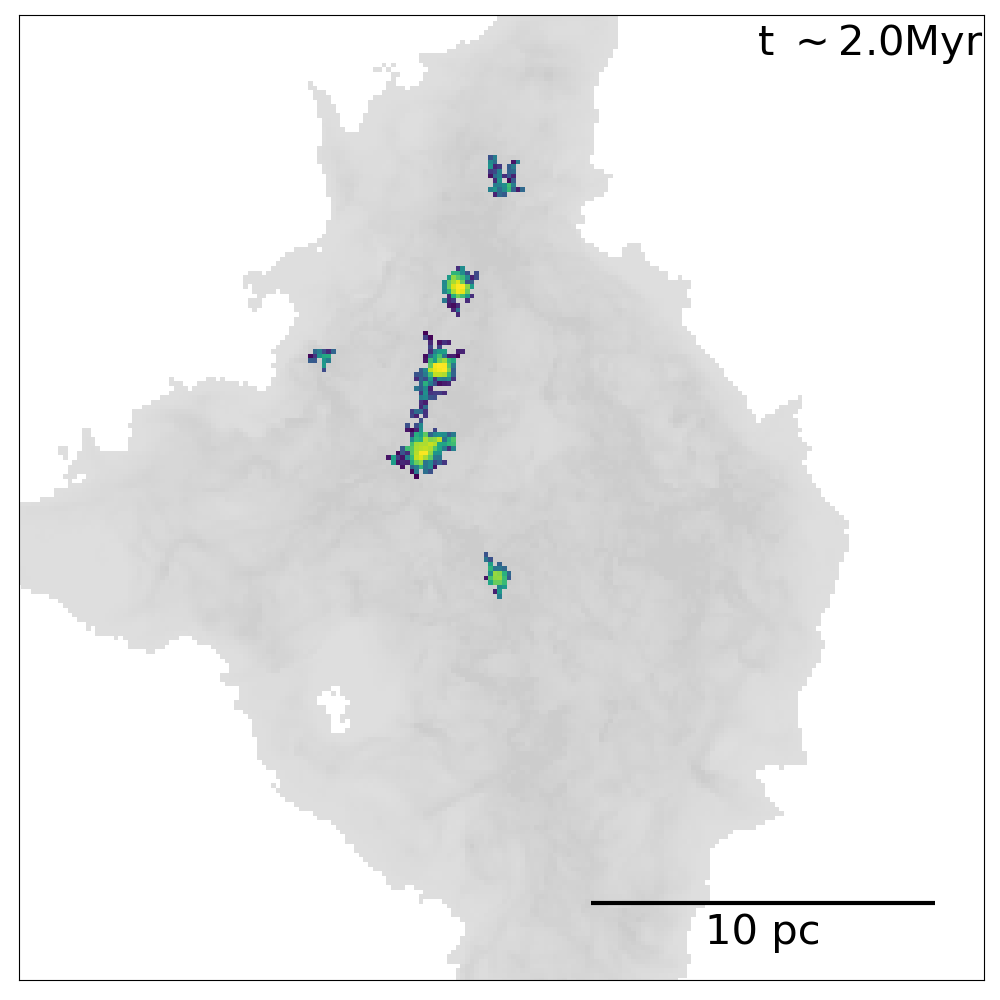}
    \includegraphics[width=0.33\linewidth]{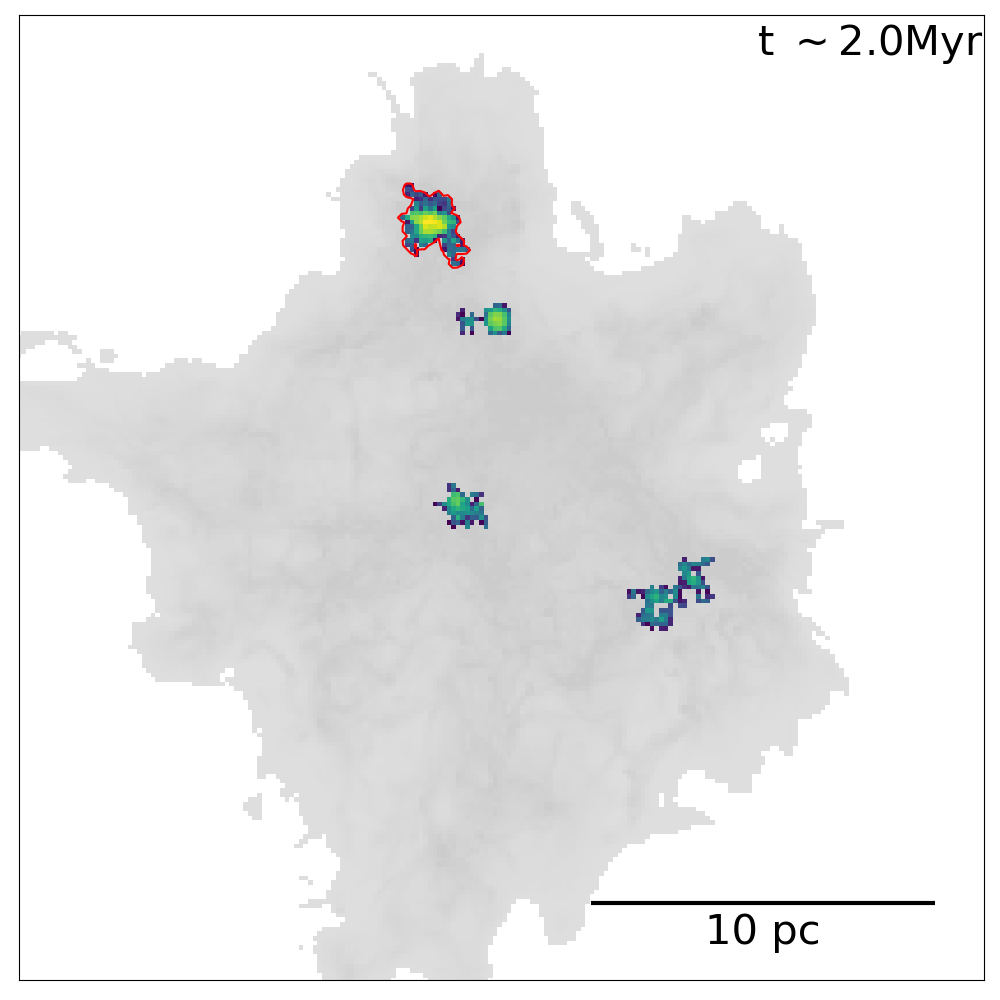}
    
    \includegraphics[width=0.33\linewidth]{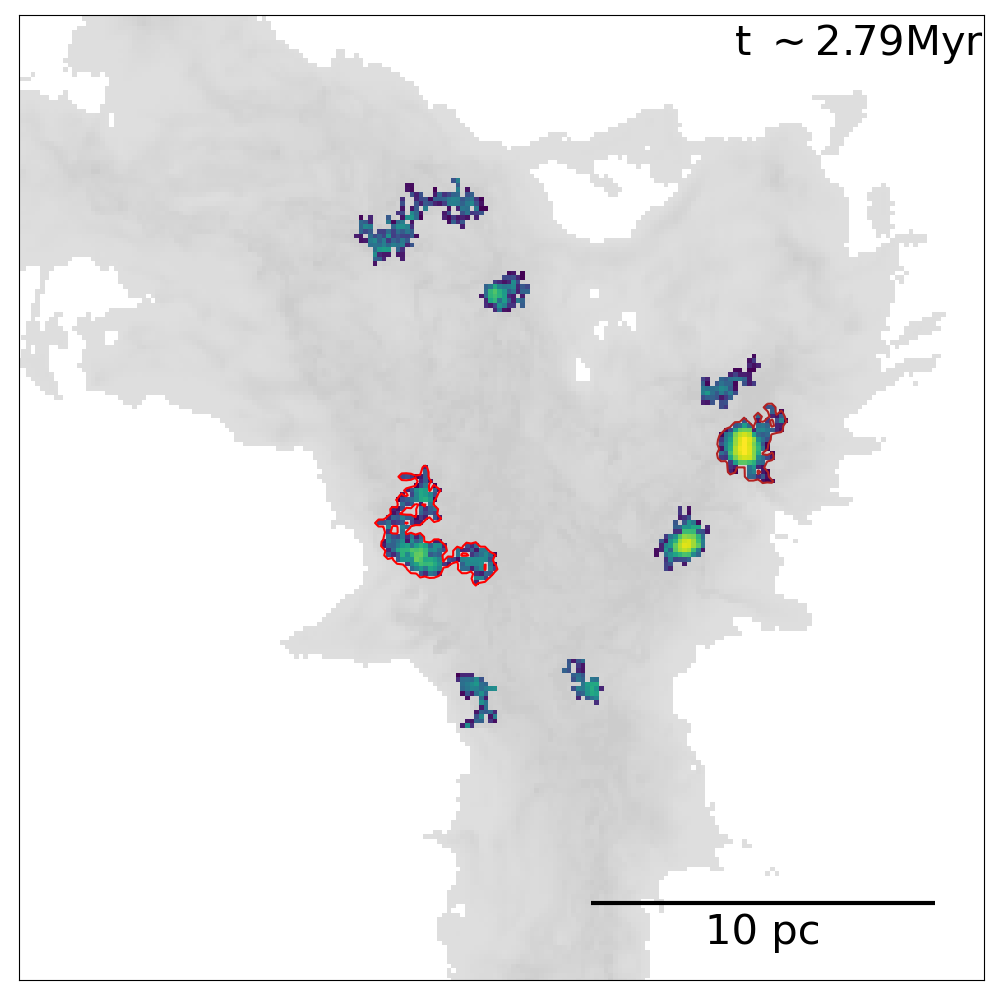}
    \includegraphics[width=0.33\linewidth]{images/trunk_contours/90_0/trunk_113_90_0.png}
    \includegraphics[width=0.33\linewidth]{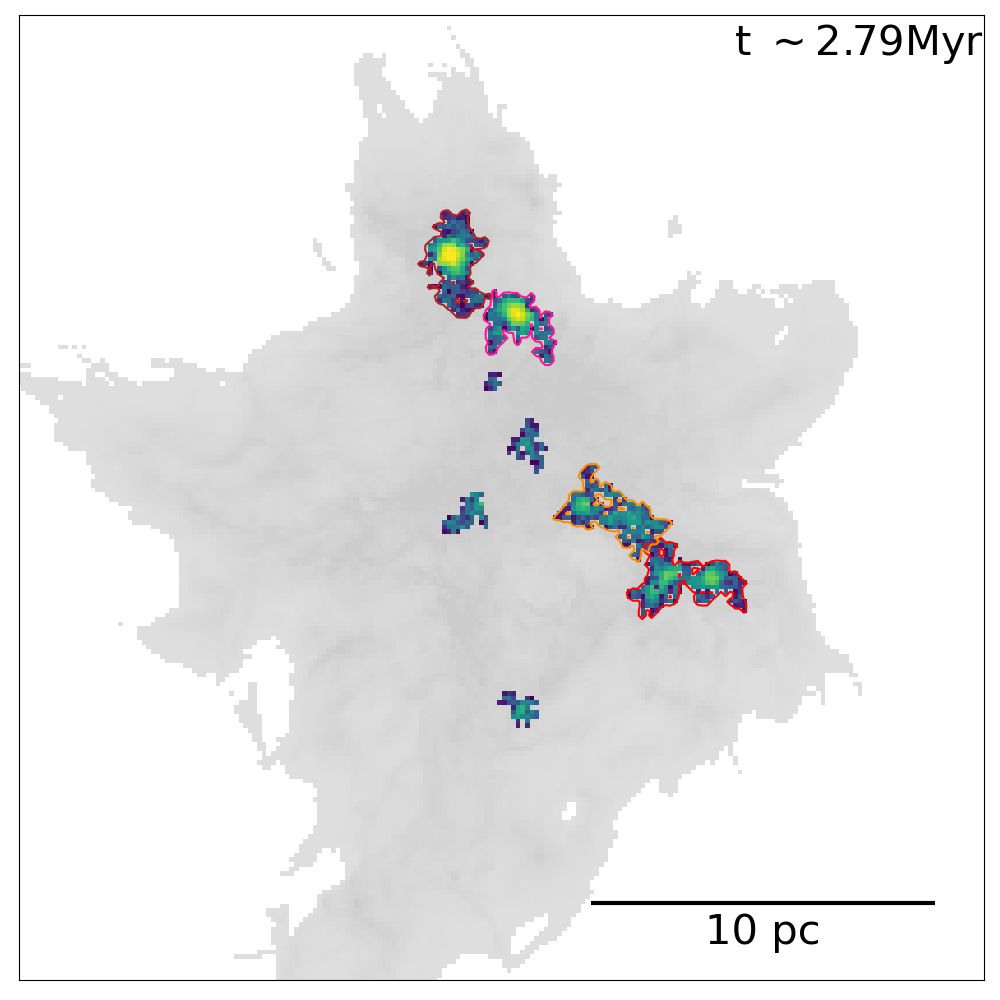}

    \caption{$^{13}$CO(2-1) moment 0 maps for different snapshots. The three columns represent the clouds projected along the x, $y$ and $z$ axes, respectively. The background greyscale represents H$_2$ gas density with $^{13}$CO(2-1) emission overlaid as viridis maps, and coloured contours represent different MCs (dendrogram trunks), with red contours representing the largest MCs ($R$) in the cube.}
    \label{fig: gmc mc 1 proj}
\end{figure*}

\begin{figure*}[htbp]
    \centering
    \includegraphics[width=0.33\linewidth]{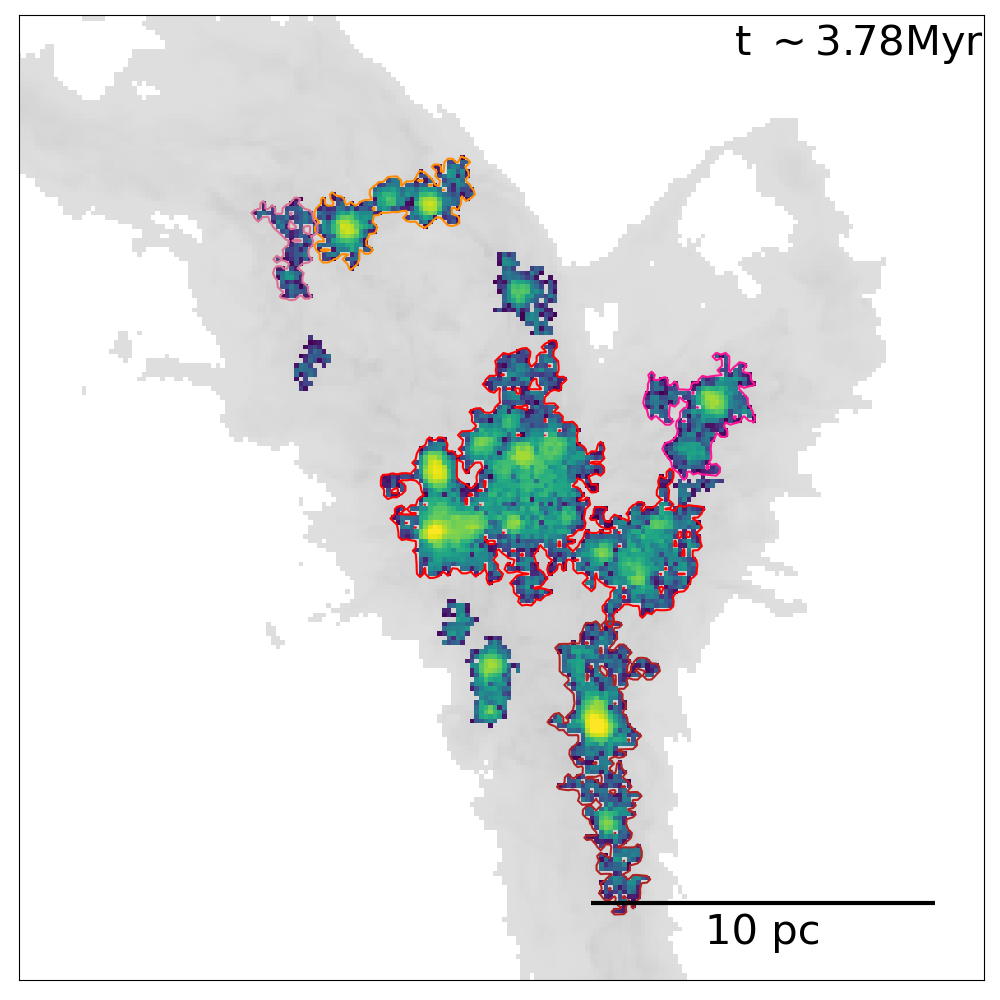}    
    \includegraphics[width=0.33\linewidth]{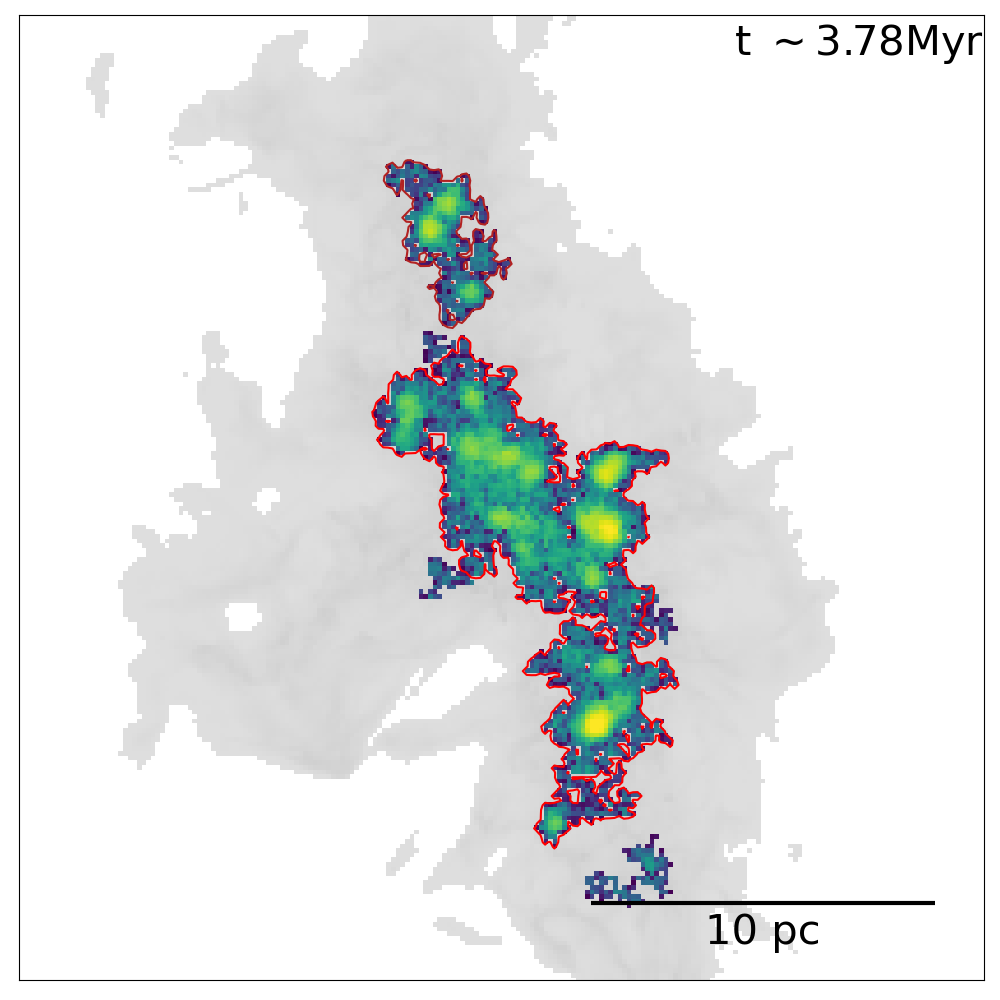}
    \includegraphics[width=0.33\linewidth]{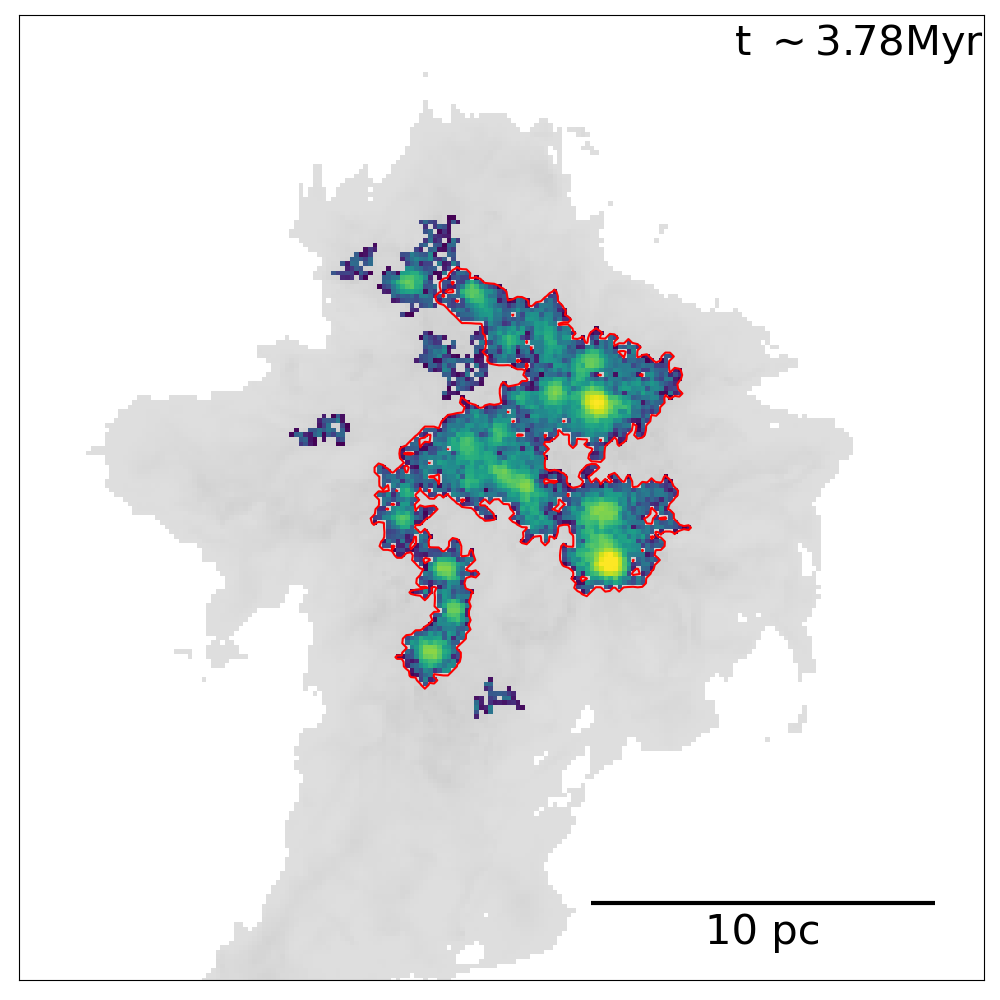}
    
    \includegraphics[width=0.33\linewidth]{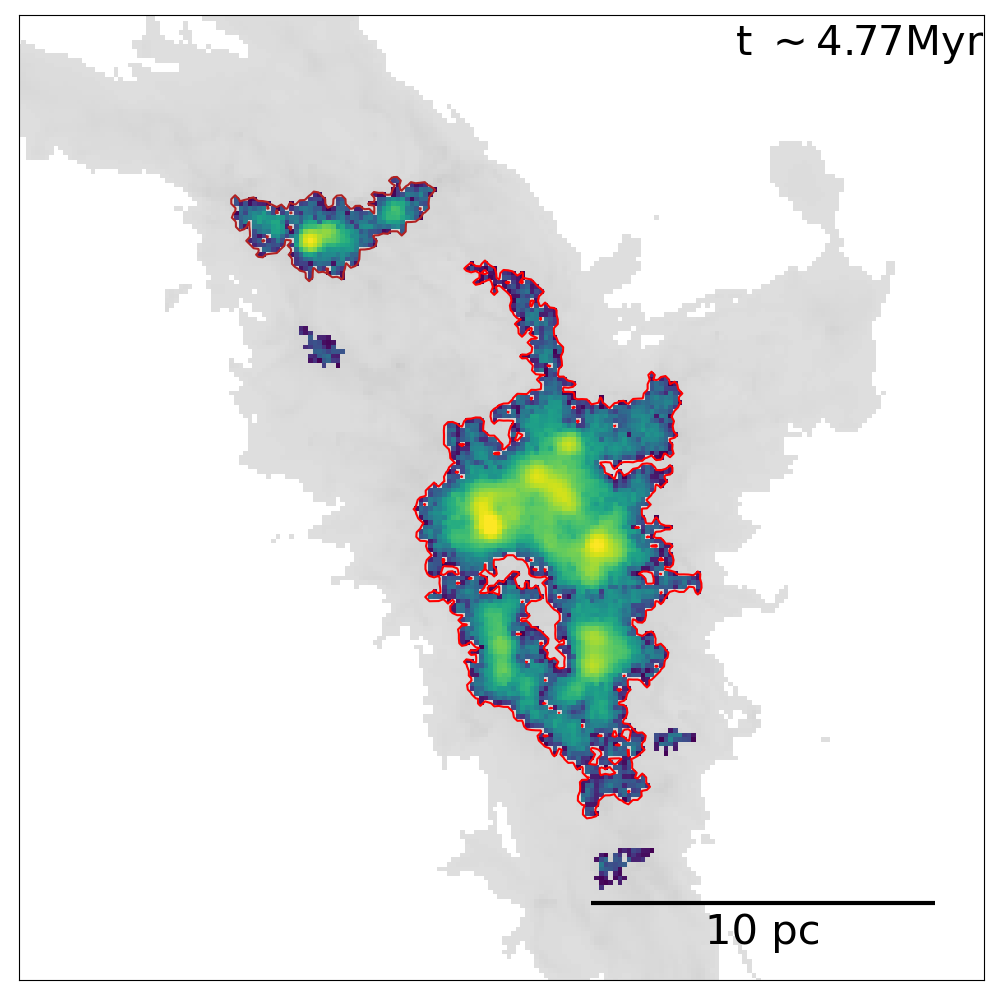}
    \includegraphics[width=0.33\linewidth]{images/trunk_contours/90_0/trunk_193_90_0.png}
    \includegraphics[width=0.33\linewidth]{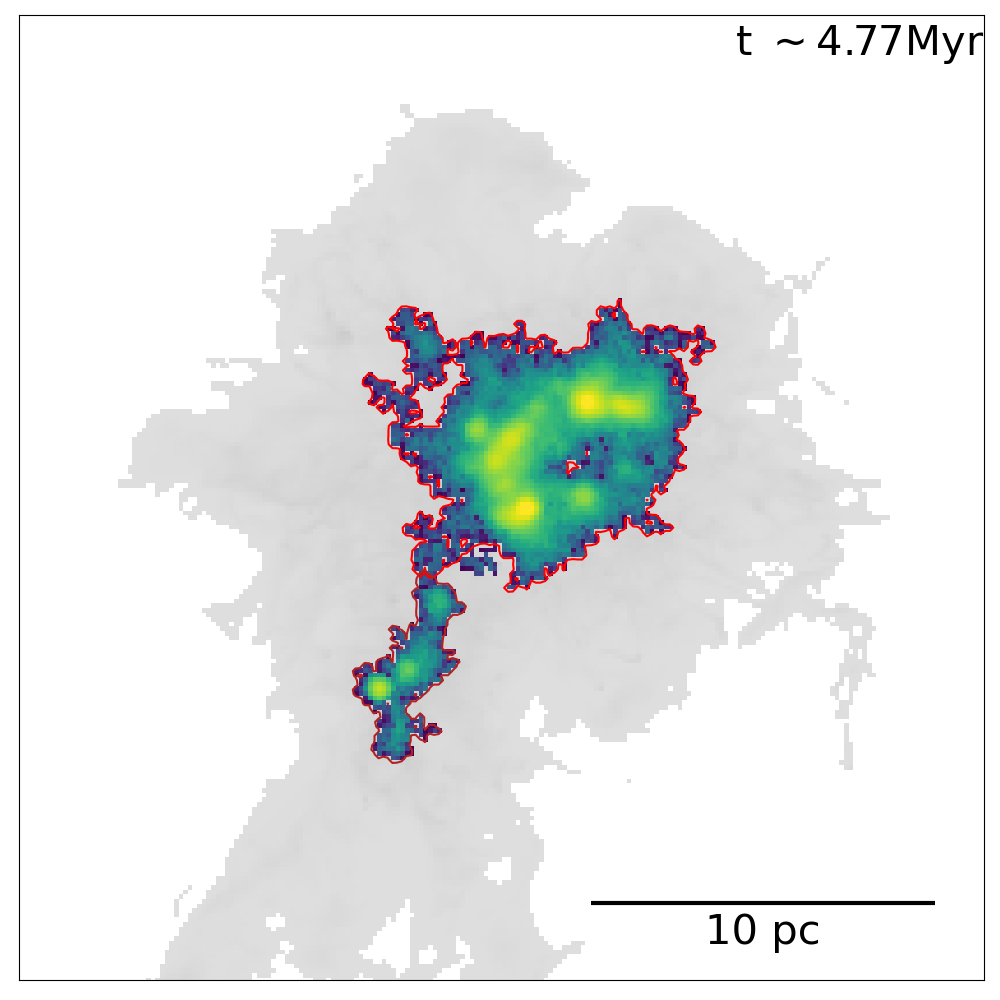}  

    \includegraphics[width=0.33\linewidth]{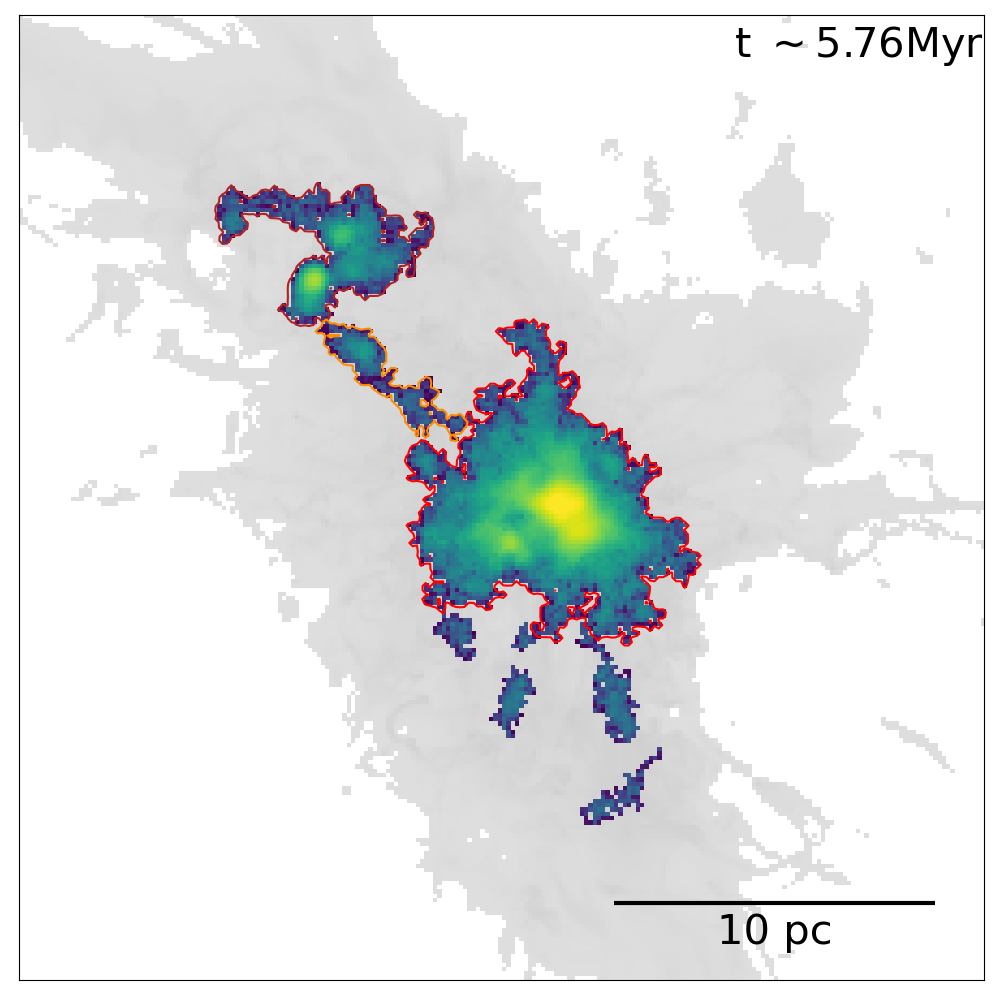}
    \includegraphics[width=0.33\linewidth]{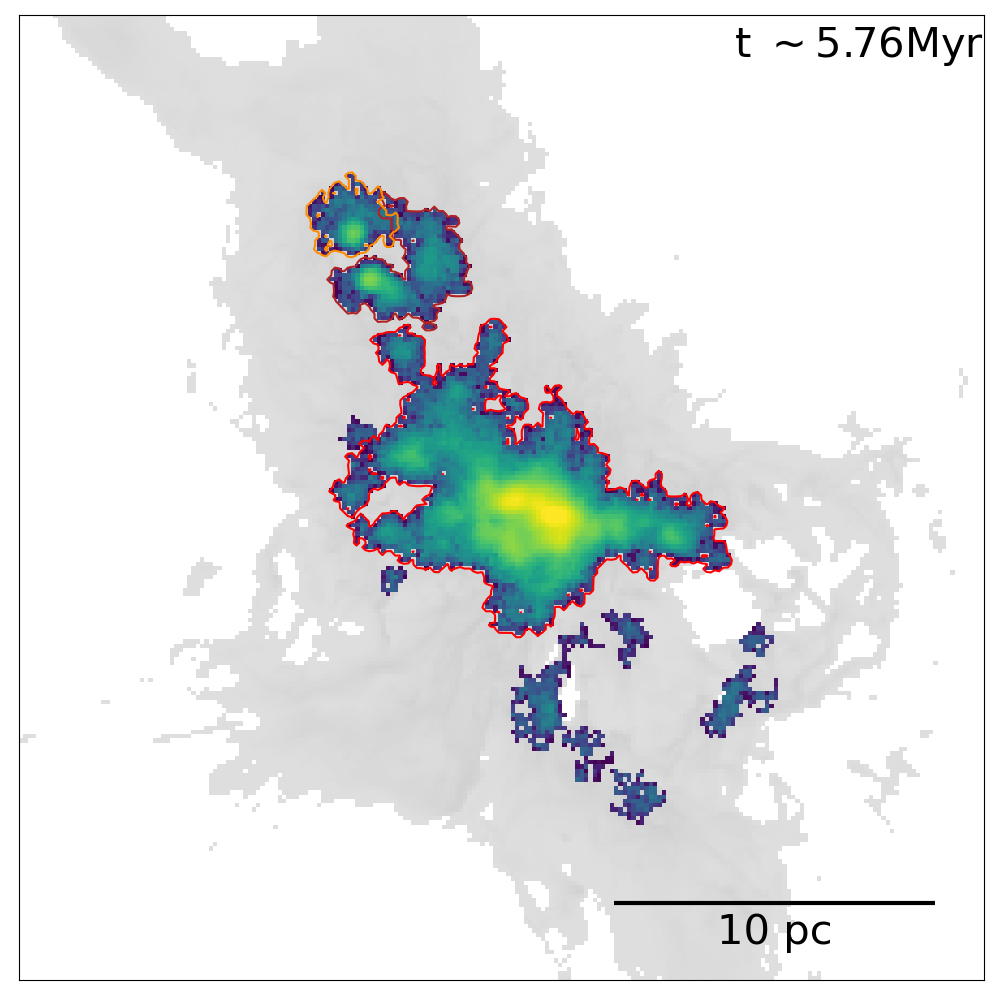}
    \includegraphics[width=0.33\linewidth]{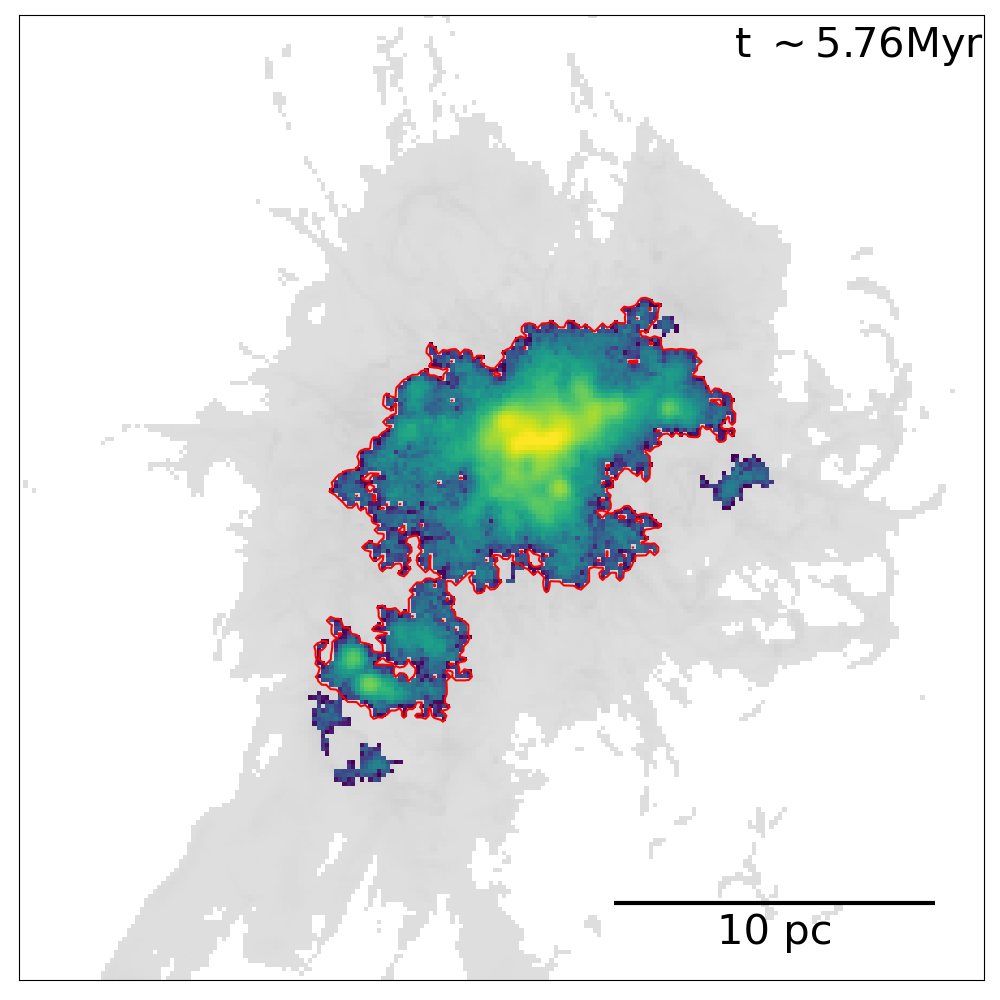}    

    \includegraphics[width=0.33\linewidth]{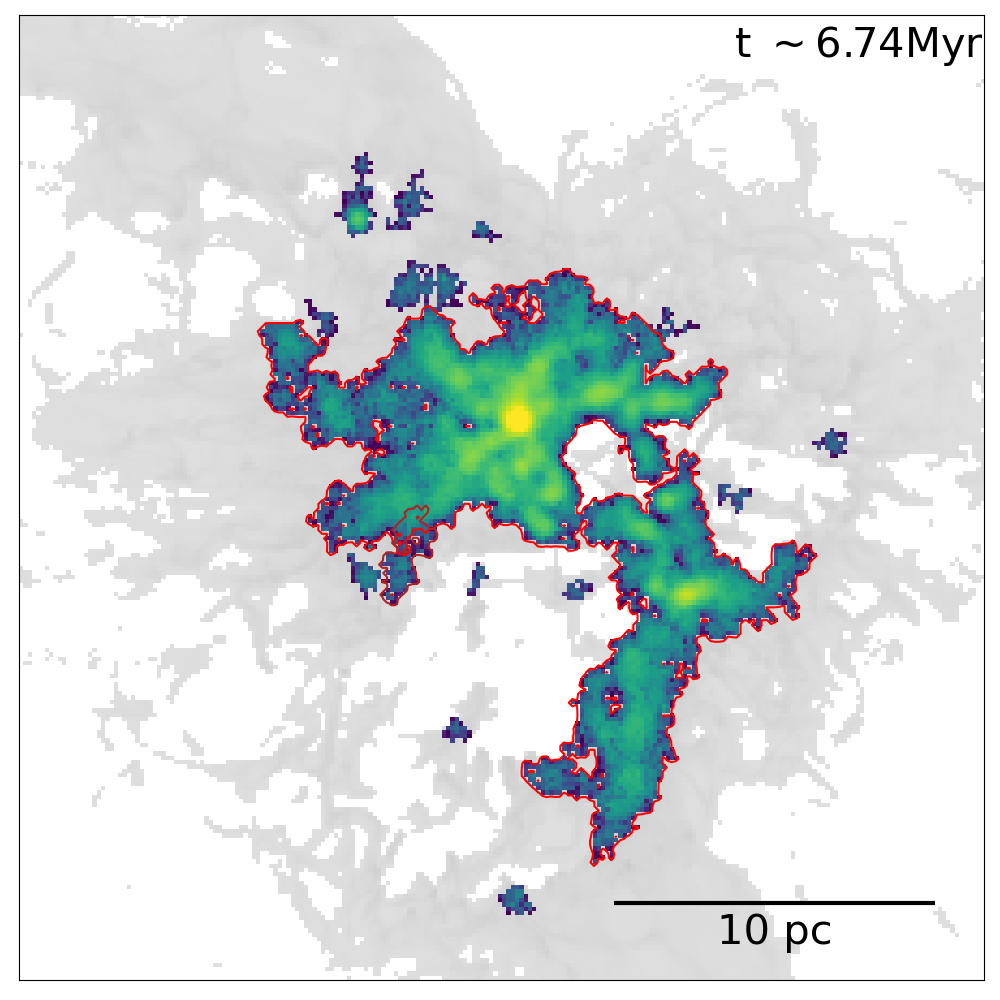}
    \includegraphics[width=0.33\linewidth]{images/trunk_contours/90_0/trunk_273_90_0.png}
    \includegraphics[width=0.33\linewidth]{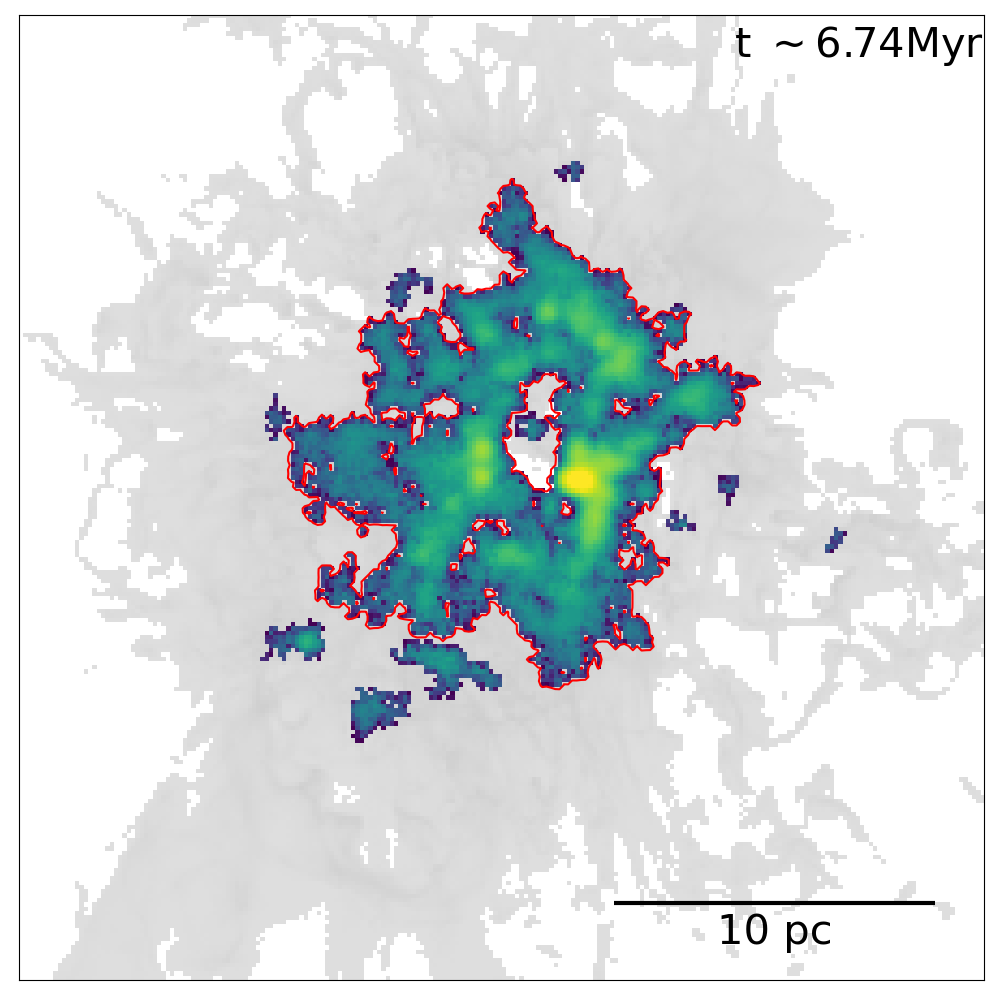}

    \caption{$^{13}$CO(2-1) moment 0 maps for different snapshots. The colors and symbols follow fig. \ref{fig: gmc mc 1 proj}.}
    \label{fig: gmc mc 2 proj}
\end{figure*}

\begin{figure*}[htbp]
    \centering
    \includegraphics[width=0.33\linewidth]{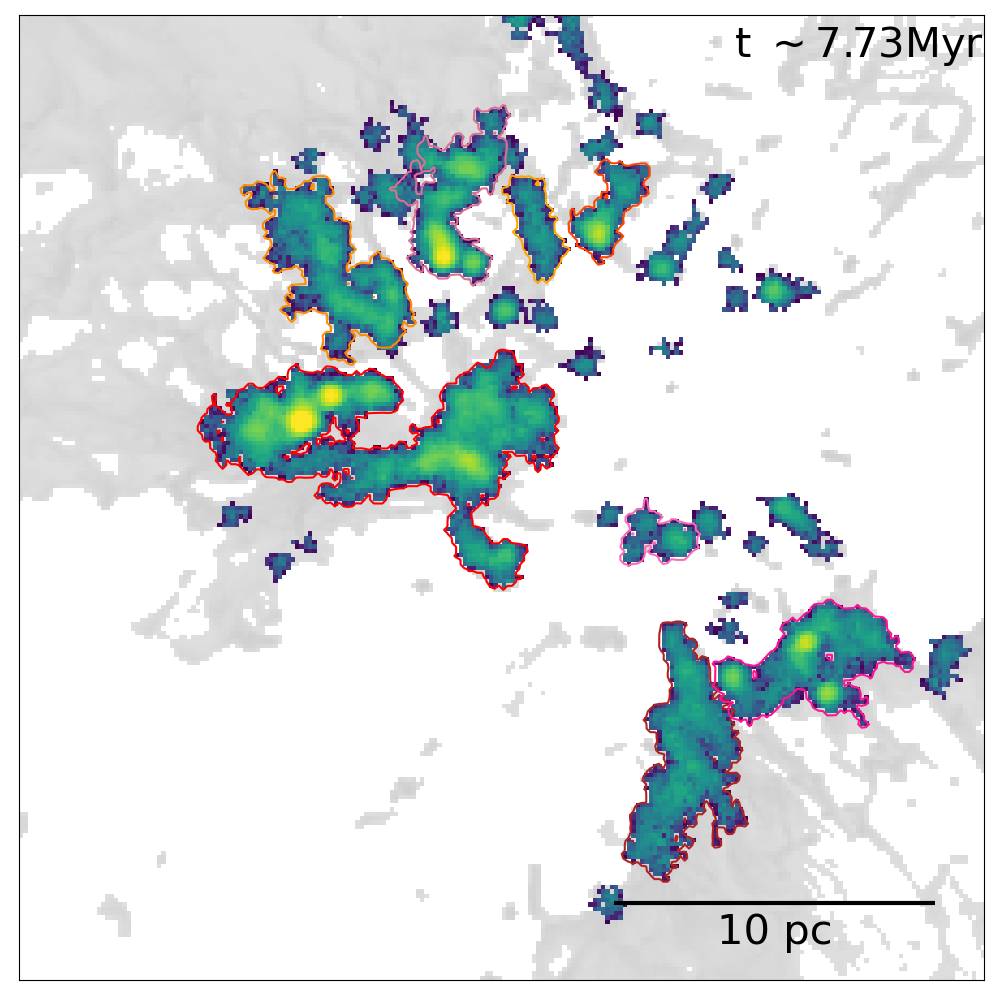}    
    \includegraphics[width=0.33\linewidth]{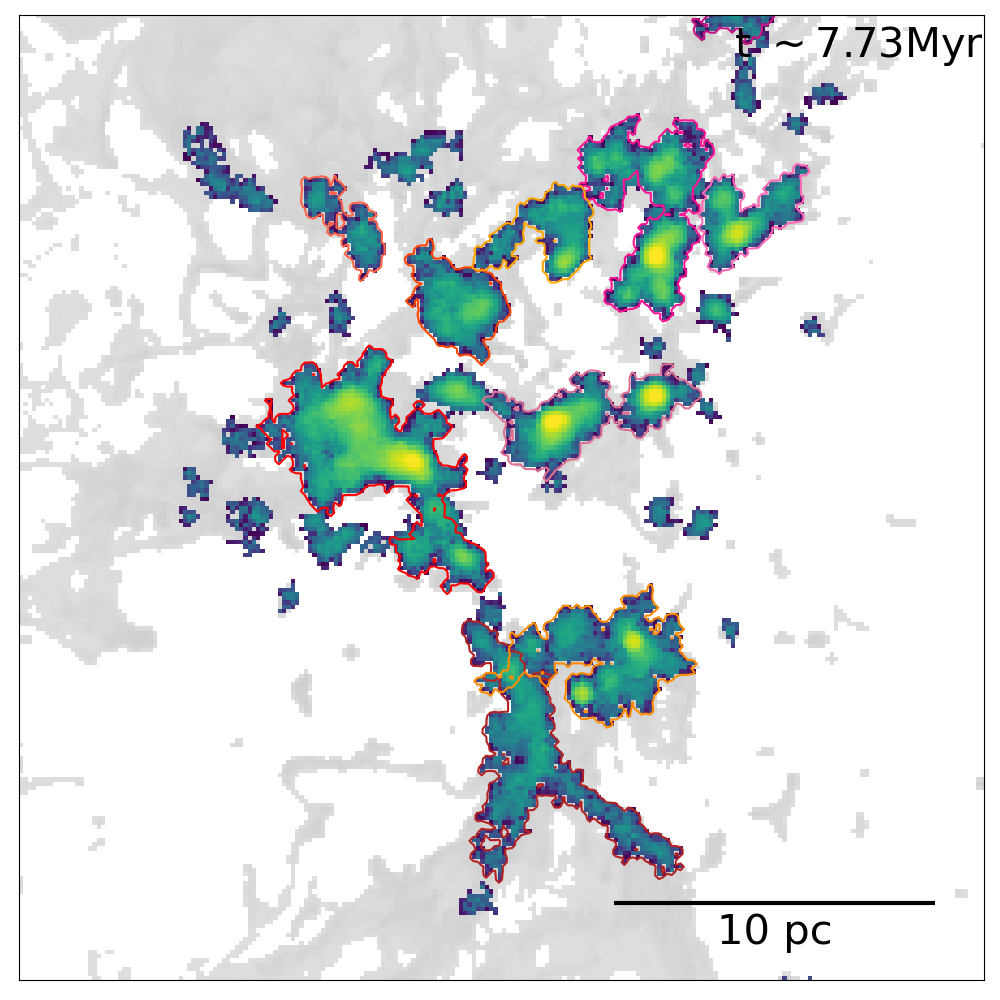}
    \includegraphics[width=0.33\linewidth]{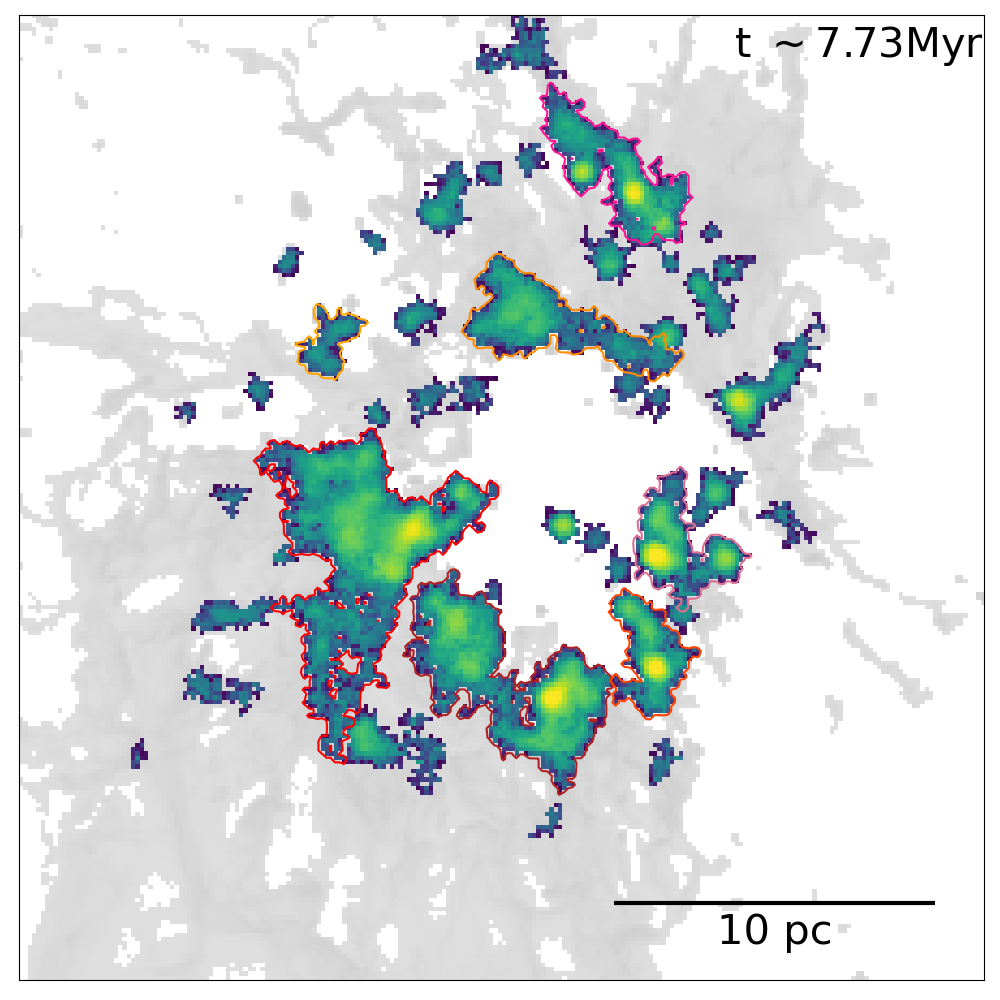}

    \includegraphics[width=0.33\linewidth]{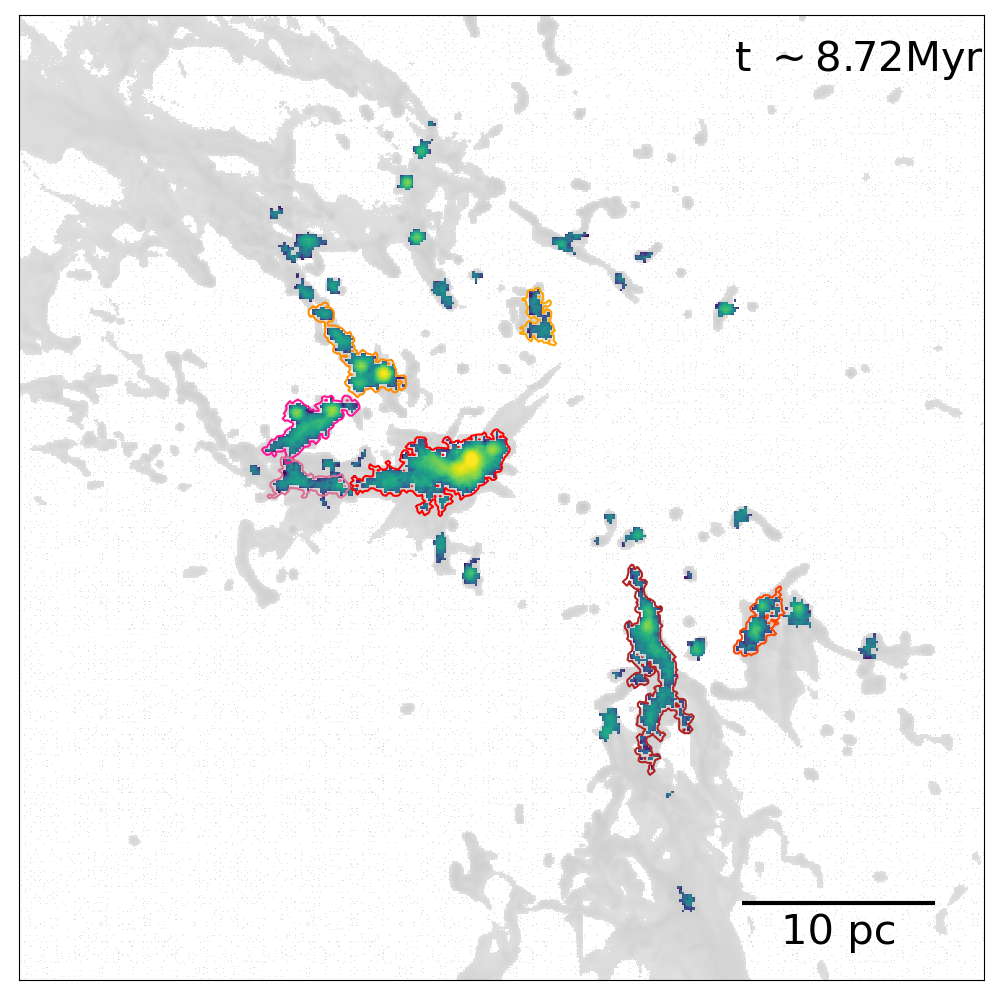}
    \includegraphics[width=0.33\linewidth]{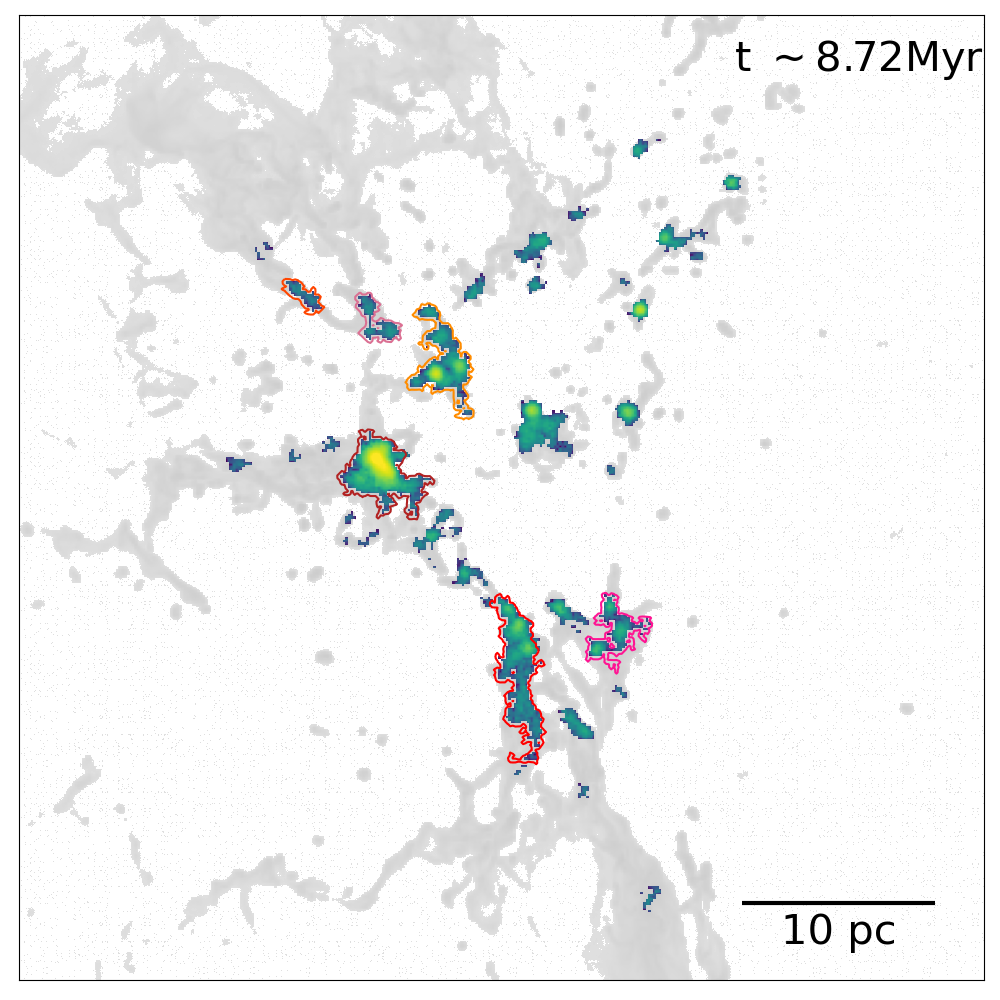}
    \includegraphics[width=0.33\linewidth]{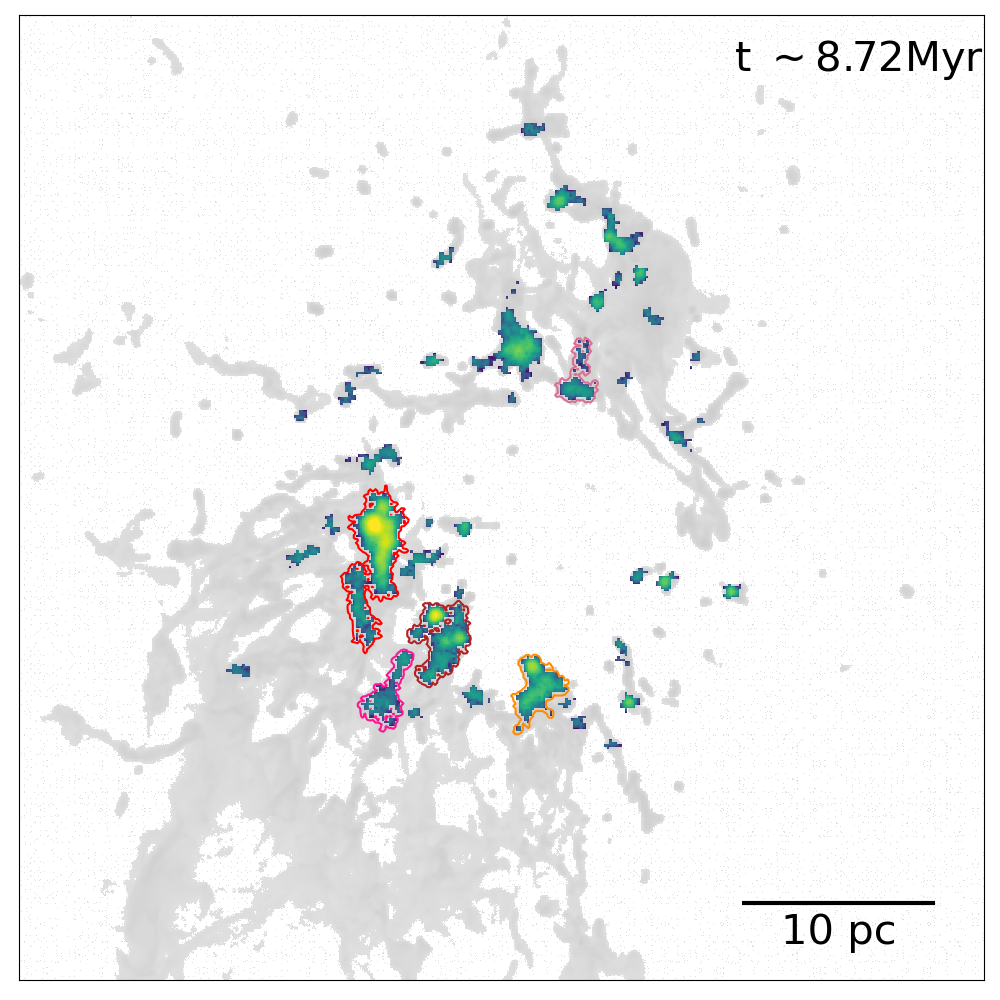}

    \includegraphics[width=0.33\linewidth]{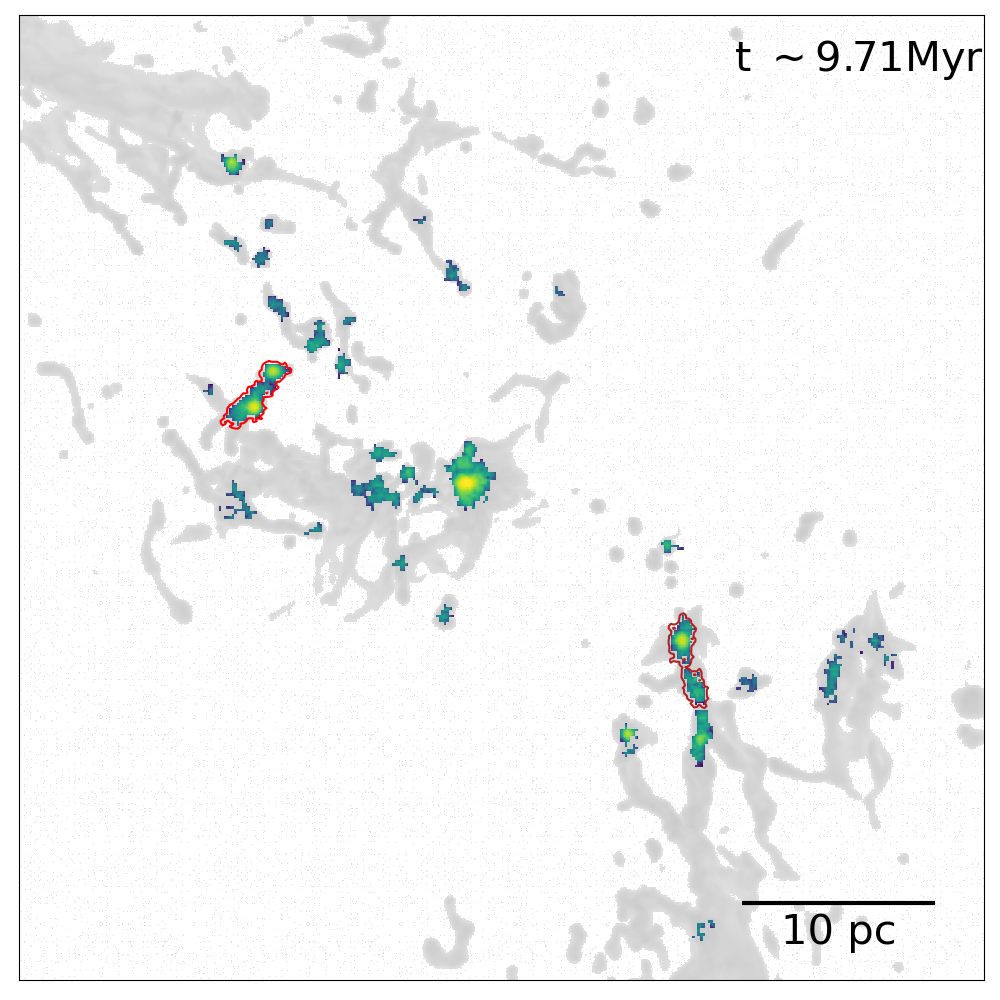}
    \includegraphics[width=0.33\linewidth]{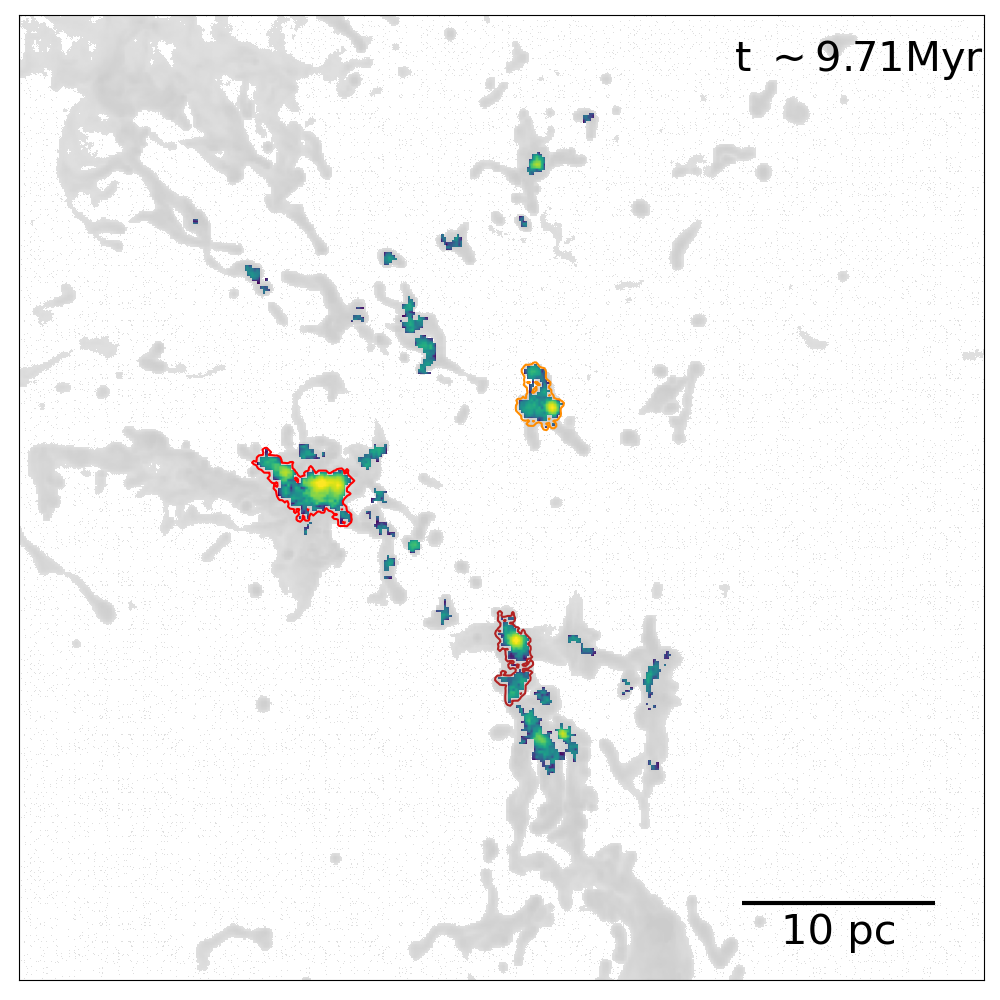}
    \includegraphics[width=0.33\linewidth]{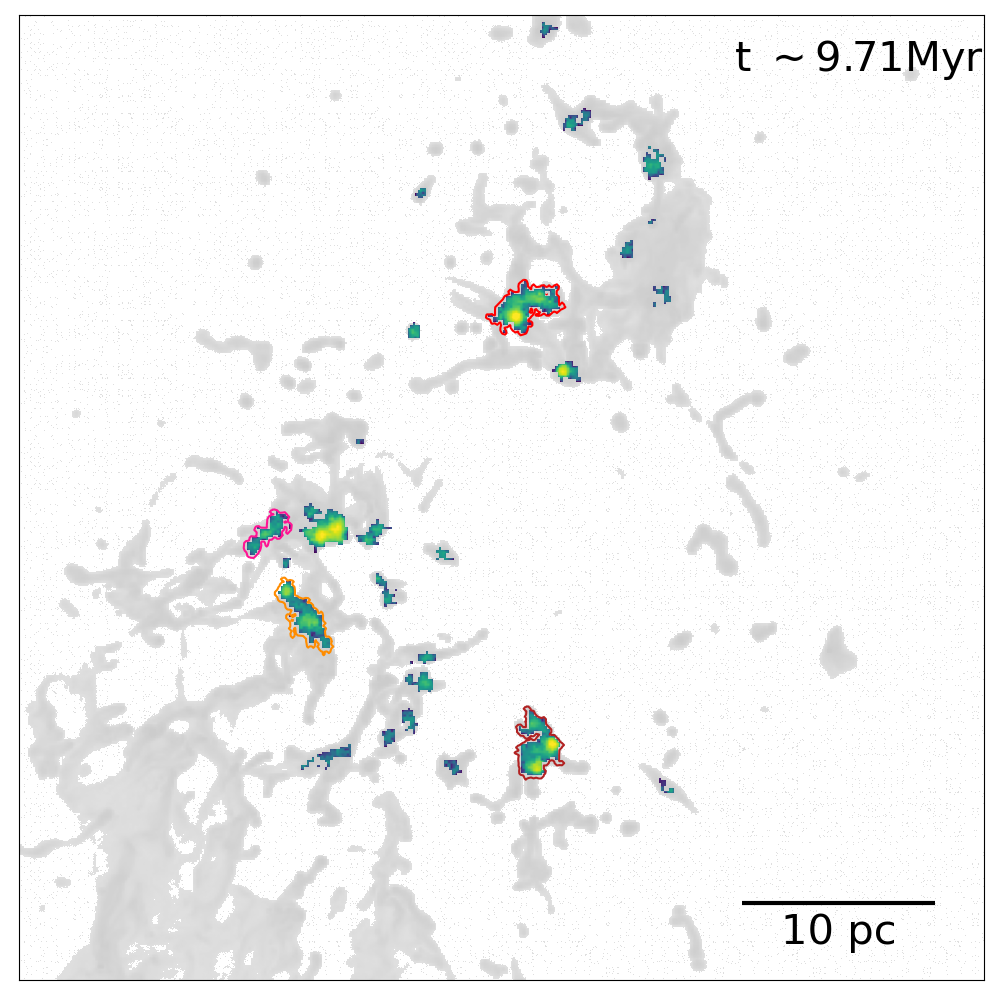}

    \includegraphics[width=0.33\linewidth]{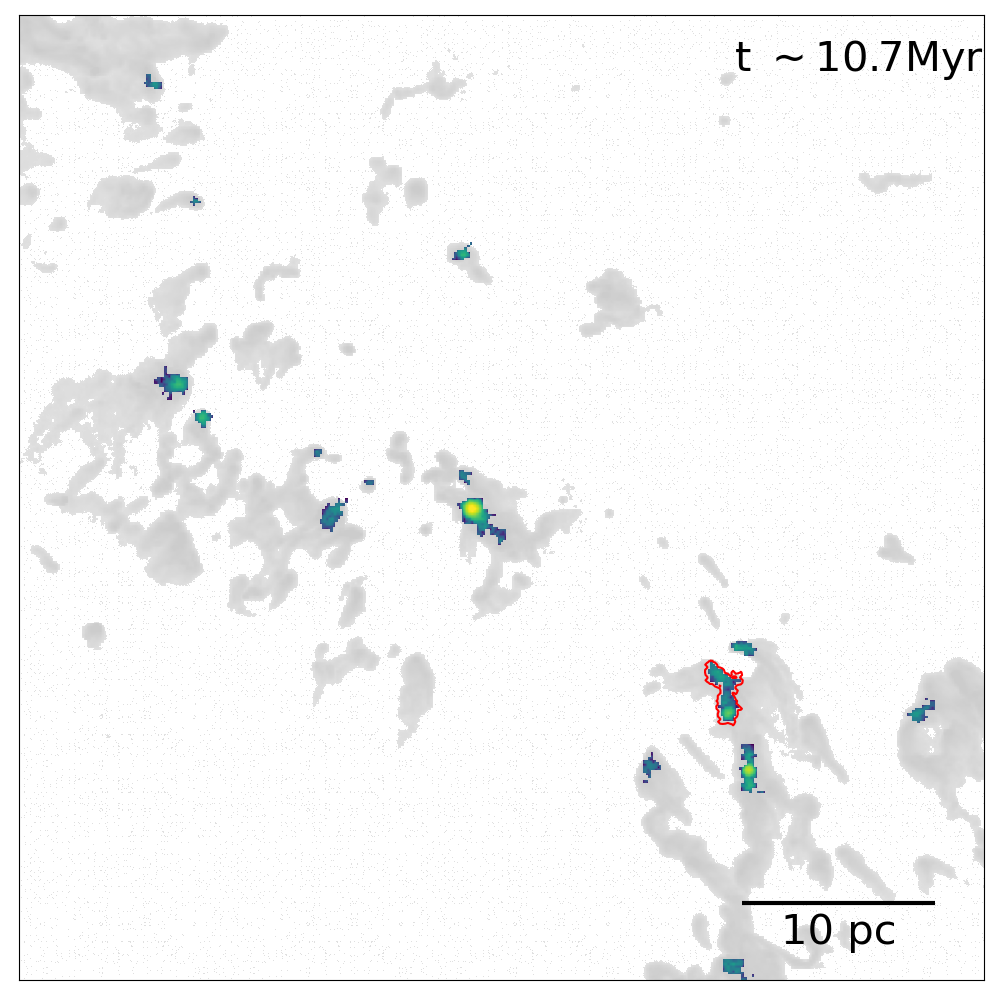}    
    \includegraphics[width=0.33\linewidth]{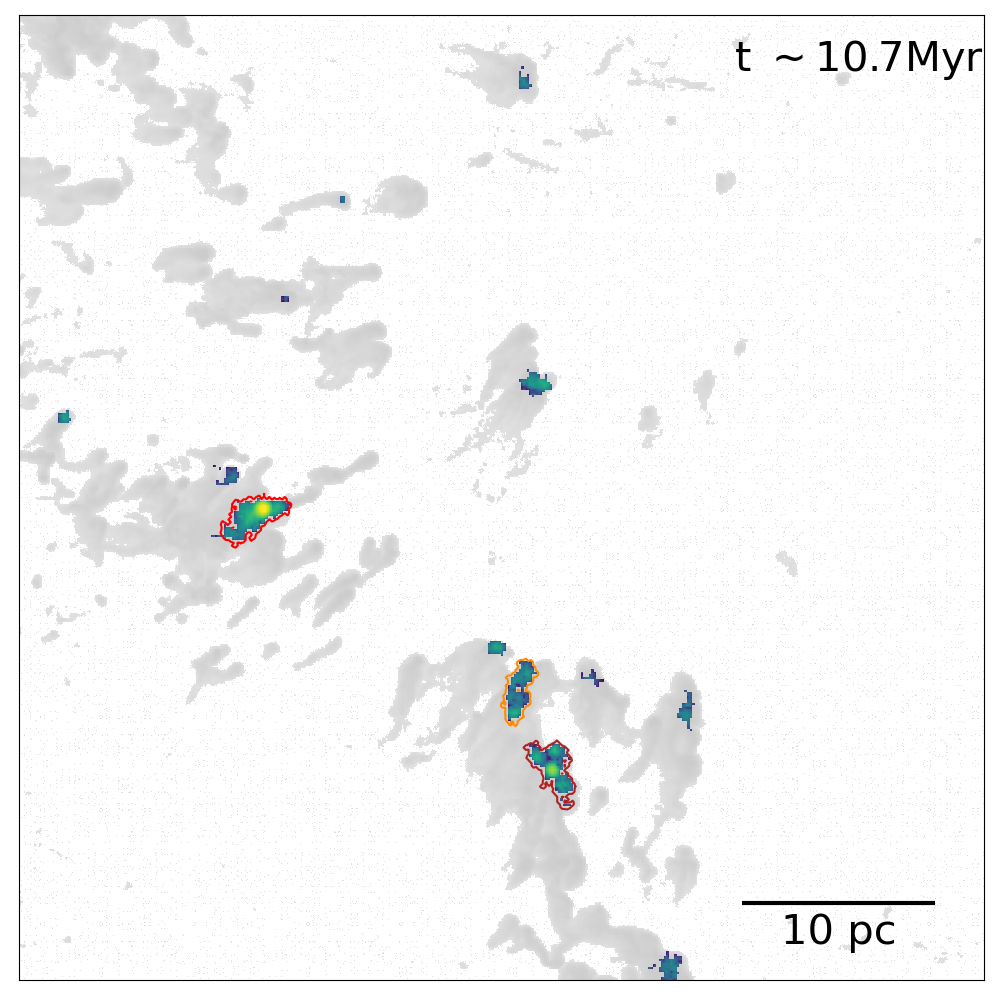}
    \includegraphics[width=0.33\linewidth]{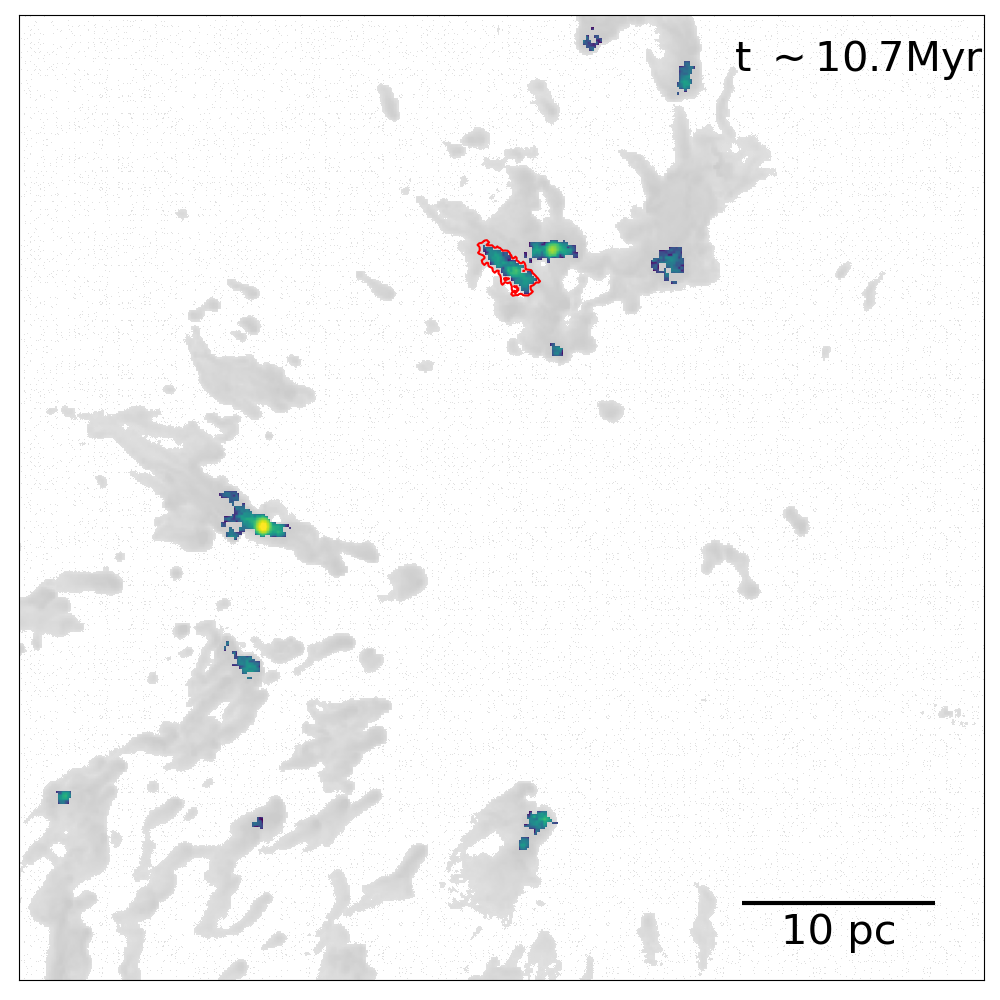}
    \caption{$^{13}$CO(2-1) moment 0 maps for different snapshots. The colors and symbols follow fig. \ref{fig: gmc mc 1 proj}.}
    \label{fig: gmc mc 3 proj}
\end{figure*}

In this section, we show a sequence of the $^{13}$CO(2-1) moment 0 maps for the GMC (i.e. the molecular gas complexes) as it evolves along with the dendrogram trunks (MCs).
The MC obtained by projecting along three orthagonal axes (Sect. \ref{subsec: radmc3d}) are shown in figures \ref{fig: gmc mc 1 proj} - \ref{fig: gmc mc 3 proj}. We also provide videos that represent all snapshots along the three projections as ancillary materials. This helps us to visualise the clouds and understand it's structure at different evolutionary stages. 
Fig. \ref{fig: gmc mc 1 proj} shows the formation of MCs as small diffuse structures. Fig. \ref{fig: gmc mc 2 proj} shows a single (or few) contour(s) that cover the entire emission in the viridis, representing most of the observed MCs. The filamentary, fractal, and complex nature of these structures is also visible in the emission maps. 
Fig. \ref{fig: gmc mc 3 proj} shows the MCs that are significantly impacted by stellar feedback processes. These lead to gas expulsion and dispersion, thus presenting the $^{13}$CO(2-1) as a 3D bubble. As more and more gas are dispersed, the number of MCs decreases. Some of these late MCs represent the early structures (Fig. \ref{fig: gmc mc 1 proj}), with the difference that the early MCs have accompanying H$_2$ gas. The absence of a molecular gas prevents the formation of MCs and ends the simulation.

\section{Inclusion of CO chemistry}\label{app: uclchem}

\begin{figure}
    \centering
    \includegraphics[width=0.7\linewidth]{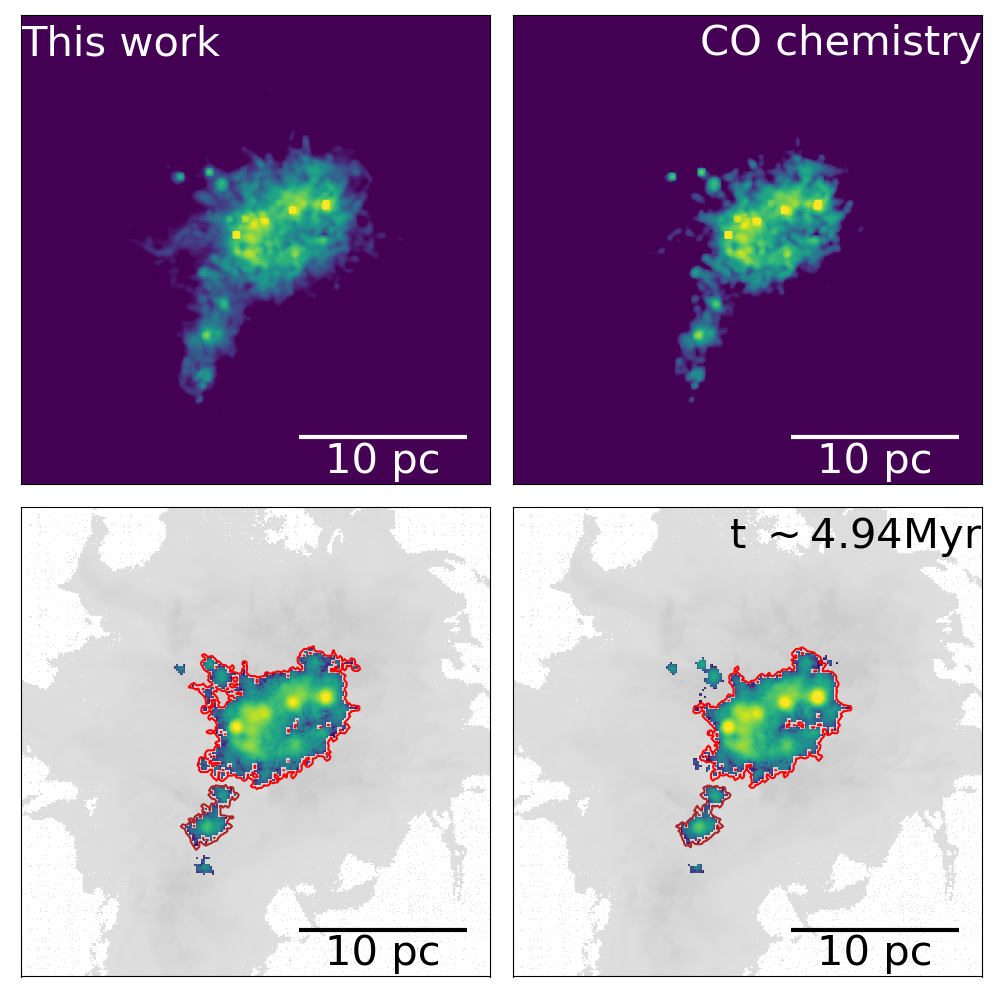}
    \caption{Comparison between the $^{13}$CO(2-1) ppv cube used in this work and created by including CO chemistry for snapshot 200 (4.94 Myr). The top rows show the RADMC-3D output and the bottom show the masked cubes with MCs (contours) overlayed on the $^{13}$CO(2-1) emission (viridis) and projected H$_2$ density (greyscale) maps.}
    \label{fig: ucl 200}
\end{figure}

\begin{figure}
    \centering
    \includegraphics[width=0.7\linewidth]{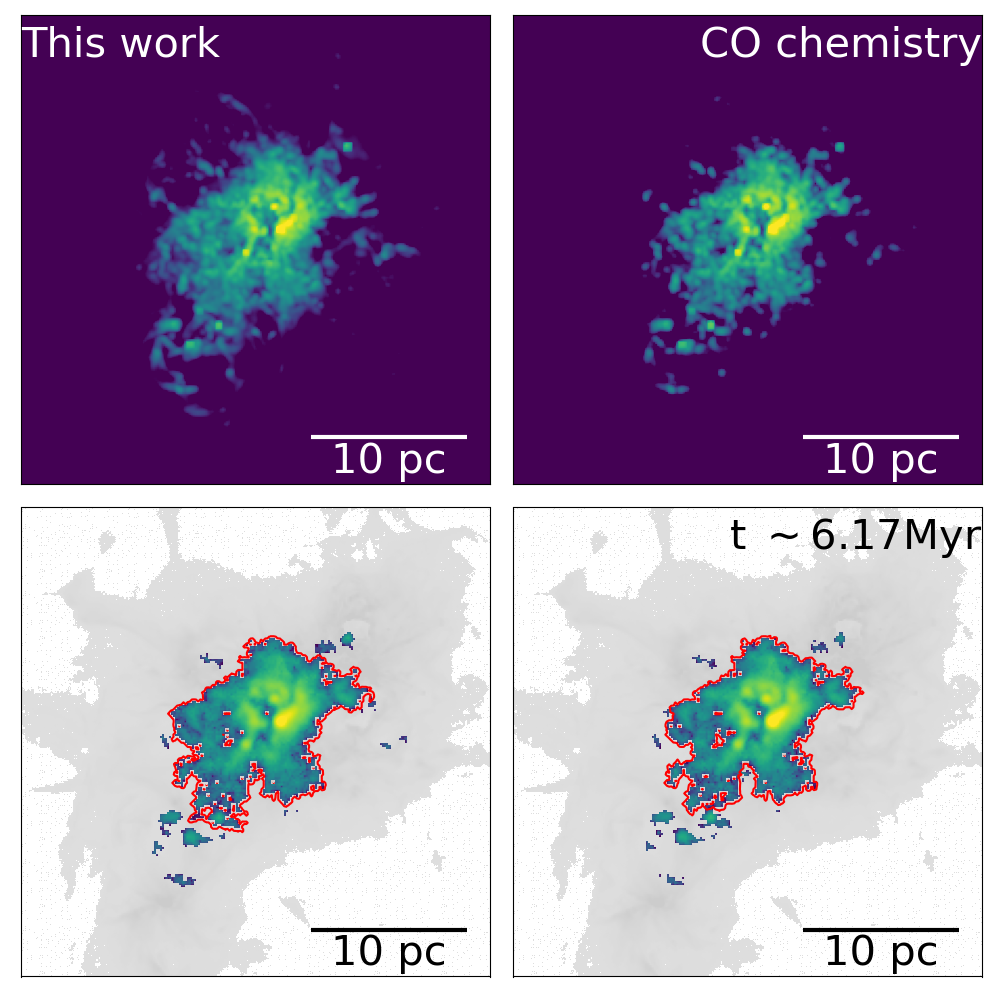}
    \caption{Comparison between the $^{13}$CO(2-1) ppv cube used in this work and created by including CO chemistry for snapshot 250 (6.17 Myr). The symbols and notations follow Fig. \ref{fig: ucl 200}.}
    \label{fig: ucl 250}
\end{figure}

\begin{figure}
    \centering
    \includegraphics[width=0.7\linewidth]{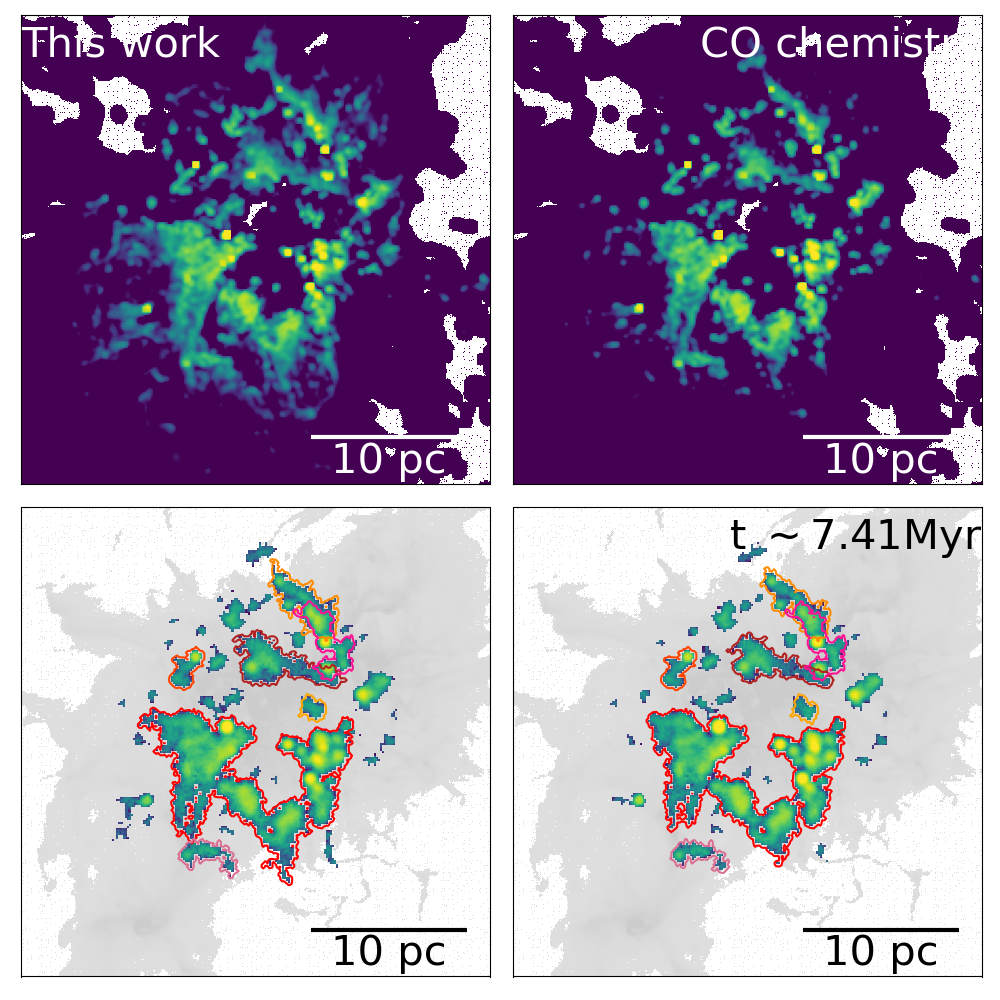}
    \caption{Comparison between the $^{13}$CO(2-1) ppv cube used in this work and created by including CO chemistry for snapshot 300 (7.41Myr). The symbols and notations follow Fig. \ref{fig: ucl 200}.}
    \label{fig: ucl 300}
\end{figure}

\begin{figure}
    \centering
    \includegraphics[width=\linewidth]{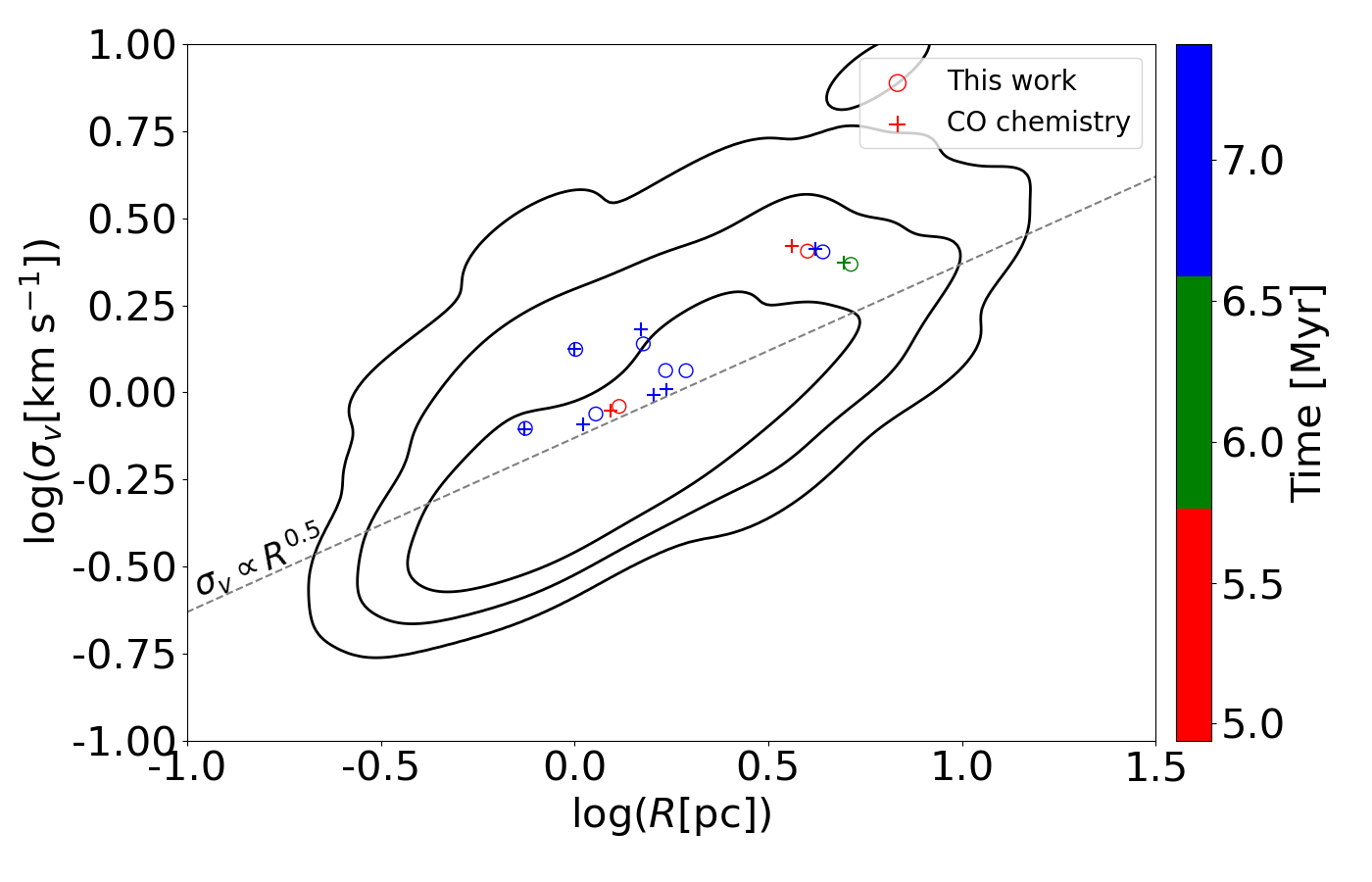}
    \caption{Size-linewidth relation ($\sigma_\varv$ versus $R$) for our MCs (circles) and those created by including CO chemistry (plus). The MCs have been extracted from three ppv cubes in Fig. \ref{fig: ucl 200} - \ref{fig: ucl 300}.
    The symbols and notations are consistent with Fig. \ref{fig:sedigism larson}.}
    \label{fig:larson ucl}
\end{figure}

\begin{figure}[htbp]
    \centering
    \includegraphics[width=\linewidth]{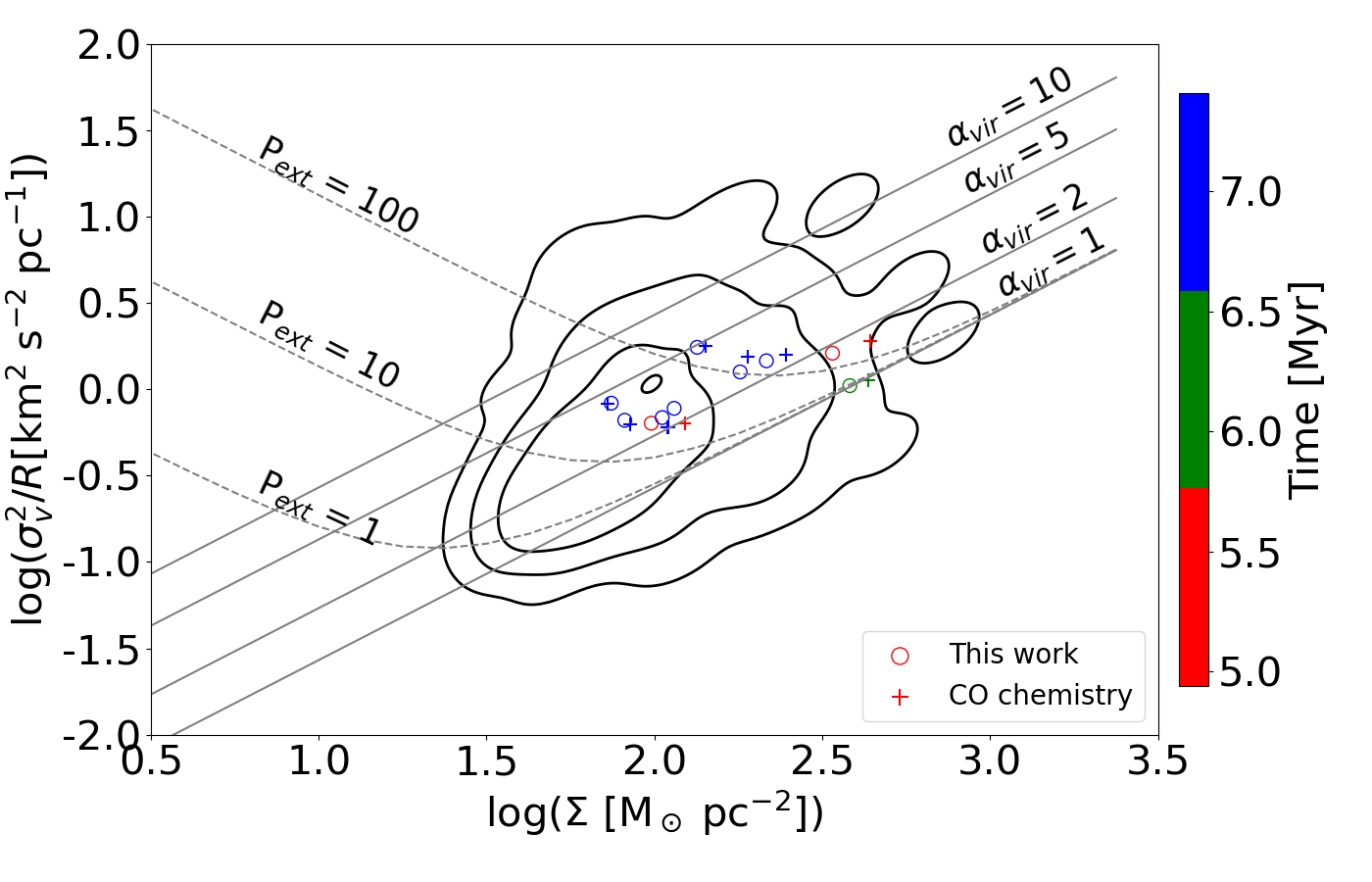}
    \caption{Scaling relation between $\sigma_\varv^2$/$R$ and surface mass density ($\Sigma$). The symbols and notations follow Fig. \ref{fig:larson ucl}. The solid gray lines represent isocontours of virial parameters. The dashed lines represent $\alpha_\mathrm{vir} = 1$ when including an external pressure P$_\mathrm{ext}$ = 1, 10, 100 $\mathrm{M_\odot \; pc ^{-3} \; km^2 \; s^{-2}}$. }
    \label{fig:heyer ucl}
\end{figure}

We post-processed the \starforge simulations with UCLCHEM \citep{holdship2017AJ....154...38H} chemical code\footnote{The pipeline is provided here: \url{https://github.com/psharda/gizmo_carver/tree/pschanges}} to estimate the abundance of CO (Sharda et al. in prep). However, due to computational cost, this has only been possible for three snapshots.
Figures \ref{fig: ucl 200} -- \ref{fig: ucl 300} show that although our fiducial approach slightly overestimates the $^{13}$CO emission, the data processing steps produce MCs of comparable size and morphology in both cases. This is further highlighted in Fig. \ref{fig:larson ucl} \& \ref{fig:heyer ucl}, which show that both sets of MCs have similar properties in the same snapshots. The inclusion of chemistry might cause a small difference in the properties of our MCs and result in smaller MCs not being detectable. However, since we study the trends in the distribution of properties over time, we expect that these are not significantly influenced by the absence of CO chemistry.

\section{Hierarchichal and isolated trunks} \label{app: hier isolated}

\begin{figure*}[htbp]
    \centering
    \includegraphics[width=\linewidth]{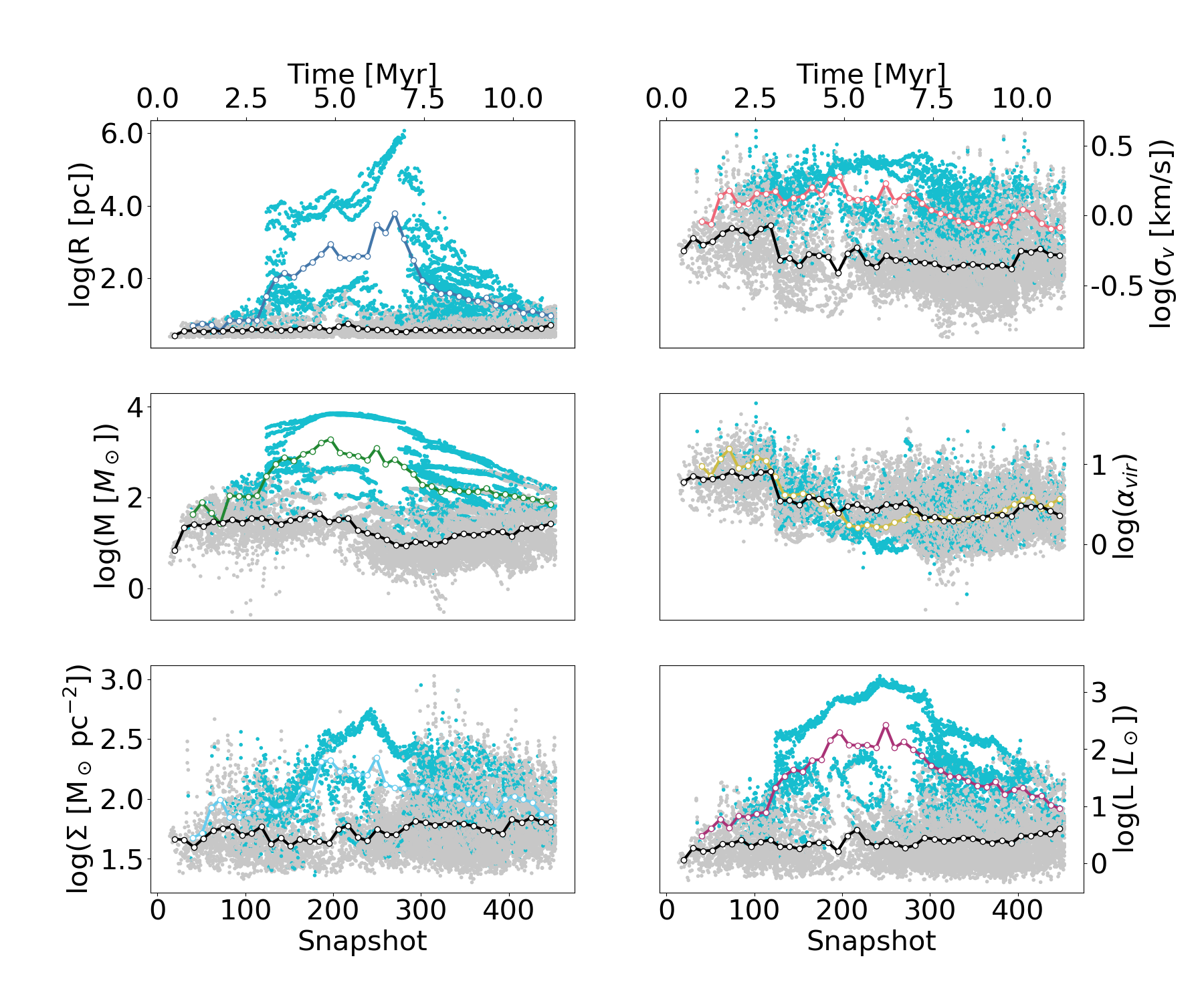}
    \caption{The cyan points represents the hierarchichal trunks (branches) and grey points represents the isolated trunks (leaves). The colored lines represents the average distribution of hierarchichal trunks, with the colors following Fig. \ref{fig: prop vs snap}. The black line represents the average distribuion for isolated trunks.}
    \label{fig: hier isol prop vs snap scatter}
\end{figure*}

This section explains the reason for selecting only trunks that are branches as molecular clouds (MCs), rather than including all trunks. In figure \ref{fig: hier isol prop vs snap scatter}, we show the distribution of properties for all dendrogram trunks. The results clearly demonstrate that only hierarchical trunks (branches) exhibit consistent trends in their properties as they evolve. In contrast, isolated trunks (leaves) have scattered distributions with no clear trends. This is likely because most of these isolated structures represent transient gas features that do not correspond to the fractal molecular clouds seen in observations (Fig. \ref{fig: gmc mc 1 proj}). Furthermore, the large sample of isolated trunks results in the average properties of the trunks (Fig. \ref{fig: hier isol prop vs snap scatter}, central line) that show no significant trends over time.

\section{Simulations with different initial turbulence}\label{sec: other simulations}

\begin{figure*}[htbp]
    \centering
    \includegraphics[width=\linewidth]{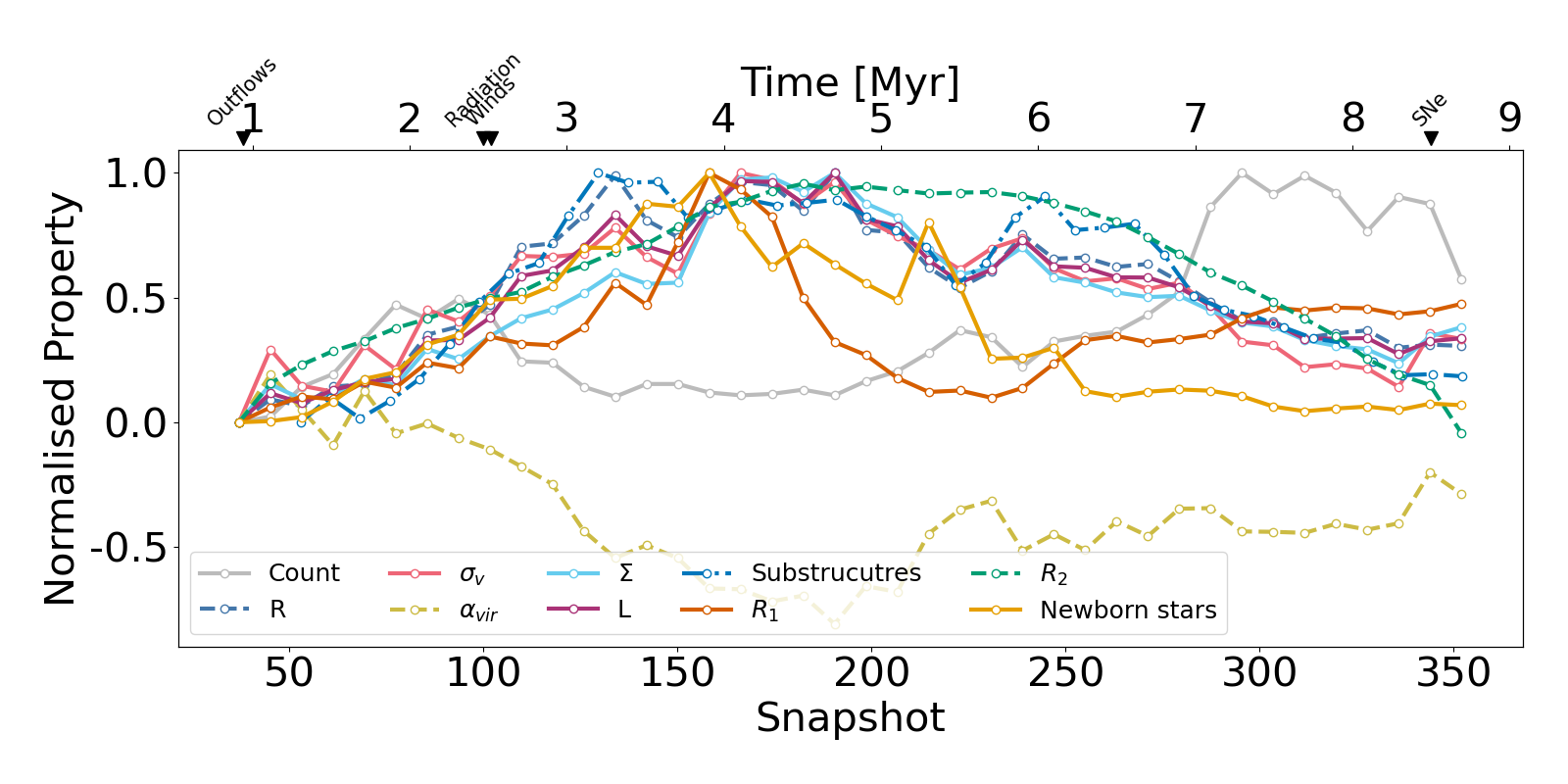}    
    \caption{Simulation with $\alpha_{vir} = 1$. Properties of the clouds at various evolutionary stages. The $R_1$ and $R_2$ values represent the morphologies of the molecular gas complexes. The symbols and colors follow fig. \ref{fig: prop vs snap} \& \ref{fig: frag vs snap}.}
    \label{fig: prop vs snap a1}
\end{figure*}

\begin{figure*}[htbp]
    \centering
    \includegraphics[width=\linewidth]{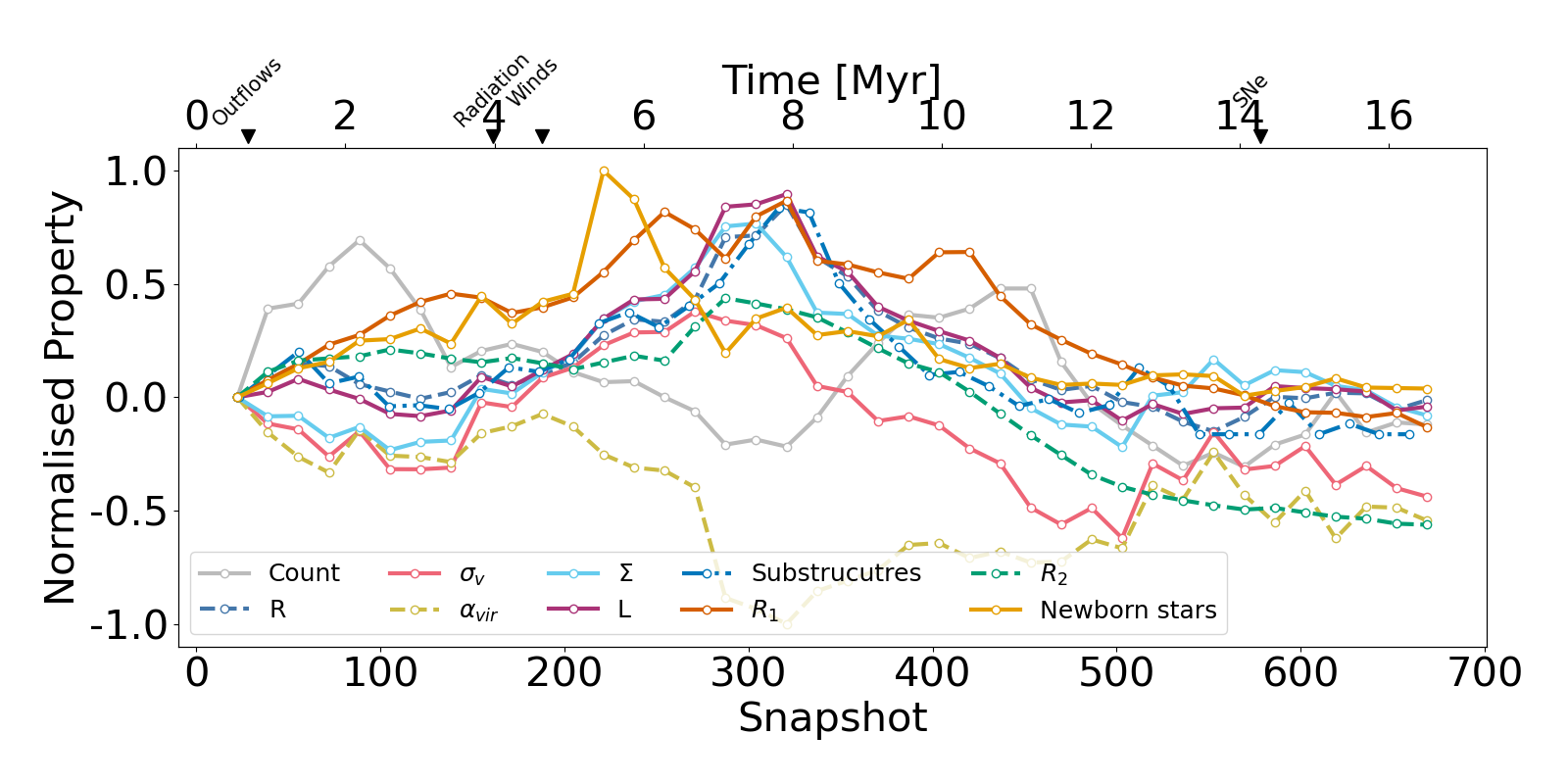}
    \caption{Simulation with $\alpha_{vir} = 4$. Properties of the clouds at various evolutionary stages. The symbols and colors follow fig. \ref{fig: prop vs snap a1}.}
    \label{fig: prop vs snap a4}
\end{figure*}

We perform our analysis on two other simulation sets. These have the same initial conditions as our fiducial runs,
with the exception of initial turbulence. These are M2e4a1 and M2e4a4 with $\alpha_{vir} = 1$ and $\alpha_{vir} = 4$, respectively. Figs. \ref{fig: prop vs snap a1} \& \ref{fig: prop vs snap a4} show the evolution of the MC properties for these two simulations, respectively. Although the GMC lifetime and the onset of different feedback mechanisms are different in the two simulations, they show a trend of increase in the MC properties representing actively growing MCs followed by a decrease in the properties due to MC dispersal by feedback. The morphology and fragementation trends of these MCs are consistent with the fiducial simulations.

\end{document}